\documentclass[preprint,12pt,authoryear]{elsarticle}

\usepackage{amssymb}
\usepackage{multirow}
\usepackage{float}
\restylefloat{figure}
\usepackage[version=3]{mhchem}
\usepackage{amsmath}
\usepackage{gensymb}
\usepackage{pifont}
\usepackage{MnSymbol}
\usepackage{subfig} 
\usepackage{dashrule}
\usepackage{bookmark}

\newcommand{\dbar}{\mbox{$D$\kern-0.8em\raise.4ex\hbox{--}}}

\journal{arXiv}

\begin{document}

\begin{frontmatter}

\title{Modelling the adsorption of fluoride onto activated alumina in the presence of other ions}

\author{Naga Samrat MVV}
\author{Gandhi K S}
\author{Kesava Rao K\corref{correspondingauthor}}
\cortext[correspondingauthor]{Corresponding author}
\ead{kesava@chemeng.iisc.ernet.in}

\address{Department of Chemical Engineering, Indian Institute of Science, Bengaluru 560012, India}

\begin{abstract}
Most of the studies on the adsorption of \ce{F-} are experimental, and have been done with synthetic solutions. Such studies rarely mimic the field situation. Therefore, selection of an adsorbent that can remove \ce{F-} from any kind of feed requires models that can predict the adsorption behavior for any given set of input conditions. From our observations and as also reported by many authors, the adsorption of \ce{F-} is affected by the presence of many ions. When modelling the adsorption of \ce{F-}, it is usually taken as a single entity getting adsorbed on the adsorbent. As this is not a proper assumption, a model was developed which takes into account all the speciation reactions that take place during adsorption, and all the species like \ce{H+}, \ce{OH-}, \ce{Na+}, \ce{Cl-}, and \ce{NO3-} present in the solution along with \ce{F-}. As an electrolyte system is involved, the Nernst-Planck equations were used to obtain the flux of each species. Using the model, the equilibrium constants and rate constants for the reactions were obtained. For one initial concentration of \ce{F-}, a reasonable fit was obtained to the batch adsorption data, except at short times. Because of an uncertainty in the amount of impurity present in the commercial adsorbent used, there was a significant discrepancy between predictions and data at higher initial concentrations of \ce{F-}. The present model can be applied to any charged adsorbent.
\end{abstract}


\end{frontmatter}

\section{Introduction}
Most of the studies on adsorption of \ce{F-} are experimental, and have been done with synthetic solutions. Such studies rarely mimic the field situation. Therefore, selection of an adsorbent that can remove \ce{F-} from any kind of feed requires models that can predict the adsorption behavior for any given set of input conditions. Data on most of the adsorbents used for the adsorption of \ce{F-} are observed to agree well with Langmuir or Freundlich isotherms \citep{Mondal15}. In the context of kinetics, they are fitted either to Langmuir kinetics, pseudo-first-order or pseudo-second-order kinetics. With the use of an analytical or numerical solution, we can predict the performance of the adsorbent without the need for laborious experiments. 

In the light of the above observations, the motivation for the present work is as follows. Many papers show that the presence of an another ion along with \ce{F-} influences the adsorption of \ce{F-} \citep{Nigussie07,Goswami12,Chatterjee14}. However, with respect to modelling the adsorption of \ce{F-} , there are about 2 to 3 other ions always present in water. These ions often affect the structural properties of the adsorbent \citep{Okamoto88,Su97,Al03} and very few theoretical studies exist on the combined effect of multiple ions on the adsorption of \ce{F-}. Most of the available models do not account for the presence of other ions. Therefore, here a model has been developed to predict the adsorption behaviour of \ce{F-} in the presence of \ce{H+}, \ce{OH-}, \ce{Cl-}, \ce{NO3-}, and \ce{Na+}.

In the papers of \cite{Fletcher06} and \cite{Tang09}, the concentration of \ce{H+} is taken as an independent variable. \cite{Hao86} assume that the surface concentration of \ce{H+} depends on the electric potential of the surface, as the alumina surface becomes charged in solution. It is also assumed that the equilibrium constants depend on the pH of the solution \citep{Hao86,Fletcher06}. In the present work, the pH is permitted to change with time, and the complexity arising because of the surface potential is neglected.

\section{Materials and methods}
\subsection{Experimental setup} \label{MM2-ES}
Adsorption was studied in the batch mode. Two setups were used. In one setup, a conical flask containing 200 mL of solution and 1 g of activated alumina adsorbent was used. The details of the adsorbent are given in Table~\ref{MM-t1}. 
\begin{table}
\begin{center}
\caption{Characteristics of AA pellets of grade OAS37, as provided by the manufacturer Oxide (India) Catalysts Pvt. Ltd., Durgapur, India and experimentally measured.}
\begin{tabular}{lcc}
\\\hline \multirow{2}{*}{Property} & \multicolumn{2}{c}{Quantity} \\
& Manufacturer & Experimental \\
\hline bulk density (kg/m$^3$) & 880 & 786\\
surface area (m$^2$/g) & 200 & 170\\
porosity & 0.2 & 0.33\\
loss on attrition (\%) & 0.10 & --\\
\ce{Al2O3} (weight \%) & 93.74 & --\\
\ce{Fe2O3} (weight \%) & 0.05 & --\\
\ce{SiO2} (weight \%) & 0.20 & --\\
\ce{Na2O} (weight \%) & 0.35 & --\\
loss on ignition (at $900\,^{\circ}\mathrm{C}$) (weight \%) & 5.66 & --\\\hline
\end{tabular}
\label{MM-t1}
\end{center}
\end{table} The flask was placed in a rotary shaker which was rotated at a speed of 100 rpm. In these experiments, it is expected that shaking will provide good mixing and hence a uniform distribution of the solution and the adsorbent. However, the adsorbent accumulated as a stationary heap near the center of the flask even with a rotation speed of 100 rpm. Thus the relative velocity between the fluid and particles is uncertain. When the shaker speed was increased beyond 100 rpm, there was fluidization of the adsorbent pellets, but after some time of operation, attrition of the pellets occurred, leading to a reduction in their size.

\begin{figure}[ht]
\begin{center}
\includegraphics[width=0.5\textheight]{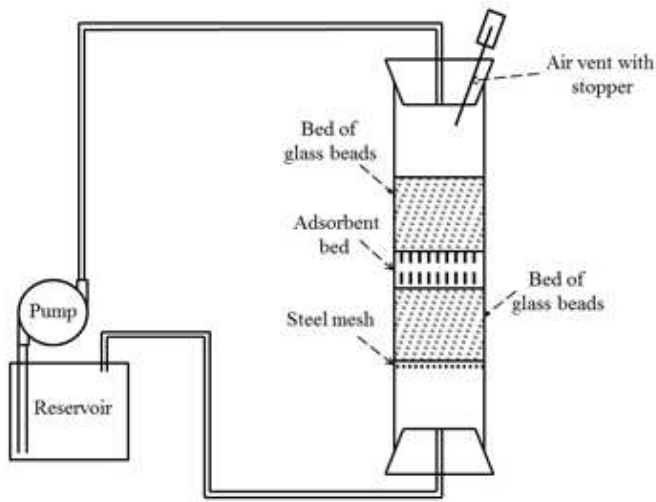}
\end{center}
\caption{Schematic view of the differential bed adsorber used for the batch experiments.}
\label{MM2-f1}
\end{figure}

In order to prevent the attrition and also to get an estimate of the relative velocity $v_{ra}$ of the fluid with respect to adsorbent, a different experimental setup was used. The need to estimate $v_{ra}$ arises because of the dependence of the external mass transfer coefficient on $v_{ra}$. Therefore, the data were acquired using a differential bed adsorber (DBA). The DBA experiments were done using a glass column of inner diameter 18 mm and the flow was controlled using a peristaltic pump (Fig.~\ref{MM2-f1}).  The flow rate of the liquid through the DBA was maintained at 1.5 mL/s and could not be increased beyond this value owing to limitations of the available peristaltic pump. The adsorbent bed of height 1 cm was sandwiched between 5 cm beds of glass beads to ensure a uniform distribution of the fluid, especially when flow was from the top to the bottom. The lower bed was supported on a steel mesh, and the contents were loaded into the column after filling it with the solution of interest. This prevented the formation of air gaps in the bed, which could cause channelling of the fluid. The adsorbent height of 1 cm corresponds to a weight of 2 g and the total volume of the circulating solution was taken such that the solids concentration or bulk density was 5 g/L.

The bed porosity $\epsilon_b$ was determined by noting the change in the level of water immediately after adding the pellets. It was found that $\epsilon_b = 0.59$. Using a microscope, the diameters of about 10 particles were measured. The masses of the group of particles was measured and hence the averaged particle density $\rho_p$ could be calculated. It was found that $\rho_p = 1910$ kg/m$^3$.

Before loading into the column, the adsorbent was soaked in deionised water for 24 h. This was performed to remove any unwanted impurities that were present on the adsorbent, as the used AA was a commercial grade alumina. Adsorption of single ions onto AA was studied at first to determine the adsorption efficiency of AA towards these ions. The concentration of ions were varied from 40 mg/L to 5 mg/L for \ce{F-}, 500 mg/L to 20 mg/L for \ce{SO4^2-}, and 1000 mg/L to 10 mg/L for \ce{HCO3-}. After these experiments, binary combinations of the ions (i) \ce{F-} and \ce{Cl-}, (ii) \ce{F-} and \ce{NO3-} were used. The change in the concentrations of the ions in the solution due to adsorption was measured by collecting samples at regular intervals of time. All the experiments were conducted at room temperature (26 - 30 $\degree$C).

\subsection{Chemicals and chemical analysis} \label{chap3:sec}
All the chemicals used in the preparation of the solutions were of analytical grade, and were used without any further purification. Deionised water having a conductivity of $0.05$ $\mu$S/cm was obtained from a Millipore unit. 

Analysis of the samples was done using an ion chromatograph (Metrohm 883 Basic IC plus) for anions and cations. The 95\% confidence limits were calculated using the data for the standards and the \textit{t}-distribution \citep{Snedecor68}. The error bars shown in the figures correspond to these limits. In order to operate the chromatographic column below the maximum number of exchange sites available in the column, it was recommended to dilute the samples. During the analysis, the samples of ROR, SRW, and SNW were diluted 5 or 10 times and samples of SFW were used without dilution. As the bicarbonate concentration cannot be measured using this method, it was calculated using titration \citep[p.~2.27]{Standards}. The concentration of total carbonate comprising \ce{CO3^2-}, \ce{HCO3-}, and \ce{CO2(aq)} was calculated from the pH of solution, the pK values of carbonic acid, and the bicarbonate concentration obtained by titration. The quality of the analysis based on the ion chromatograph was checked using international standards like ION 915, ION 96.4, and MISSIPPI - 03. It was found that the concentrations of all ions quantified were within 5\% of the certified values of these standards.

\subsection{Surface studies} \label{MM2-SS}
The determination of the possible reactions that can occur on the adsorbent can be determined by a study of its surface. It is known that the surface of AA changes with time upon soaking in water \citep{Wijnja99,Lefevre02,Ryazanov04,Carrier07}. This is mainly attributed to the hydration of the alumina surface, which was observed to change to bayerite or gibbsite based on the pH of the solution \citep{Carrier07}. As the hydration reactions involve the use of \ce{H+} ions, the use of titration can give us an insight into the equilibrium constants of these reactions.

Potentiometric titration experiments were done using a Mettler Toledo auto titrator (DL50 Rondolino). This was equipped with an automatic microburet and the pH of the solution after the subsequent addition of the titrant was measured by a Mettler Toledo pH electrode (DG111 - SC) containing 3 M \ce{KCl} solution. The pH meter was calibrated using Merck standard solutions of pH 4.0 and 7.0. For all the experiments, 0.02 N \ce{HCl} solution was used as the titrant. It was calibrated using 5 mL of 0.02 N tris-hydroxy methyl amino methan (TRIZMA) solution. The solution to be titrated consisted of 0.25 g of the adsorbent suspended in 10 mL of de-ionized water. This solution was made alkaline by the addition of 1 mL of 1 N \ce{NaOH}, and the pH was 12.0.


\section{A model for titration and batch adsorption} \label{chap2_th}
A good model is needed to predict the adsorption behaviour for various concentrations of the ions. Isotherms like Langmuir, Freundlich, and Sips have been used to predict the equilibrium behaviour for the adsorption of \ce{F-} \citep{Mondal15}, but it is assumed that there is only one species in the solution. It is also assumed that the adsorbent consists of only occupied and vacant sites. In practice, there is more than one species in the solution and there is a competition between the species for adsorption. Further, the adsorbent consists of  different kinds of sites which can form varied complex species. Therefore, it is essential to include the concentrations of the other species in the model as well as different products formed by different sites. With these assumptions one can predict the equilibrium composition on the adsorbent and of the solution for any given initial condition.

The papers which show a decreased uptake of \ce{F-} in the presence of other ions do not mention or model how the concentrations of these ions change with time \citep{Islam07,Tang09,Dou11,Chatterjee14}. During the adsorption process, the following reactions have been proposed \citep{Hao86}
\begin{align}
\ce{\bond{3}AlOH} + \ce{H+ <=>} & \ce{\bond{3}AlOH2+} \label{Th-r1a}\\
\ce{\bond{3}AlOH}   \ce{<=>} &\ce{\bond{3}AlO-} + \ce{H+} \label{Th-r1b}\\
\ce{\bond{3}AlOH2+} + \ce{F- <=>} & \ce{\bond{3}AlF} + \ce{H2O} \label{Th-r1c}
\end{align} where \ce{\bond{3}} denotes a surface group. Here the reaction \begin{equation} \nonumber
\ce{\bond{3}AlOH} + \ce{F- <=>} \ce{\bond{3}AlF} + \ce{H2O} 
\end{equation} is not considered because it can be obtained by the addition of reaction (\ref{Th-r1a}), (\ref{Th-r1c}), and the water reaction. \begin{equation} \nonumber
\ce{H2O <=>} \ce{H+} + \ce{OH-} 
\end{equation}

In the most commonly used modelling approaches \citep{Nigussie07,Mondal15} only (\ref{Th-r1c}) has been used. From (\ref{Th-r1a}) - (\ref{Th-r1c}) it can be seen that along with the adsorbed species \ce{\bond{3}AlF}, there are also \ce{\bond{3}AlOH}, \ce{\bond{3}AlOH2+}, and \ce{\bond{3}AlO-} present on the adsorbent. So there is need to consider these species also. The groundwater usually contains many other species like \ce{HCO3-}, \ce{SO4^{2-}}, etc. along with \ce{F-}. If there is competitive adsorption, then a few more species will also be produced \citep{Wijnja99,Appel13}. 
\begin{align}
\ce{\bond{3}AlOH2+} + \ce{HCO3- <=>} & \ce{\bond{3}AlOCO2-} + \ce{H2O} + \ce{H+} \label{Th-r2a}\\
\ce{\bond{3}AlOH2+} + \ce{SO4^{2-} <=>} &\ce{\bond{3}AlOSO3-} + \ce{H2O} \label{Th-r2b}
\end{align}

\begin{figure}[ht]
\begin{center}
\includegraphics[width=0.4\textheight]{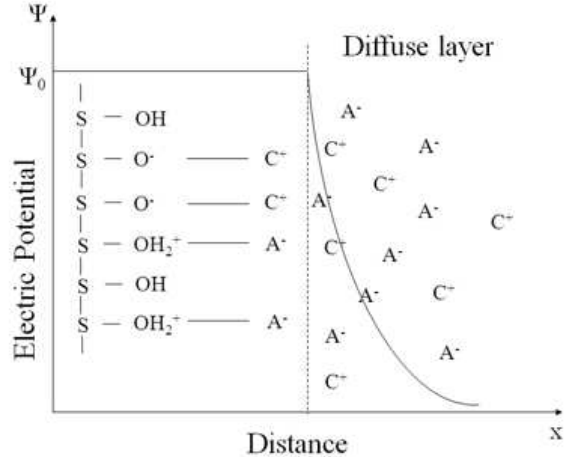}
\end{center}
\caption{Schematic view of the possible distribution of the ions at the surface of a charged solid. Here \ce{C+} and \ce{A-} denote cations and anions, respectively, and S denotes a surface site.}
\label{MM2-f1.1}
\end{figure}

In the present section, the assumptions and governing equations for the multicomponent adsorption are discussed. First, it is assumed that the charge neutrality is separately obeyed by the adsorbent and the solution. In existing models, the surface of the adsorbent is assumed to be charged, and the charge is balanced by an electrical double layer of ions in the solution \citep{Bockris06} (Fig.~\ref{MM2-f1.1}).
The surface concentration $c_{i,s}$ and bulk concentration $c_i$ of an ion $i$ at equilibrium, i.e. when there is no mass tranfer, are related by \begin{equation}
\displaystyle{c_{i,s} = c_i \: e^{\dfrac{-z_i F \psi _0}{RT}}} \label{Th-r3}
\end{equation} where $\psi _0$ is the surface potential, or potential at the outer Helmholtz plane, $F$ is Faraday constant, $R$ is universal gas constant, and $T$ is the absolute temperature of the system \citep{Hao86}. However, accounting for the diffuse double layer makes models of continuous contacting unwieldy. Further (\ref{Th-r3}) is not strictly valid when the rate of adsorption is non-zero. The assumption of charge neutrality for the solid phase is equivalent to assuming a sharp double layer equivalent to a group of capacitors in parallel. Hence the concentrations at the surface and the bulk are assumed to be equal. 

\begin{figure}[ht]
\begin{center}
\includegraphics[width=0.6\textheight]{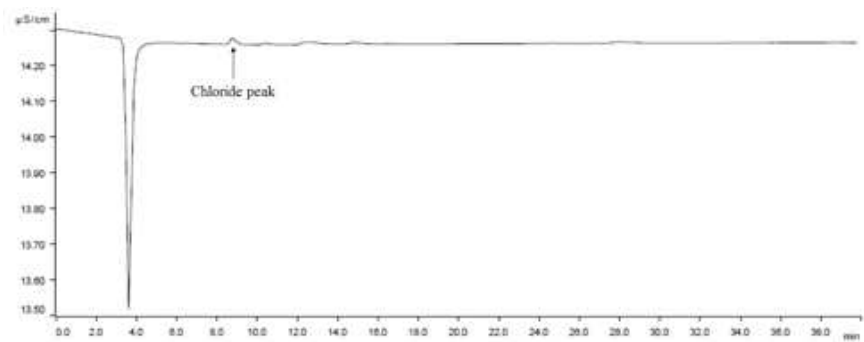}
\end{center}
\caption{Chromatogram from an ion chromatograph for anions in deionized water. Here the x-axis represents the time and the y-axis represents the conductivity.}
\label{MM2-f1.2}
\end{figure}

The charged species formed on the solid surface are assumed to be neutralised by non-reactive ions present in the solution. In the present study, chloride, nitrate and sodium were found to not complex with the adsorbent i.e., they were non-reactive. Similarly \cite{Nagashima99} found that \ce{Na+} did not affect the adsorption of \ce{H+} onto $\gamma - $ alumina. Thus, it is assumed that \ce{\bond{3}AlOH2+} is neutralized by either \ce{Cl-} or \ce{NO3-}. In the present case, the basis for taking \ce{Cl-} in a solution made from \ce{NaF} and deionized (DI) water is that the pH of the DI water was about 6.0, but the water without any ions should have a pH of 7.0. Therefore, upon analysis of this DI water for anions using an ion chromatograph, it was also observed that there was a minute peak for \ce{Cl-} which is below the limit of detection (Fig.~\ref{MM2-f1.2}). 
 Similarly, for \ce{\bond{3}AlO-}, \ce{Na+} is the counter ion, as it is added along with \ce{F-}. Hence (\ref{Th-r1a}) - (\ref{Th-r1c}) are replaced by
\begin{align}
\ce{\bond{3}AlOH} + \ce{H+} + \ce{Cl- <=>} & \ce{\bond{3}AlOH2+Cl-} \label{Th-r4a}\\
\ce{\bond{3}AlOH} + \ce{Na+ <=>} &\ce{\bond{3}AlO^-Na+} + \ce{H+} \label{Th-r4b}\\
\ce{\bond{3}AlOH2+Cl-} + \ce{F- <=>} & \ce{\bond{3}AlF} + \ce{H2O} + \ce{Cl-} \label{Th-r4c}
\end{align} Equilibrium constants for (\ref{Th-r4a}) and (\ref{Th-r4b}) were obtained by titration, and for (\ref{Th-r4c}) by batch adsorption.

\subsection{Titration}
The reactions that occur during titration are (\ref{Th-r4a}) and (\ref{Th-r4b}) and the equilibrium constants for these reactions are given by 
\begin{align}
K_1 = \dfrac{\hat{a}_1}{\hat{a}_2 a_1} \label{Th-r5.0a}\\
K_2 = \dfrac{\hat{a}_3 a_1}{\hat{a}_2} \label{Th-r5.0b}
\end{align}
where $\hat{a}_1$ - $\hat{a}_3$, and $a_1$ are the electrochemical activities \citep[pg.~598]{Benjamin02} of \ce{\bond{3}AlOH2+Cl-}, \ce{\bond{3}AlOH}, \ce{\bond{3}AlO^-Na+}, and \ce{H+}, respectively. As the double layer is collapsed onto a single plane in our model, $\psi_0 = 0$. Hence, the electrochemical activities may be replaced by the chemical activites. Considering the adsorbed phase to be an ideal solution, the activities may be replaced by the mole fractions $\{y_i\}$ in the adsorbed phase. Similarly, considering a dilute solution $a_1 = \frac{c_1}{c_{1,0}}$ where $c_{1,0} = 1.0$ mol/L is a reference concentration. Therefore, with these assumptions (\ref{Th-r5.0a}) and (\ref{Th-r5.0b}) can be rewritten as
\begin{align}
K_1 = \dfrac{q_1}{q_2 c_1} \label{Th-r5a}\\
K_2 = \dfrac{q_3 c_1}{q_2} \label{Th-r5b}
\end{align}
where $q_j = y_j c_T$ is the concentration of adsorbed species $j$ (moles of $j$ per unit mass of the adsorbent), $c_T$ is the total number of sites on the adsorbent (moles per unit mass of the adsorbent), and $c_i$ is the concentration of solute species $i$ (moles of $i$ per unit volume of the solution). 

For (\ref{Th-r4a}) and (\ref{Th-r4b}), $q_j$ with $j$ varying from 1-3 corresponds to the species \ce{\bond{3}AlOH2+Cl-}, \ce{\bond{3}AlOH}, \ce{\bond{3}AlO^-Na+}, respectively, and $c_1$ corresponds to the concentration of \ce{H+} in the solution. The expressions for the equilibrium equations (\ref{Th-r5a}) and (\ref{Th-r5b}) are consistent with simple mass action kinetics for the special case where the charge-neutralising counter ions do not participate in the adsorption or complexing process. Neutralisation occurs just based on Coulombic attraction. Similar mass action kinetics were found suitable by \cite{Winkler71} for the removal of phosphate by activated alumina activated with \ce{HNO3}, where the measured rate was independent of the concentration of nitrate in the solution. 

In the titration experiment, the titrant was \ce{HCl} and the solution to be titrated consisted of AA in \ce{NaOH}. The mass balances for \ce{Cl-} and \ce{Na+} are given by 
\begin{align}
c_{3i} V_a + c_{imp} V_a = c_3 (V_a + V_b) + m_p q_1 \label{Th-r6a}\\
c_{5i} V_b = c_5 (V_a + V_b) + m_p q_3 \label{Th-r6b}
\end{align}
where $c_{3}$ and $c_{5}$ are the concentrations of \ce{Cl-} and \ce{Na+}, respectively, $c_{3i}$ is the concentration of the titrant (acid), $c_{5i}$ is the initial concentration of the base, $V_b$ is the volume of the base taken, $V_a$ is the volume of the acid added, $m_p$ is the mass of the adsorbent taken, and $c_{imp}$ is the total concentration of impurity present in the solution after addition of the acid. As mentioned earlier, the AA used in the experiments was a commercial adsorbent which contained some impurities. In view of the leaching of \ce{NO3-} from the fresh adsorbent when it is soaked in deionized water, it was assumed that the impurity was \ce{NO3-}. The term $c_{imp} V_a$ represents an ad hoc attempt to include the leaching of nitrate into the solution during titration. More satisfactory alternatives will be explored in the future. As shown later, there was no adsorption of \ce{Cl-} and \ce{NO3-} on the adsorbent. Also, becasue of the same charge of these ions, \ce{Cl-} and \ce{NO3-} are treated interchangeably in modelling. In addition, the solution must be electrically neutral, or
\begin{equation} \label{Th-r7.1}
c_1 - c_3 - \dfrac{K_w}{c_1} + c_5 = 0 
\end{equation} Here $c_1$ is the concentration of \ce{H+} and $K_w$ is ionic product of water. In (\ref{Th-r7.1}), $K_w/c_1$ gives the concentration of \ce{OH-} in solution. There can be a loss of aluminium in the form of soluble complexes such as \ce{AlF_n(H2O)_{6-n}^{(3-n)}}, \ce{AlF_n^{3-n}}, and \ce{AlOH_n^{3-n}} \citep{Nordin99,George10,Jin10}. This factor has been ignored in the present work. 

The mass balance for the adsorbates is given by \begin{equation} \label{Th-r7}
q_1 + q_2 + q_3 = c_T
\end{equation} where $q_i$ are the concentrations of species on the adsorbent. There are six equations (\ref{Th-r5a}) - (\ref{Th-r7}) and seven unknowns $c_1$, $c_3$, $c_5$, $q_1$, $q_2$, $q_3$, and $V_a$. Therefore, in order to obtain a unique solution, we need to specify one of the seven unknowns. In the titration experiment, the pH of the solution is measured as a function of the volume of the titrant added $V_a$. Equations (\ref{Th-r5a}) to (\ref{Th-r7}) can be solved to obtain $V_a$ in terms of $c_1$ as \begin{equation} \label{Th-r8}
V_a = \frac{{\displaystyle\left(c_{5i} + c_1 - \frac{K_w}{c_1}\right)V_b + \left(1 - \frac{K_2}{K_1 c^2_1}\right)\frac{{\displaystyle m_p c_T}}{{\displaystyle 1 + \frac{1}{K_1 c_1} + \frac{K_2}{K_1 c_1^2}}}}}{{\displaystyle c_{3i} + c_{imp} - \left(c_1 - \dfrac{K_w}{c_1}\right)}}
\end{equation} The above equation can be rewritten in terms of the pH as \begin{equation} \label{Th-r8.1}
V_a = \frac{{\displaystyle \left( c_{5i} + 10^{-pH} - \frac{K_w}{10^{-pH}} \right)V_b + \left( 1 - \frac{K_2}{K_1 10^{-2pH}} \right) \dfrac{m_p c_T}{1 + \dfrac{1}{K_1 10^{-pH}} + \dfrac{K_2}{K_1 10^{-2pH}}}}}{{\displaystyle c_{3i} + c_{imp} - \left( 10^{-pH} - \dfrac{K_w}{10^{-pH}} \right)}}
\end{equation} Equation (\ref{Th-r8.1}) can be used to estimate the values of $K_1$, $K_2$, $c_T$, and $c_{imp}$ by fitting it to the data of pH vs. $V_a$.

\subsection{Batch adsorption: equilibrium}
The above methodology can be followed for relating the variables involved in the batch adsorption of \ce{F-}. An equation for the mass balance of \ce{F-} has to be added, and the other mass balances have to be modified as titrant is not added to the solution. Equations (\ref{Th-r6a}) - (\ref{Th-r6b}) are modified to 
\begin{align} 
c_{2i} = c_2 + \rho _b q_4 \label{Th-r9a}\\
c_{3i} + \rho _b q_{1i} = c_3 + \rho _b q_1 \label{Th-r9b}\\
c_{5i} = c_5 + \rho _b q_3 \label{Th-r9c}
\end{align}
where $c_2$ and $q_4$ are the concentrations of \ce{F-} and \ce{\bond{3}AlF}, respectively, and $\rho_b$ is the mass of adsorbent pellets per unit volume of the solution in the bed and the stirred vessel (Fig.~\ref{Th-f1}). Here the mass balance for \ce{Cl-} is changed because during the batch experiments, the adsorbent was taken after soaking in DI water for 24 h and the solution which contained the leachate was discarded, whereas the titration experiments were done along with the leachate. Here $q_{1i}$ is the initial concentration of the impurity/leachate present on the adsorbent. The equilibrium constant for the reaction (\ref{Th-r4c}) is defined as \begin{equation} \label{Th-r10}
K_3 = \dfrac{q_4}{q_1 c_2}
\end{equation} The charge neutrality equation for the solution is \begin{equation} \label{Th-r11}
c_1 - c_2 - c_3 - \frac{K_w}{c_1} + c_5 = 0
\end{equation} Therefore, there are eight unknowns $c_1$, $c_2$, $c_3$, $c_4$, $q_1$, $q_2$, $q_3$, $q_4$ and eight equations (\ref{Th-r5a}, \ref{Th-r5b}, \ref{Th-r9a} - \ref{Th-r11}). 

\subsection{Batch adsorption: kinetics}
\begin{figure}[ht]
\begin{center}
\includegraphics[width=0.6\textheight]{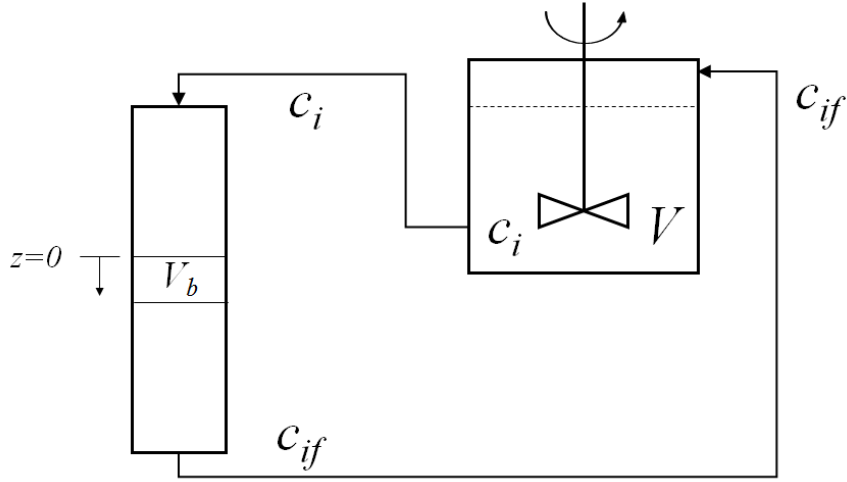}
\end{center}
\caption{Sketch of the differential adsorber used for the batch experiments.}
\label{Th-f1}
\end{figure}

Till now, the equations governing equilibrium were discussed. Modelling batch adsorption or continuous processing in a column requires rate data, in addition to equilibrium data. The setup shown in Fig.~\ref{MM2-f1} was used to obtain data on adsorption kinetics, and the notation for the concentrations is shown in Fig.~\ref{Th-f1}.  The volume of the adsorber bed is very small compared to the volume of the stirred vessel.

For convenience, the species in the solution are numbered such that 1 denotes \ce{H+}, $n-1$ denotes \ce{OH-}, and $n$ denotes any species whose concentration is eliminated using the electroneutrality condition. For example, for a 5-species system containing \ce{H+}, \ce{Cl-}, \ce{OH-}, \ce{F-}, and \ce{Na+}, we have 1 = \ce{H+}, 2 = \ce{F-}, 3 = \ce{Cl-}, 4 = \ce{OH-}, and 5 = \ce{Na+}. One of the important variables is the pH. As it affects the adsorption of \ce{F-}, special effort was made to account for the changes in pH, by treating the reaction between \ce{H+} and \ce{OH-} as instantaneous.

The mass balance for species $i$ in a well-mixed stirred vessel is given by \begin{equation}
V \dfrac{d c_i}{dt} = \dot{Q} (c_{i,f} - c_i) + V \dot{R}_i;\quad i = 1,n
\end{equation} where $V$ is the volume of the liquid in the vessel, $c_{i,f}$ and $c_i$ are the inlet and outlet concentrations, respectively, $\dot{Q}$ is the volumetric flowrate of the liquid, and $\dot{R}_i$ is the molar rate of production of $i$ per unit volume by chemical reactions. We have
\begin{align}
\dot{R}_i = 0; \quad i\:\neq \:1\:\& \:n-1 \label{Th-r12a}\\
\dot{R}_1 = \dot{R}_{n-1} = k_f c_{\ce{H_2O}} - k_b c_{\ce{H+}} c_{\ce{OH-}} \label{Th-r12b}
\end{align}
where $k_f$ and $k_b$ are the rate constants for the reaction \begin{equation} \label{Th-r13}
\ce{H2O} \ce{<=>} \ce{H+} + \ce{OH-}
\end{equation} It is assumed that (\ref{Th-r13}) is always close to equilibrium. Hence $\dot{R}_1\approx 0$ and \begin{equation} \label{Th-r14}
K_w = \dfrac{a_{\ce{H+}} a_{\ce{OH-}}}{a_{\ce{H2O}}} \approx c_{\ce{H+}} c_{\ce{OH-}}
\end{equation} where $a_i$ is the activity of species $i$ and $c_{\ce{H+}}$ is the concentration of \ce{H+} in mol/L. Hence, $c_{\ce{OH-}} = c_{n-1} = K_w/c_1$. To eliminate $\dot{R_1}$, we subtract the mass balance for $i=n-1$ from the balance for $i=1$ to obtain \begin{equation} \label{Th-r15}
V \left( \dfrac{dc_1}{dt} - \dfrac{dc_{n-1}}{dt} \right) = \dot{Q} \left[ (c_{1,f} - c_1)-(c_{n-1,f}-c_{n-1}) \right]
\end{equation} In (\ref{Th-r15}), $c_{n-1}$ is replaced by $K_w/c_1$.

The charge neutrality condition is used to eliminate the concentration of the $n^{th}$ species. Thus \begin{equation} \label{Th-r16}
\sum_{i=1}^n z_i c_i = 0 \Rightarrow c_n = -\sum_{i=1}^{n-1} \frac{z_i c_i}{z_n}
\end{equation} where $z_i$ is the charge of species $i$, expressed in multiples of the charge of the electron. Using (\ref{Th-r14}) and noting that $z_1 = 1$, and $z_{n-1}=-1$, (\ref{Th-r16}) reduces to \begin{equation} \label{Th-r17}
c_n = -\frac{1}{z_n}\left( c_1 - \frac{K_w}{c_1}\right) - \sum_{i=2}^{n-2} \frac{z_i}{z_n} c_i
\end{equation}

Therefore, it is necessary to solve only $n-1$ mass balances, and $c_n$ can be obtained from (\ref{Th-r17}). Hence the mass balances for species $i$ in the stirred vessel are given by
\begin{align}
V \frac{d}{dt} \left( c_1 - \dfrac{K_w}{c_{1}} \right) &= \dot{Q} \left[(c_{1,f}-c_1) - \left( \dfrac{K_w}{c_{1,f}} - \dfrac{K_w}{c_1} \right) \right] \label{Th-r18a} \\
V \frac{dc_i}{dt} &= \dot{Q} (c_{i,f} - c_i);\quad i=2,n-2 \label{Th-r18b}
\end{align}

\begin{figure}[ht]
\begin{center}
\includegraphics[width=0.65\textheight]{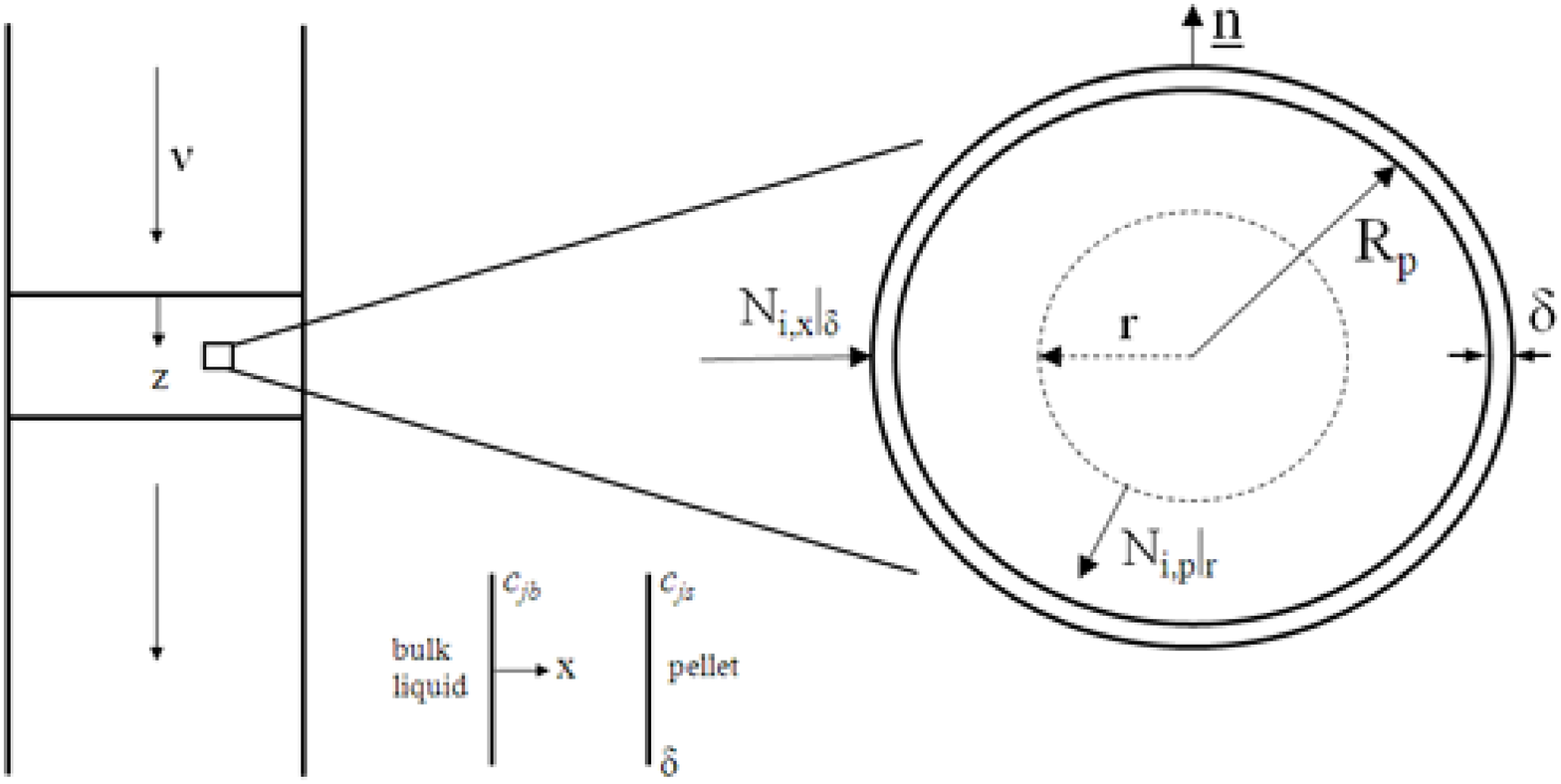}
\end{center}
\caption{Sketch of the different fluxes into the pellet and the liquid film surrounding it. Here $R_p$ is the radius of the pellet, and \underline{n} is the unit outward normal to the external surface of the pellet.}
\label{Th-f2}
\end{figure}

Let us now consider the model for the differential packed bed. The external mass transfer rate to each pellet is calculated using a film model with the assumption that a liquid film of  thickness $\delta$ surrounds the pellet. As is customary, convection and accumulation are neglected in the film. The dissociation of water is instantaneous and is permitted to occur in the film. The other reactions need adsorbent and occur only in the pellets. As before, the solution is assumed to remain electrically neutral. The film thickness is assumed to be small compared to the particle size, and the curvature of the film is neglected. Let $\tilde{N}_{ix}$ denote the molar flux of $i$ in the x-direction (negative radial direction) 
(Fig.~\ref{Th-f2}). 
The mass balance for species $i$ in the bulk liquid is given by \begin{equation}
\frac{\partial}{\partial t}(\epsilon_b c_{i,b}) + \frac{\partial}{\partial z} N_{i,z} = - a_p \tilde{N}_{i,x}|_{x=0} + \dot{R}_{i,b};\quad i=1,n
\end{equation} where $\dot{R_{ib}}$ is the molar rate of production of $i$ per unit volume of the bed, $\epsilon _b$ is the porosity of the bed, and $a_p = 3(1-\epsilon _b)/R_p$ is the external surface area of the pellets per unit volume of the bed. As in the case of the stirred vessel, in the bulk liquid, we have
\begin{align}
\dot{R}_{i,b} = 0; \quad i\:\neq \:1\:\& \:n-1 \label{Th-r19a}\\
\dot{R}_{1,b} = \dot{R}_{n-1,b} = k_f c_{\ce{H_2O}} - k_b c_{\ce{H^+}} c_{\ce{OH^-}} \label{Th-r19b}
\end{align}
Proceeding as in the case of the stirred vessel, we obtain
\begin{align}
\frac{{\displaystyle\partial}}{{\displaystyle\partial t}}\left[\epsilon _b \left(c_{1,b} - \frac{K_w}{c_{1,b}}\right)\right] &+ \frac{\partial}{\partial z}\left[ N_{1,z} - N_{n-1,z}\right] = - a_p \left[ \tilde{N}_{1,x} - \tilde{N}_{n-1,x} \right]\rvert_{x=0} \label{Th-r20a} \\
\frac{\partial}{\partial t}(\epsilon _b c_{i,b}) &+ \frac{\partial}{\partial z} N_{i,z} = - a_p \tilde{N}_{i,x}\rvert_{x=0};\quad i=2,n-2 \label{Th-r20b}\\
c_{n,b} &= -\frac{1}{z_n} \left( c_{1,b} - \frac{K_w}{c_{1,b}} \right) - \sum_{i=2}^{n-2} \frac{z_i}{z_n} c_i \label{Th-r20c}
\end{align}

For isothermal diffusion, the fluxes are related to the driving forces such as the gradient of the chemical potential by the generalized Maxwell-Stefan equations \citep{Krishna97} \begin{equation} \label{Th-r20.1} \begin{split}
\underline{d}_i \equiv - \dfrac{x_i}{RT} \nabla_{T,p} \mu_i &- \dfrac{1}{cRT}(c_i \bar{V}_i - \omega_i) \nabla p + \dfrac{1}{cRT}\left(c_i\underline{F}_i - \omega_i \sum_{k=1}^n c_k \underline{F}_k \right) \\ 
& = \sum^n_{j=1,i} \dfrac{x_j\underline{N}_i - x_i\underline{N}_j}{c \dbar\;_{ij}}, \: i = 1,2,\ldots ,n. \\ 
\end{split}
\end{equation} where $\underline{d}_i$ is the driving force acting on $i$ per unit volume of the mixture, $\underline{F}_i$ is the body force acting per unit mass of $i$, $\underline{N}_i$ is the molar flux of $i$, $x_i$ is the mole fraction, $c_i \bar{V}_i$ is the volume fraction of species $i$, $\omega_i$ is the mass fraction of $i$, and $\mu_i$ is the chemical potential of $i$ and $\nabla_{T,p} \mu_i$ is the gradient of the chemical potential at a constant temperature $T$ and pressure $p$. Here $c, R$, and $\dbar\;_{ij}$ are the total concentration, the gas constant, and the Maxwell-Stefan diffusivity, respectively and $n$ is the total number of species. As discussed in \citet[pp.~765-768]{Bird02}, the first of (\ref{Th-r20.1}) arises naturally in the expression for the entropy production rate $\dot{\sigma}$ in a multicomponent fluid mixture. The term $\underline{d}_i$ is called a ``driving force", even though it does not have the dimensions of a force. However, $cRT\underline{d}_i$ may be regarded as a force per unit volume. The expression for $T\dot{\sigma}$ contains terms of the form $\sum_{i=1}^N \underline{j}_i \dfrac{cRT}{\rho_i} \underline{d}_i$, where $\underline{j}_i$ is the mass flux of $i$ relative to the mass average velocity and $\rho_i$ is the density of $i$. Assuming that each flux $j_i$ is a linear function of all the driving forces $\{\underline{d}_k\}$, the flux relations can be inverted to obtain the second of (\ref{Th-r20.1}) \citep{Curtiss99}. Predictions of (\ref{Th-r20.1}) coupled with the mass balances agree fairly well with data obtained from many systems \citep{Krishna97}.

For isobaric diffusion in a dilute electrolyte solution, (\ref{Th-r20.1}) can be simplified to obtain \citep{Krishna87} \begin{equation} \label{Th-r21}
\displaystyle\underline{N}_i = -D_i \nabla c_i - \frac{z_i c_i D_i F \nabla \phi_e}{RT} + c_i \underline{v} 
\end{equation} 
where $\underline{v}$ is the velocity vector of the solution, $F$ is the Faraday constant, and $\phi_e$ is the electrostatic potential generated because of the ions. For a dilute system, the diffusivity $\dbar\;_{ij}$ in (\ref{Th-r20.1}) may be replaced by $\dbar\;_{iw}$ i.e. the diffusivity of species $i$ with respect to water. This quantity is defined as $D_i$ in (\ref{Th-r21}). Equation (\ref{Th-r21}) is called the Nernst-Planck equation and has been used by many authors to model ion-exchange onto resins \citep{Frey86,Jia04,Bachet14}. In the context of defluoridation with an oxide adsorbent, this approach has not been used earlier. During adsorption, it is assumed that the adsorbent and solution are electrically neutral separately. 

The current density is given by 
\begin{equation} \label{Th-r22}
\underline{I} = \sum_{i=1}^n z_i \underline{N}_i F 
\end{equation} or, using (\ref{Th-r21}), \begin{equation} \label{Th-r23}
\underline{I} = \sum_{i=1}^n \left( z_i c_i \underline{v} - \frac{z_i^2 F^2 c_i D_i \nabla \phi_e}{RT} - z_i F D_i \nabla c_i \right) 
\end{equation} As the system is electrically neutral, (\ref{Th-r16}) and (\ref{Th-r23}) imply that \begin{equation} \label{Th-r24}
\underline{I} = - \sum_{i=1}^n \left( \frac{z_i^2 F^2 c_i D_i \nabla \phi_e}{RT} + z_i F D_i \mathop{\nabla} c_i \right) 
\end{equation} 

As no current passes through the system $I = 0$. Hence (\ref{Th-r24}) reduces to \begin{equation} \label{Th-r25}
\nabla \phi_e = -\; \frac{{\displaystyle RT \sum_{i=1}^n z_iD_i\nabla c_i}}{{\displaystyle F \sum_{i=1}^n z_i^2 c_iD_i}}
\end{equation} Using standard electrochemical engineering terminology \citep{Gu97,Newman12}, the conductivity number $\kappa$ and transport numbers $t_i$ are defined by 
\begin{align}
\kappa &\equiv \frac{{\displaystyle F^2}}{{\displaystyle RT}} \sum z_i^2 c_i D_i \label{Th-r26a}\\
t_i &\equiv \frac{{\displaystyle z_i^2c_iD_i}}{{\displaystyle \sum_{k=1}^n z_k^2c_kD_k}} \label{Th-r26b}
\end{align} Hence \begin{equation} \label{Th-r26c}
\nabla \phi_e = -\; \dfrac{RT}{F} \dfrac{\sum_{i=1}^n z_iD_i\nabla c_i}{\kappa RT/F^2} \quad \text{and} \quad \frac{t_i}{z_i} = \frac{z_ic_iD_i}{\kappa RT/F^2}
\end{equation}

Substituting (\ref{Th-r25}) and (\ref{Th-r26c}) in (\ref{Th-r21}), we obtain\begin{equation*} 
\underline{N}_i = -D_i\nabla c_i + \frac{t_i}{z_i}\left[\sum_{k=1}^n z_kD_k\nabla c_k \right] + c_i \underline{v}
\end{equation*} or
\begin{equation} \label{Th-r27}
\displaystyle\underline{N}_i = -\sum_{j=1}^n \mathfrak{D}_{ij} \nabla c_j + c_i\underline{v}
\end{equation} where \begin{equation} \label{Th-r27a}
\mathfrak{D}_{ij} \equiv \delta_{ij}D_i - \frac{{\displaystyle t_i}}{{\displaystyle z_i}}z_jD_j
\end{equation} and $\delta_{ij}$ is the Kronecker delta.

It may be noted that the fluxes are coupled to the driving forces of all the species due to enforcement of charge neutrality. Ignoring the axial diffusion and axial dispersion in the bulk liquid, \begin{equation} \label{Th-r28}
N_{i,z} = c_{i,b} v_z = \epsilon_b c_{i,b} \bar{v}_z
\end{equation} where $N_{i,z}$ is the z-component of the molar flux and $\bar{v}_z$ is the interstitial velocity. 

If we ignore convection in the liquid film, and if diffusion occurs only in the x-direction (Fig.~\ref{Th-f2}), (\ref{Th-r27}) reduces to \begin{equation} \label{Th-r30}
\tilde{N}_{i,x} = - \sum_{j=1}^n \mathfrak{D}_{ij} \frac{\partial c_j}{\partial x}
\end{equation}

Ignoring the accumulation in the liquid film, and if the only reaction in the film is the dissociation of water, the mass balances are given by 
\begin{align} 
\frac{d\tilde{N}_{i,x}}{dx} &= 0;\quad i=2,n-2 \label{Th-r31a} \\
\frac{d\tilde{N}_{i,x}}{dx} &= \dot{R}_{i,f};\quad i=1\: \& \:i=n-1 \label{Th-r31b}
\end{align}
Following the procedure used earlier, equations (\ref{Th-r31b}) are combined to eliminate the rate of the decomposition of water and may be rewritten as \begin{align} 
\nonumber \frac{d\tilde{N}_{1,x}}{dx} - \frac{d\tilde{N}_{n-1,x}}{dx} = 0
\end{align} The above equation can be integrated to obtain \begin{equation} \label{Th-r32}
\tilde{N}_{1,x} - \tilde{N}_{n-1,x} = \text{constant}
\end{equation} Assuming that the concentration varies linearly across the film, and using (\ref{Th-r30}) in (\ref{Th-r32}) we obtain \begin{equation} \label{Th-r33}
\tilde{N}_{1,x} - \tilde{N}_{n-1,x} = \sum_{j=1}^n (\mathfrak{D}_{1j} - \mathfrak{D}_{n-1,j}) \frac{{\displaystyle (c_{j,b} - c_{j,s})}}{{\displaystyle \delta}}
\end{equation} where $c_{j,b}$ is the concentration in the bulk liquid, $c_{j,s}$ is the concentration at the surface of the pellet, and $\delta$ is the thickness of the film (Fig.~\ref{Th-f2}). Note that the $\mathfrak{D}_{ij}$'s are functions of the concentrations, but have been treated as constants while deriving (\ref{Th-r33}) from (\ref{Th-r30}). Similarly, \begin{equation} \label{Th-r33a}
\tilde{N}_{i,x} = \sum_{j=1}^n \mathfrak{D}_{ij} \dfrac{(c_{j,b} - c_{j,s})}{\delta};\quad i=2,n-2
\end{equation}

The expressions derived for fluxes to the surface of the pellet can be substituted into the balances (\ref{Th-r20a}) and (\ref{Th-r20b}) for the bed. These reduce to \begin{equation} \label{Th-r34}
\frac{{\displaystyle \partial}}{{\displaystyle \partial t}}\left[ \epsilon _b \left(c_{1,b} - \frac{{\displaystyle K_w}}{{\displaystyle c_{1,b}}}\right)\right] + \frac{\partial}{\partial z}\left[ \epsilon_b \bar{v}_z\left(c_{1,b} - \frac{K_w}{c_{1,b}}\right)\right] = - \;a_p \sum_{j=1}^n \left( \mathfrak{D}_{1j} - \mathfrak{D}_{n-1,j}\right) \frac{{\displaystyle (c_{j,b}-c_{j,s})}}{{\displaystyle\delta}}
\end{equation} 
\begin{equation} \label{Th-r35}
\frac{{\displaystyle \partial}}{{\displaystyle \partial t}}( \epsilon _b c_{i,b}) + \frac{\partial}{\partial z}( \epsilon_b \bar{v}_z c_{i,b}) = -\; a_p \sum_{j=1}^n \mathfrak{D}_{ij} \frac{{\displaystyle (c_{j,b}-c_{j,s})}}{{\displaystyle\delta}};\quad i=2,n-2
\end{equation}

Let us now consider the liquid inside the pellets. The mass balance are given by \begin{equation} \label{Th-r36}
\frac{\partial}{\partial t} (\epsilon_p c_{i,p}) + \nabla . \underline{N}_{i,p} = \rho_p \dot{R}_{i,p} + \dot{R}_{i,x}
\end{equation} where $\epsilon_p$ is the porosity of the pellet, $c_{i,p}$ is the molar concentration of $i$ per unit volume of the fluid, $\underline{N}_{i,p}$ is the molar flux of species $i$ in the pellet, $\dot{R}_{i,p}$ is the molar rate of production of $i$ per unit mass of the pellet, $\rho_p$ is the density of the pellet, and $\dot{R}_{i,x}$ is the rate of production of $i$ due to the water reaction (\ref{Th-r13}). Thus
\begin{align}
\dot{R}_{i,x} &= 0; \quad i\neq1\: \& \:i\neq n-1 \label{Th-r37a}\\
\dot{R}_{1,x} &= \dot{R}_{n-1,x} \label{Th-r37b}
\end{align}
Subtracting the mass balance equation for $i=n-1$ from that of $i=1$, we obtain 
\begin{align} 
\frac{\partial}{\partial t} \left[\epsilon_p\left(c_{1,p} - \frac{K_w}{c_{1,p}}\right)\right] &+ \nabla . \left[\underline{N}_{1,p} - \underline{N}_{n-1,p}\right] = \rho_p (\dot{R}_{1,p} - \dot{R}_{n-1,p}) \label{Th-r38a} \\
\frac{\partial}{\partial t}(\epsilon_p c_{i,p}) &+ \nabla . \underline{N}_{i,p} = \rho_p \dot{R}_{i,p}; \quad i=2,n-2 \label{Th-r38b}
\end{align}

To simplify the analysis, (\ref{Th-r38a}) and (\ref{Th-r38b}) are integrated over the the volume $V_p$ of the pellet. Thus \begin{align}
\frac{\partial}{\partial t}\left[\epsilon_p V_p \left(\bar{c}_{1,p} - \frac{K_w}{\bar{c}_{1,p}}\right) \right] + \underline{n} . \left[ A_p (\underline{N}_{1,p} - \underline{N}_{n-1,p})\right]_{r=R_p} = \rho_p V_p (\bar{\dot{R}}_{1,p} - \bar{\dot{R}}_{n-1,p}) \label{Th-r39a}\\
\frac{\partial}{\partial t}\left(\epsilon_p V_p \bar{c}_{i,p}\right) + \underline{n} . A_p \left. \underline{N}_{i,p} \right|_{r=R_p} = \rho_p V_p \bar{\dot{R}}_{i,p}; \quad i=2,n-2 \label{Th-r39b}
\end{align} where overbars denote volume-averaged values, $A_p$ is the external surface area of the pellet, $\underline{n}$ is the unit outward normal, and $R_p$ is the radius of the pellet (Fig.~\ref{Th-f2}).

The pellets that are normally used for adsorption are porous materials, and hence the diffusion process may occur by bulk, Knudsen, and surface diffusion \citep{Ruthven84}. Therefore, inside the pellets there will be an effective diffusion. Ignoring convection within the pellets, the flux of a species $i$ in dilute electrolyte solution is given by an equation analogous to (\ref{Th-r27}), i.e. \begin{equation} \label{Th-r40}
\underline{N}_{i,p} = - \sum_{j=1}^n \mathfrak{D}_{ij}^e \nabla c_{j,p}
\end{equation} where $\mathfrak{D}_{ij}^e$ denotes an effective diffusion coefficient. The above equation is similar to the dusty gas model for a dilute mixture \citep{Krishna97}, but with the incorporation of an electrical force. Therefore 
\begin{align}
\mathfrak{D}_{ij}^e \equiv D_i^e \delta_{ij} &- \frac{t_i}{z_i} z_j D_j^e \label{Th-r41a} \\
\frac{1}{D_i^e} = \frac{1}{D_{iw}^e} + \frac{1}{D_{im}^e}&;\quad t_i = \frac{z_i^2 D_i^e c_{i,p}}{\sum_{j=1}^n z_j^2 D_j^e c_{j,p}} \label{Th-r41b}  
\end{align}
where $D_{iw}^e$ is an effective ion-water bulk diffusion coefficient, and $D_{im}^e$ is an effective Knudsen diffusion coefficient of species $i$ in the porous medium. In the present model, the Knudsen diffusion is neglected. For the porous medium, $D_i^e = D^e_{iw} = \epsilon_p^{3/2} D_i$ \citep{Tjaden16}. The radial component of the flux is given by \begin{equation} \label{Th-r42}
\underline{n}. \underline{N}_{i,p} = \left. N_{i,p}\right|_r = - \sum_{j=1}^n \mathfrak{D}_{ij}^e \frac{\partial c_{j,p}}{\partial r}
\end{equation}

In the spirit of the linear driving force model \citep{Gleuckauf55,Sircar00,Moreira06,Tefera14}, we approximate $\left. \frac{\partial c_{j,p}}{\partial r}\right|_{r=R_p}$ by \begin{equation} \label{Th-r43}
\left. \frac{\partial c_{j,p}}{\partial r}\right|_{r=R_p} = \frac{c_{j,p,s} - \bar{c}_{j,p}}{\delta '} = \frac{c_{j,s} - \bar{c}_{j,p}}{\delta '}
\end{equation} where $\bar{c}_{j,p}$ is the volume averaged concentration of $j$ in the pellet, $c_{j,p,s} = c_{j,s}$ is the value of $c_j$ at the external surface of the pellet, and $\delta '$ is a constant. Hence the fluxes are given by 
\begin{align}
\left. N_{1,p}\right|_{r=R_p} - \left. N_{n-1,p}\right|_{r=R_p} &= - \sum_{j=1}^n (\mathfrak{D}_{1j}^e - \mathfrak{D}_{n-1j}^e) \dfrac{(c_{j,s}-\bar{c}_{j,p})}{\delta '} \label{Th-r43.1a}\\
\left. N_{i,p}\right|_{r=R_p} &= - \sum_{j=1}^n \mathfrak{D}_{ij}^e \dfrac{(c_{j,s}-\bar{c}_{j,p})}{\delta '}; \quad i = 2,n-2 \label{Th-r43.1b}
\end{align}
Using (\ref{Th-r43}) and (\ref{Th-r42}), (\ref{Th-r39a}) and (\ref{Th-r39b}) can be rewritten as 
\begin{align} 
\frac{\partial}{\partial t}\left[ \epsilon_p V_p \left( \bar{c}_{1,p} - \frac{K_w}{\bar{c}_{1,p}} \right) \right] &- A_p \sum_{j=1}^n \left(\mathfrak{D}_{1j}^e - \mathfrak{D}_{n-1j}^e \right) \frac{(c_{j,s} - \bar{c}_{j,p})}{\delta '} = \rho_p V_p (\bar{\dot{R}}_{1,p} - \bar{\dot{R}}_{n-1,p}) \label{Th-r44a} \\
\frac{\partial}{\partial t}( \epsilon_p V_p \bar{c}_{i,p}) &- A_p \sum_{j=1}^n \mathfrak{D}_{ij}^e \frac{(c_{j,s} - \bar{c}_{j,p})}{\delta '} = \rho_p V_p \bar{\dot{R}}_{i,p}; \quad i=2,n-2  \label{Th-r44b}
\end{align}

The concentration $c_{j,s}$ in the above equations may be eliminated by ensuring the continuity of fluxes at the surface of the pellet. Thus 
\begin{align}
\tilde{N}_{1,x} - \tilde{N}_{n-1,x} = - \left. (N_{1,p} - N_{n-1,p})\right|_{r=R_p}  \label{Th-r45a} \\
\tilde{N}_{i,x} = - \left. N_{i,p}\right|_{r=R_p}; \quad i=2,n-2  \label{Th-r45b}
\end{align}
Substituting the fluxes we obtain
\begin{align} 
\sum_{j=1}^n \left(\mathfrak{D}_{1j} - \mathfrak{D}_{n-1,j} \right) \frac{(c_{j,b} - c_{j,s})}{\delta} &= \sum_{j=1}^n \left(\mathfrak{D}_{1j}^e - \mathfrak{D}_{n-1,j}^e \right) \frac{(c_{j,s} - \bar{c}_{j,p})}{\delta '} \label{Th-r46a} \\
\sum_{j=1}^n \mathfrak{D}_{ij} \frac{(c_{j,b} - c_{j,s})}{\delta} &= \sum_{j=1}^n \mathfrak{D}_{ij}^e \frac{(c_{j,s} - \bar{c}_{j,p})}{\delta '};\quad i=2,n-2 \label{Th-r46b}
\end{align} 
Equations (\ref{Th-r46a}) and (\ref{Th-r46b}) involve $n$ unknowns $c_{j,s}, j=1,n$. However, as discussed earlier, the final form of the expressions for the fluxes in the bulk and the pellet can be obtained by eliminating the species $n-1$ and $n$, and can be written as (see~\ref{app5})
\begin{align}
\begin{split} 
\tilde{N}_{1,x} - \tilde{N}_{n-1,x} = \sum_{j=1}^{n-2} k_{b,1j} (c_{j,b} - c_{j,s}),\\ \tilde{N}_{i,x} = \sum_{j=1}^{n-2} k_{b,ij} (c_{j,b} - c_{j,s});\quad i=2,n-2\\
\end{split} \label{Th-r47a} \\
\begin{split} 
\left. (N_{1,p} - N_{n-1,p})\right|_{r=R_p} = - \sum_{j=1}^{n-2} k_{p,1j} (c_{j,s} - \bar{c}_{j,p}),\\ 
\left. N_{i,p}\right|_{r=R_p} = -\sum_{j=1}^{n-2} k_{p,ij} (c_{j,s} - \bar{c}_{j,p});\quad i=2,n-2 
\end{split} \label{Th-r47b}
\end{align}
where $k_{b,ij}$ and $k_{p,ij}$ are effective mass transfer coefficients for transfer from the bulk solution to the surface of the pellet, and from the surface of the pellet to the interior, respectively. They are given by \begin{equation} \label{Th-r48}
\left.\begin{aligned}
k_{b,ij} &= \begin{cases}
 \lambda_{i1} - \frac{{\displaystyle \phi}}{{\displaystyle c_{1,s}}} \lambda_{i,n-1}& j=1\\
 \lambda_{ij}& j\neq1
 \end{cases}\\
k_{p,ij} &= \begin{cases}
 \lambda_{i1}' - \frac{{\displaystyle \phi '}}{{\displaystyle c_{1,s}}} \lambda_{i,n-1}'& j=1\\
 \lambda_{ij}'& j\neq1
 \end{cases}
\end{aligned}
\right\}
\quad i=1,n-2;\\ j=1,n-2
\end{equation} where \begin{align}
\left.\begin{aligned}
\lambda_{ij} &= \begin{cases}
 \mathbb{D}_{ij} & i\neq1\\
 \mathbb{D}_{1j} - \mathbb{D}_{n-1,j} & i=1
 \end{cases}\\
\lambda_{ij}' &= \begin{cases}
\mathbb{D}_{ij}' & i\neq1\\
 \mathbb{D}_{1j}' - \mathbb{D}_{n-1,j}' & i=1
 \end{cases}\\
\end{aligned}
\right\}
\quad i=1,n-2;j=1,n-1 \label{Th-r49a} \\ \nonumber \\
\left.\begin{aligned}
\mathbb{D}_{ij} &= \frac{{\displaystyle (\mathfrak{D}_{ij} - \xi_{jn}\mathfrak{D}_{in})}}{{\displaystyle \delta}} \\
\mathbb{D}_{ij}' &= \frac{{\displaystyle (\mathfrak{D}_{ij}^e - \xi_{jn}\mathfrak{D}_{in}^e)}}{{\displaystyle \delta'}}
\end{aligned}
\right\}
\quad i=1,n-1;j=1,n-1 \label{Th-r49b}
\end{align}  
A detailed derivation is given in \ref{app5}. The quantities $\mathfrak{D}_{ij}$, $\mathfrak{D}_{ij}^e$ are defined in (\ref{Th-r27a}) and (\ref{Th-r41a}), respectively, and $\phi$, $\phi'$, and $\xi_{ij}$ in (\ref{Th-r48}) - (\ref{Th-r49b}) are defined by
\begin{align}
\nonumber \phi = \frac{K_w}{c_{1,b}},\; \phi' = \frac{K_w}{\bar{c}_{1,p}},\; \xi_{ij} = \frac{z_i}{z_j}
\end{align}
Equations (\ref{Th-r45a}) - (\ref{Th-r47b}) imply that \begin{equation} \label{Th-r50.1}
\sum_{j=1}^{n-2} k_{b,ij}(c_{j,b} - c_{j,s}) = \sum_{j=1}^{n-2} k_{p,ij}(c_{j,s} - \bar{c}_{j,p}); \quad i = 1,n-2
\end{equation} 

Let us relate the balances in the stirred vessel to the balances in the bed. Integrating (\ref{Th-r34}) over the volume of the bed and using (\ref{Th-r47a}), we obtain \begin{equation} \label{Th-r50}
\begin{split}
V_b \frac{d}{d t}\left[\epsilon_b \left(\bar{c}_{1,b} - \frac{K_w}{\bar{c}_{1,b}}\right)\right] + &A\epsilon_b \bar{v}_z \left[\left( c_{1,b} - \frac{K_w}{c_{1,b}}\right)_{z=L} - \left(c_{1,b} - \frac{K_w}{c_{1,b}}\right)_{z=0} \right]\\
&=-a_p A \sum_{j=1}^{n-2} \int_{0}^{L} k_{b,1j} (c_{j,b} - c_{j,s})dz
\end{split}
\end{equation} where $A$ is the area of cross section of the bed, and $\bar{c}_{1,b}$ is the volume averaged bulk concentration. As the solution is recirculated at a volumetric flow rate $\dot{Q}$, we have $\dot{Q} = A \epsilon_b \bar{v}_z$, $\left.c_{1,b}\right|_{z=L} = c_{1,f}$, and $\left.c_{1,b}\right|_{z=0} = c_1$ (see Fig.~\ref{Th-f1}). Equation (\ref{Th-r50}) can be rewritten as \begin{equation} \label{Th-r51}
\begin{split}
\dot{Q}\left[\left(c_{1,f} - \frac{K_w}{c_{1,f}}\right) - \left(c_{1} - \frac{K_w}{c_{1}}\right)\right] &= -V_b \frac{d}{d t}\left[\epsilon_b\left(\bar{c}_{1,b} - \frac{K_w}{\bar{c}_{1,b}}\right)\right] \\ 
&-a_p A \sum_{j=1}^{n-2} \int_{0}^{L} k_{b,1j} (c_{j,b} - c_{j,s})dz
\end{split}
\end{equation} Similarly, the balances (\ref{Th-r35}) for $i=2,n-2$ in the bulk liquid may be integrated over the volume of the bed to obtain \begin{equation} \label{Th-r52}
\dot{Q} (c_{i,f} - c_i) = - V_b \frac{d}{d t} (\epsilon_b \bar{c}_{i,b}) - a_p A \sum_{j=1}^{n-2} \int_{0}^{L} k_{b,ij} (c_{j,b} - c_{j,s})\;dz; \quad i=2,n-2
\end{equation}

Using (\ref{Th-r51}) and (\ref{Th-r52}), the mass balances (\ref{Th-r18a}) and (\ref{Th-r18b}) for the stirred vessel take the form
\begin{align}
\begin{split}
V&\left(\frac{dc_{1}}{dt} - \frac{dc_{n-1}}{dt}\right) = -V_b \frac{d}{dt} \left[\epsilon_b\left(\bar{c}_{1,b} - \frac{K_w}{\bar{c}_{1,b}}\right)\right] - a_p A \sum_{j=1}^{n-2} \int_{0}^{L} k_{b,1j} (c_{j,b} - c_{j,s})\;dz
\end{split} \label{Th-r53a} \\
\begin{split}
V&\frac{dc_{i}}{dt} = -V_b \frac{d}{dt} (\epsilon_b \bar{c}_{i,b}) - a_p A \sum_{j=1}^{n-2} \int_{0}^{L} k_{b,ij} (c_{j,b} - c_{j,s})\;dz; \quad i=2,n-2
\end{split} \label{Th-r53b}
\end{align}

Using (\ref{Th-r43.1a}) - (\ref{Th-r44b}) and (\ref{Th-r47b}), we obtain 
\begin{align}
\sum_{j=1}^{n-2} k_{p,1j} (c_{j,s} - \bar{c}_{j,p}) &= \frac{V_p}{A_p} \frac{\partial}{\partial t}\left[\epsilon_p\left(\bar{c}_{1,p} - \frac{K_w}{\bar{c}_{1,p}}\right)\right] - \frac{\rho_p V_p}{A_p} (\bar{\dot{R}}_{1,p} - \bar{\dot{R}}_{n-1,p}); \label{Th-r54a} \\
\sum_{j=1}^{n-2} k_{p,ij} (c_{j,s} - \bar{c}_{j,p}) &= \frac{V_p}{A_p} \frac{\partial}{\partial t}(\epsilon_p \bar{c}_{i,p}) - \frac{\rho_p V_p}{A_p} \bar{\dot{R}}_{i,p};\quad i=2,n-2 \label{Th-r54b}
\end{align}
Substituting (\ref{Th-r54a}) and (\ref{Th-r54b}) in (\ref{Th-r53a}) and (\ref{Th-r53b}), respectively, and using (\ref{Th-r50.1}) the final form of the overall mass balance is given by \begin{equation} \label{Th-r55}
V \frac{dc_i '}{dt} = - V_b \frac{d}{d t}(\epsilon_b \bar{c}_{i,b}') -\frac{a_p A V_p}{A_p} \int_0^L \left[\frac{\partial}{\partial t}(\epsilon_p \bar{c}_{i,p}') - \rho_p\bar{\dot{R}}_{i,p}'\right]dz; \quad i =1,n-2
\end{equation} where $c_1' = c_1 - K_w/c_1, \bar{\dot{R}}_{1,p}' = \bar{\dot{R}}_{1,p} - \bar{\dot{R}}_{n-1,p}$ and $c_i' = c_i, \bar{\dot{R}}_{i,p}' = \bar{\dot{R}}_{i,p};\; i=2,n-2$, subscripts $p$ and $b$ represent concentrations in the solution in the particle and the bulk, respectively, and an overbar represent the average concentration. As the volume of the bed $V_b$ is $\ll$ $V$, the volume of the stirred vessel, (\ref{Th-r55}) may be approximted by \begin{equation} \label{Th-r56}
V \frac{dc_i'}{dt} = -\frac{a_p A V_p}{A_p} \int_0^L \left[\frac{\partial}{\partial t}(\epsilon_p \bar{c}_{i,p}') - \rho_p\bar{\dot{R}}'_{i,p}\right]dz; \quad i =1,n-2
\end{equation} and the mass balance on the adsorbate is \begin{equation} \label{Th-r57}
\dfrac{\partial \bar{q}_j}{\partial t} = \bar{\dot{R}}'_{j,p}; \quad j =1,m
\end{equation} 

\section{Results and discussion}
\subsection{Adsorption of single components}
\begin{figure}[ht!]
\begin{center}
\includegraphics[width=0.45\textheight]{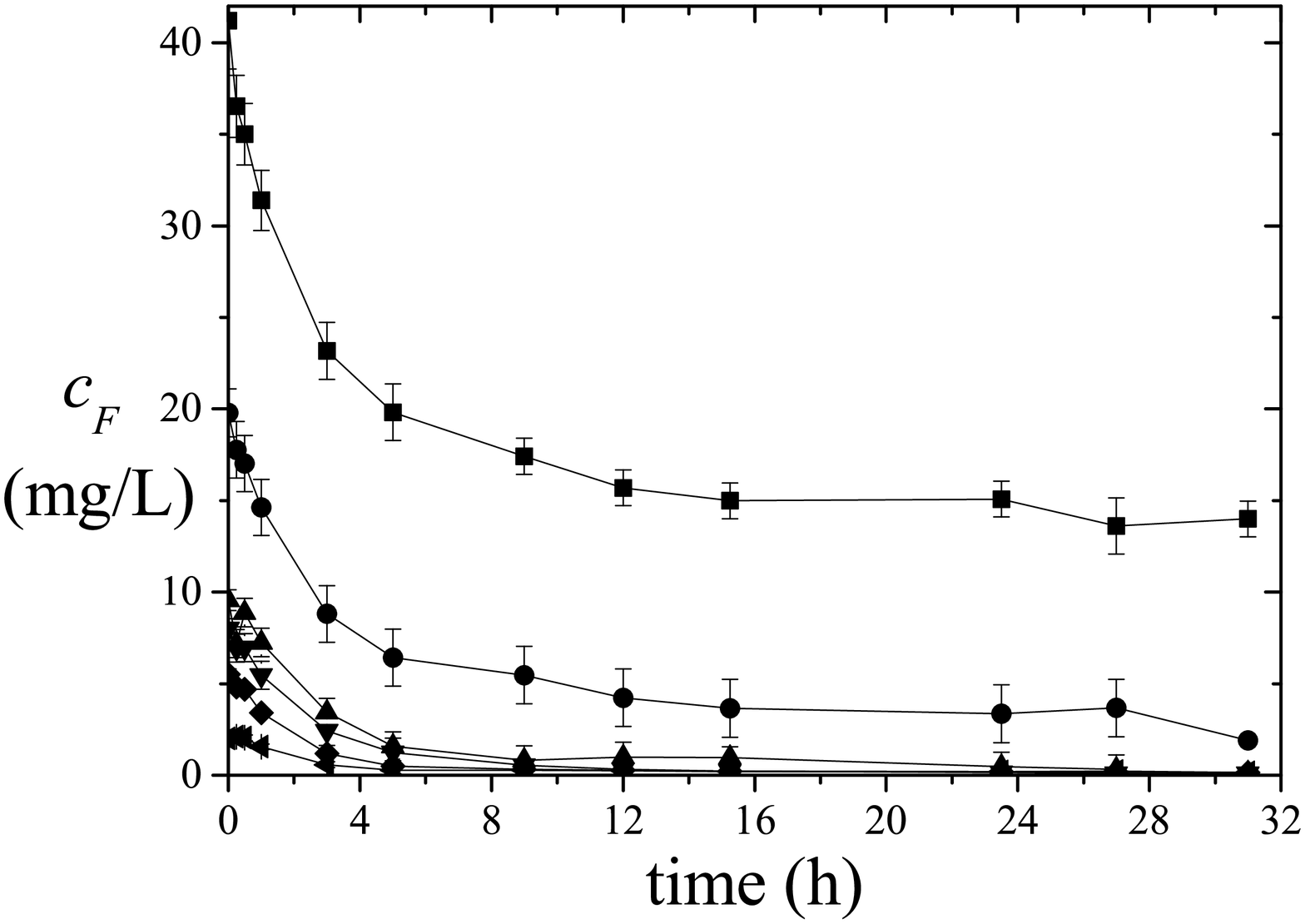}
\end{center}
\caption{Variation of the concentration of \ce{F-}, $c_{F}$ with time of operation for different initial concentrations of \ce{F-} in mg/L: $\filledmedsquare$, 40; {\huge$\fontdimen16\textfont2=2.5pt _\bullet$}, 20; $\filledmedtriangleup$, 10; $\filledmedtriangledown$, 8; {\huge $\fontdimen16\textfont2=2pt _\filleddiamond$}, 5; $\filledmedtriangleleft, 2$. Mass of the adsorbent used $m_{AA} = 1.0$ g and volume of solution taken $V_s = 200$ mL.}
\label{Re2-f1}
\end{figure}

\begin{figure}[ht!]
\begin{center}
\includegraphics[width=0.45\textheight]{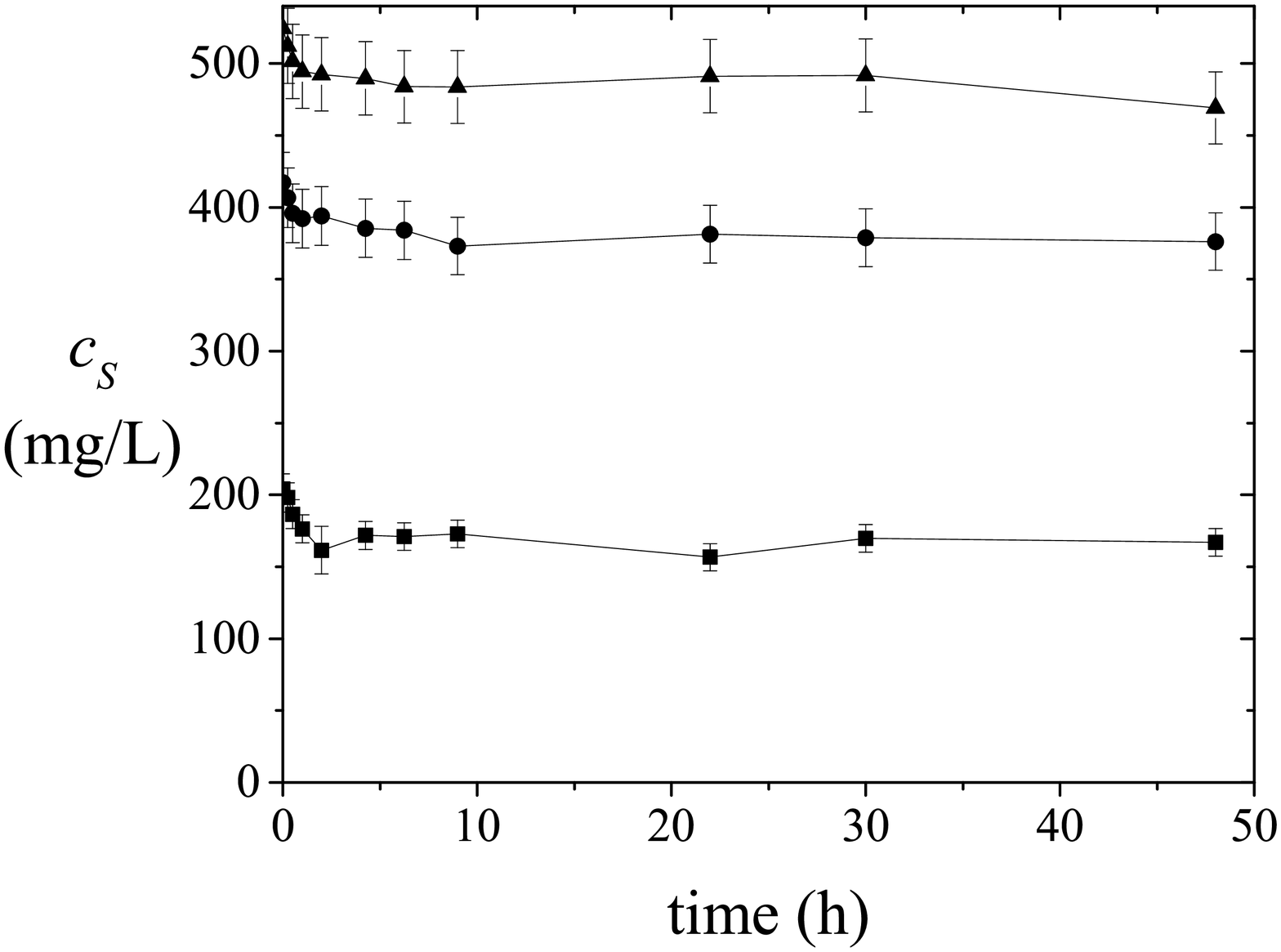}
\end{center}
\caption{Variation of the concentration of \ce{SO4^{2-}}, $c_{S}$ with the time of operation for different initial concentrations of \ce{SO4^{2-}} in mg/L : $\filledmedtriangleup$, 524; {\huge$\fontdimen16\textfont2=2.5pt _\bullet$}, 417; $\filledmedsquare$, 204. Parameter values are as in Fig.~\ref{Re2-f1}.}
\label{Re2-f2}
\end{figure}

\begin{figure}[ht!]
\begin{center}
\includegraphics[width=0.4\textheight]{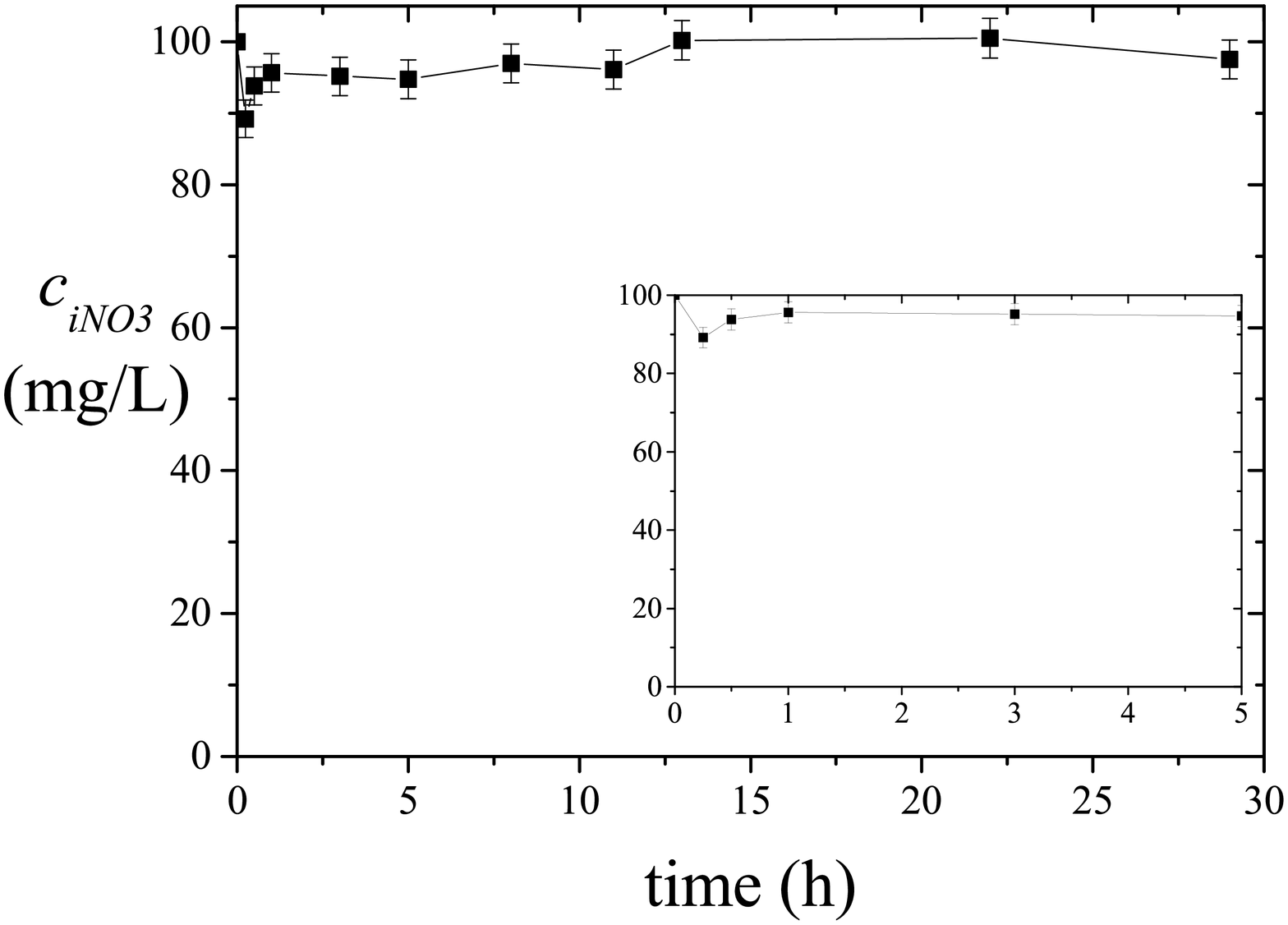}
\end{center}
\caption{Variation of the concentration of \ce{NO3-}, $c_{N}$ with the time of operation for an initial concentration of 100 mg/L of \ce{NO3-}. The inset shows the variation of $c_N$ in the first few hours of operation. Parameter values are as in Fig.~\ref{Re2-f1}.}
\label{Re2-f3}
\end{figure}

\begin{figure}[ht!]
\begin{center}
\includegraphics[width=0.45\textheight]{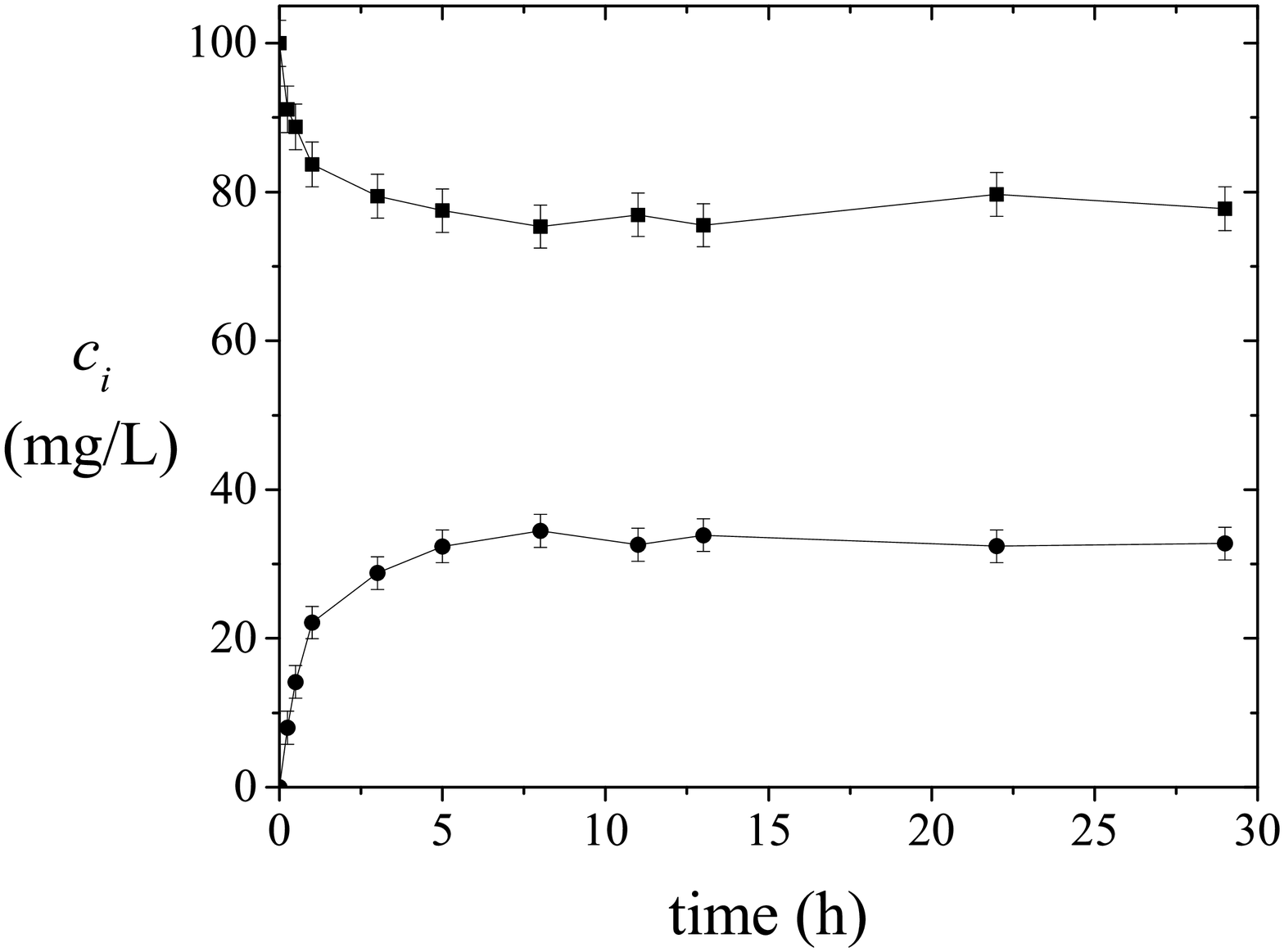}
\end{center}
\caption{Variation of the concentration of species $i$, $c_{i}$ with the time of operation for an initial concentration of 100 mg/L of \ce{Cl-}: $\filledmedsquare$, \ce{Cl-}; {\huge$\fontdimen16\textfont2=2.5pt _\bullet$}, \ce{NO3-}. Parameter values are as in Fig.~\ref{Re2-f1}.}
\label{Re2-f4}
\end{figure}

\begin{figure}[ht]
\begin{center}
\includegraphics[width=0.4\textheight]{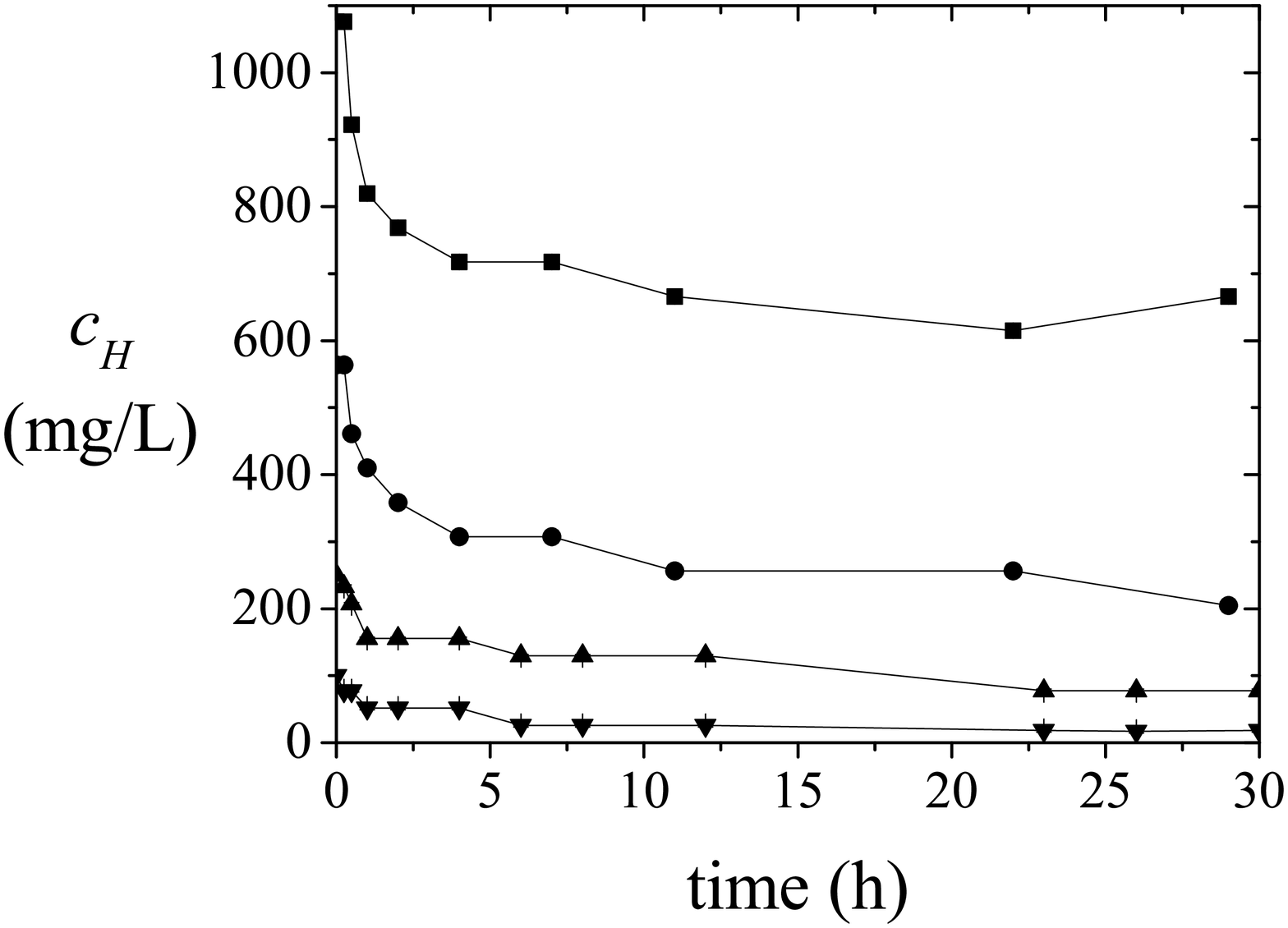}
\end{center}
\caption{Variation of the concentration of \ce{HCO3-}, $c_{H}$ with the time of operation for different initial concentrations of \ce{HCO3-} in mg/L: $\filledmedsquare$, 1076; {\huge$\fontdimen16\textfont2=2.5pt _\bullet$}, 563; $\filledmedtriangleup$, 250; $\filledmedtriangledown$, 100. Parameter values are as in Fig.~\ref{Re2-f1}.}
\label{Re2-f5}
\end{figure}

\begin{figure}[ht!]
\begin{center}
\includegraphics[width=0.35\textheight]{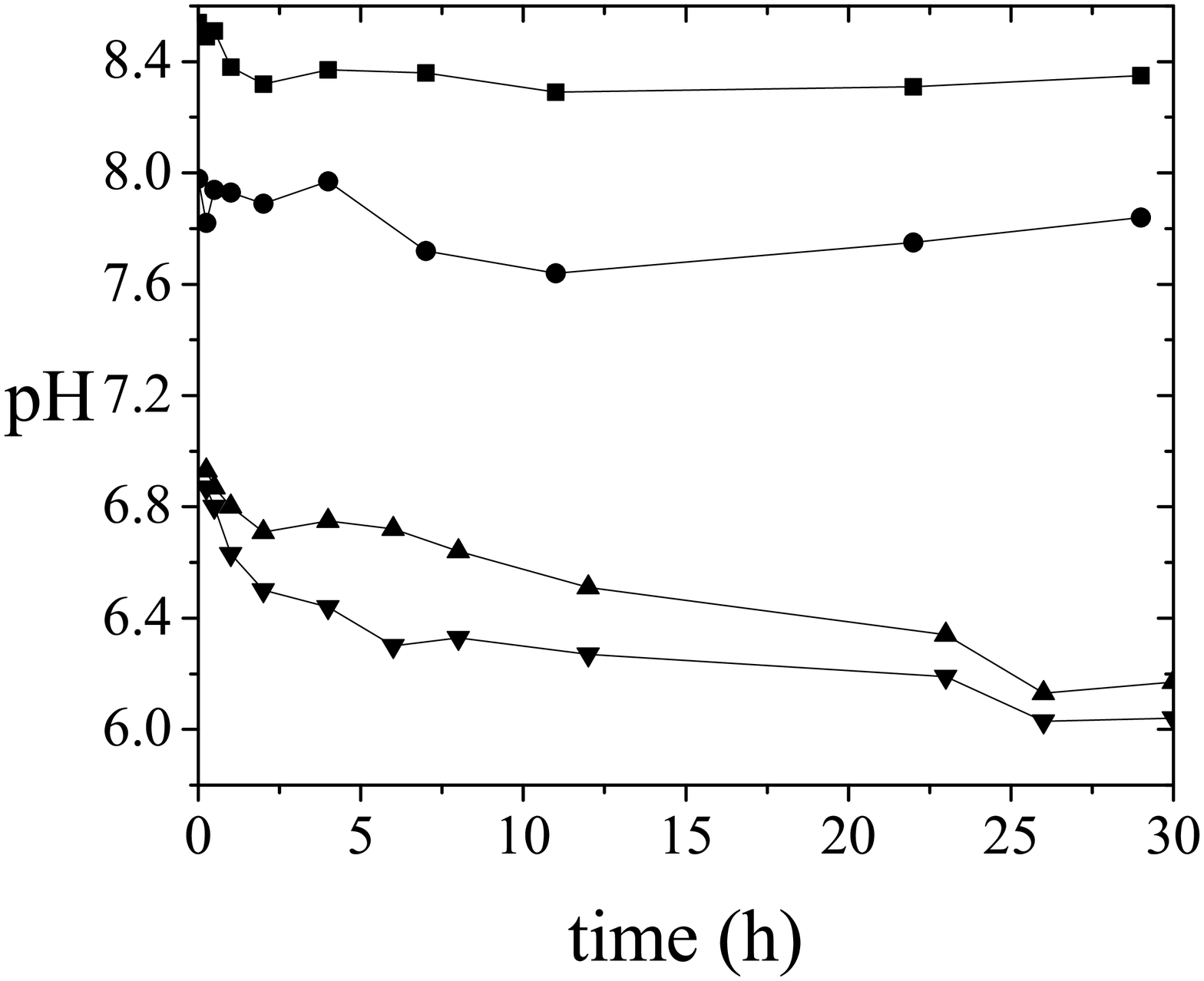}
\end{center}
\caption{Variation of pH with the time of operation for different initial concentrations of \ce{HCO3-}: $\filledmedsquare$, 1076; {\huge$\fontdimen16\textfont2=2.5pt _\bullet$}, 563; $\filledmedtriangleup$, 250; $\filledmedtriangledown$, 100. Parameter values are as in Fig.~\ref{Re2-f1}.}
\label{Re2-f6}
\end{figure}

Activated alumina (AA) is an amphoteric oxide because of its ability to accept and donate protons. During this process, based on the affinity of the sites on the adsorbent towards the ions, they are adsorbed on the surface. Therefore, in order to know the affinity of different ions for the adsorbent, experiments were conducted with \ce{F-}, \ce{HCO3-}, \ce{SO4^{2-}}, \ce{NO3-}, and \ce{Cl-}. Fluoride, \ce{HCO3-}, and \ce{SO4^{2-}} were adsorbed on AA with an increase in the pH of the solution for \ce{F-} and \ce{SO4^{2-}}, but a decrease in the pH for \ce{HCO3-}. An equilibrium was attained in 15-20 h for \ce{F-} (Fig.~\ref{Re2-f1}), and 5 h for \ce{SO4^{2-}} (Fig.~\ref{Re2-f2}).  The uptake of \ce{SO4^{2-}} was very low. With respect to \ce{NO3-}, there was a slight decrease in the concentration of \ce{NO3-} in the solution at the start of the experiment, but after 1 h it was constant (Fig.~\ref{Re2-f3}). Therefore, there it can be assumed that there was no adsorption of \ce{NO3-} onto AA. For \ce{Cl-}, there was a decrease in the concentration for 2-4 h, along with a leaching of \ce{NO3-} into the solution (Fig.~\ref{Re2-f4}). This was probably because \ce{HNO3} was used as a binding agent during the manufacture of AA (Oxide India Pvt. Ltd., private communication 2014). Even after soaking in DI water, some amount of \ce{NO3-} might have remained on the adsorbent. Hence, the drop in \ce{Cl-} can be caused by an exchange of \ce{Cl-} and \ce{NO3-} to maintain an electrically neutral solution. However, this argument does not hold during the adsorption of \ce{SO4^{2-}} and \ce{F-}. The leaching of \ce{NO3-} has attained equilibrium after 1 h during the adsorption of \ce{SO4^{2-}} and after 4 h during the adsorption of \ce{F-}. Adsorption of \ce{HCO3-} attained an equilibrium at about 10 h (Fig.~\ref{Re2-f5}) and there was a faster drop in the pH at lower concentrations of \ce{HCO3-} compared to that with higher concentrations (Fig.~\ref{Re2-f6}).  

In the case of ions which were adsorbed, there was either release or uptake of \ce{H+} or \ce{OH-}, which causes a change in the pH. From some of the surface studies with the help of FTIR, it was predicted that the possible reactions of these ions with the adsorbent can be \citep{Hao86}
\begin{align}
\ce{AlOH2+} + \ce{F-} &\ce{<=>} \ce{AlF} + \ce{H2O}\\
\ce{AlOH} + \ce{F-} &\ce{<=>} \ce{AlF} + \ce{OH-}
\end{align}

\begin{figure}[ht]
\begin{center}
\includegraphics[width=0.4\textheight]{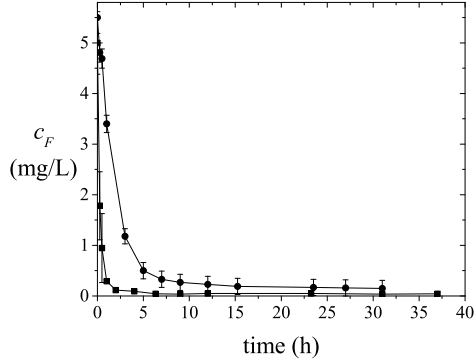}
\end{center}
\caption{Variation of the concentration of \ce{F-}, $c_{F}$ with the time of operation for an initial concentration of 5 mg/L of \ce{F-}: $\filledmedsquare$, differential bed adsorber; {\huge$\fontdimen16\textfont2=2.5pt _\bullet$}, shaker. The bulk density in both the experiments was kept at a constant value of 5 g/L. Volumetric flow rate for the differential bed adsorber = 1.5 mL/s.}
\label{Re2-f7}
\end{figure}

In the experiments with the shaker setup, discussed in section~\ref{MM2-ES}, there was no movement of the particles from the center of the flask. Therefore, to avoid this and get more reliable data, the differential bed adsorber setup was used for the experiments. It was observed that there was a considerable decease in the external mass transfer resistance (Fig.~\ref{Re2-f7}). This can help in obtaining a better prediction of the kinetic results. However, no attempt was made to eliminate the mass transfer resistance completely in the experiments.

\subsection{Negligible adsorption of NO3-, Cl-, and Na+ in the presence of F-}
For the selective removal of \ce{F-} from solutions, it is necessary to check for the dependence of the removal of \ce{F-} on the presence of other ions. These ions can either increase or decrease the uptake of \ce{F-} by the adsorbent. As the surface of AA is positively charged when the pH of the solution is below the pH of zero point charge (pH$_{\ce{zpc}}$), it is possible for the surface to adsorb or attract anions. The converse is true if pH $>$ pH$_{\ce{zpc}}$. So the adsorption of \ce{F-} onto AA was checked in the presence of ions such as \ce{NO3-}, \ce{Cl-}, and \ce{Na+}.

\begin{figure}[ht]
\begin{center}
\includegraphics[width=0.4\textheight]{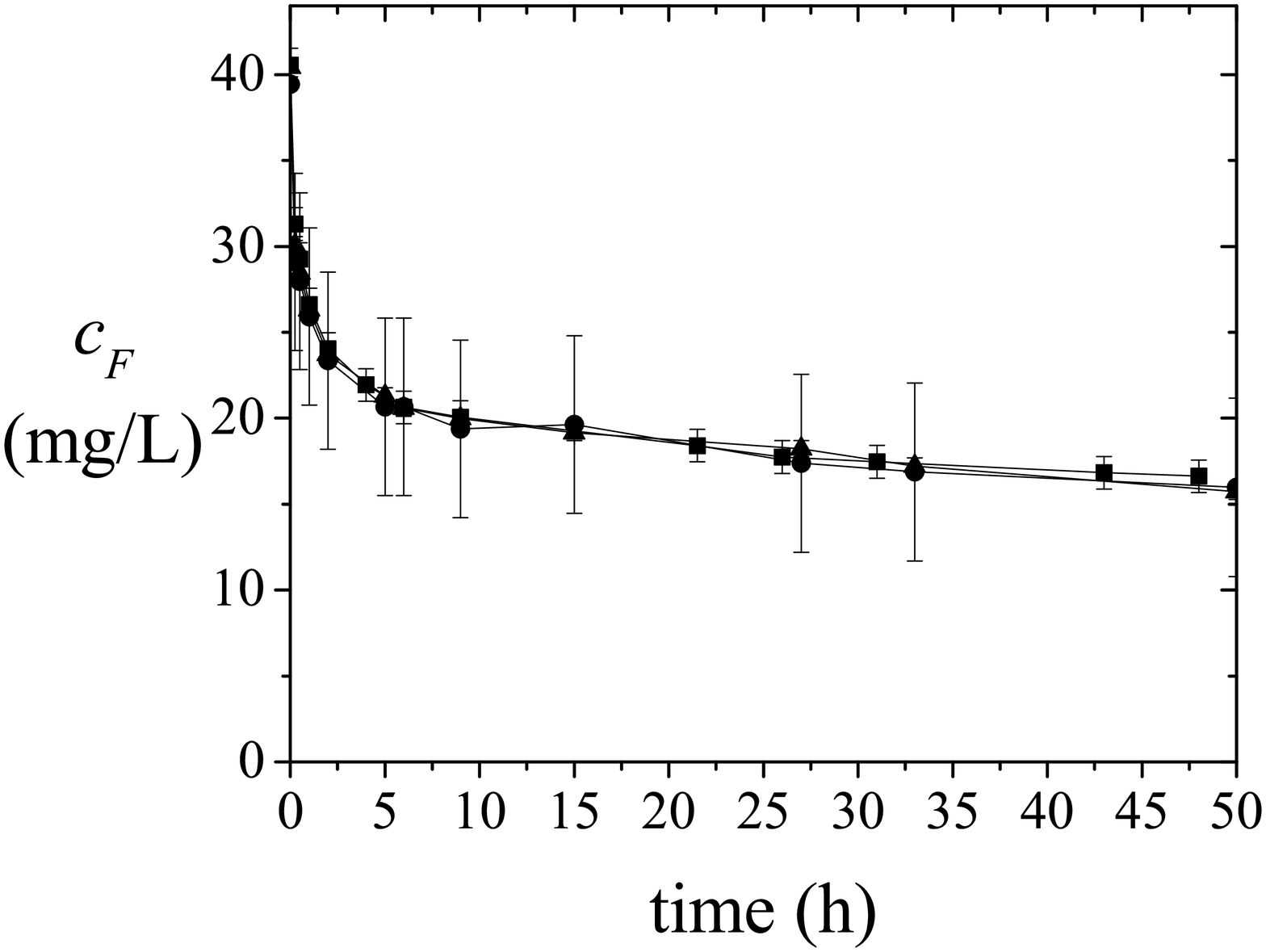}
\end{center}
\caption{Variation of the concentration of \ce{F-}, $c_{F}$ with the time of operation for an initial concentration of 40 mg/L of \ce{F-} and different initial concentrations of \ce{Cl-} in mg/L: {\huge$\fontdimen16\textfont2=2.5pt _\bullet$}, 800; $\filledmedsquare$, 200; $\filledmedtriangleup$, 0. The bulk density was 5 g/L and the volumetric flow rate was 1.5 mL/s.}
\label{Re2-f8}
\end{figure}

\begin{figure}[ht]
\begin{center}
\includegraphics[width=0.4\textheight]{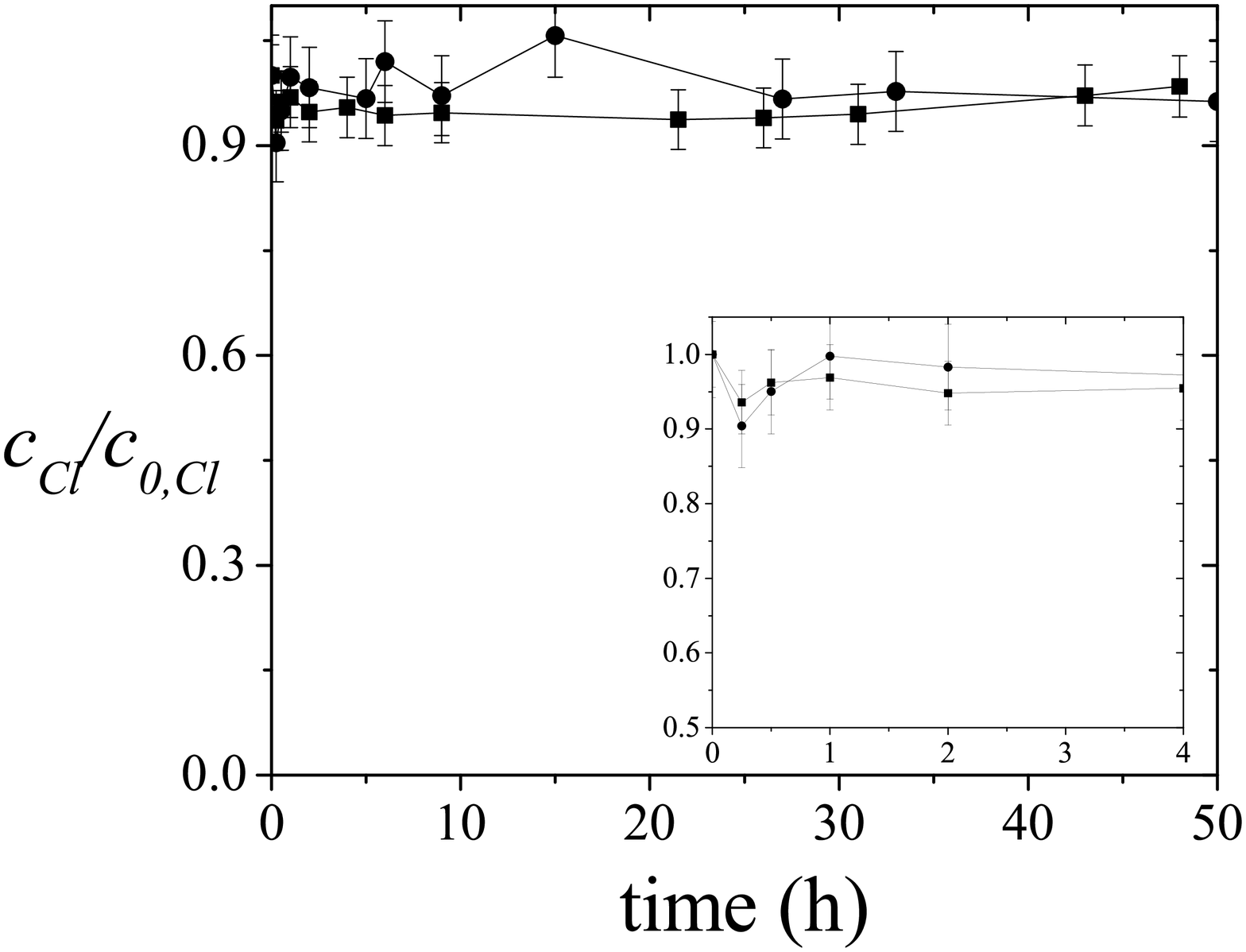}
\end{center}
\caption{Variation of the concentration of \ce{Cl-}, $c_{Cl}$ with the time of operation for an initial concentration of 40 mg/L of \ce{F-} and different initial concentrations of \ce{Cl-} ($c_{0,Cl}$) in mg/L: {\huge$\fontdimen16\textfont2=2.5pt _\bullet$}, 800; $\filledmedsquare$, 200. Parameter values are as in Fig.~\ref{Re2-f8}.}
\label{Re2-f9}
\end{figure}

\begin{figure}[ht]
\begin{center}
\includegraphics[width=0.4\textheight]{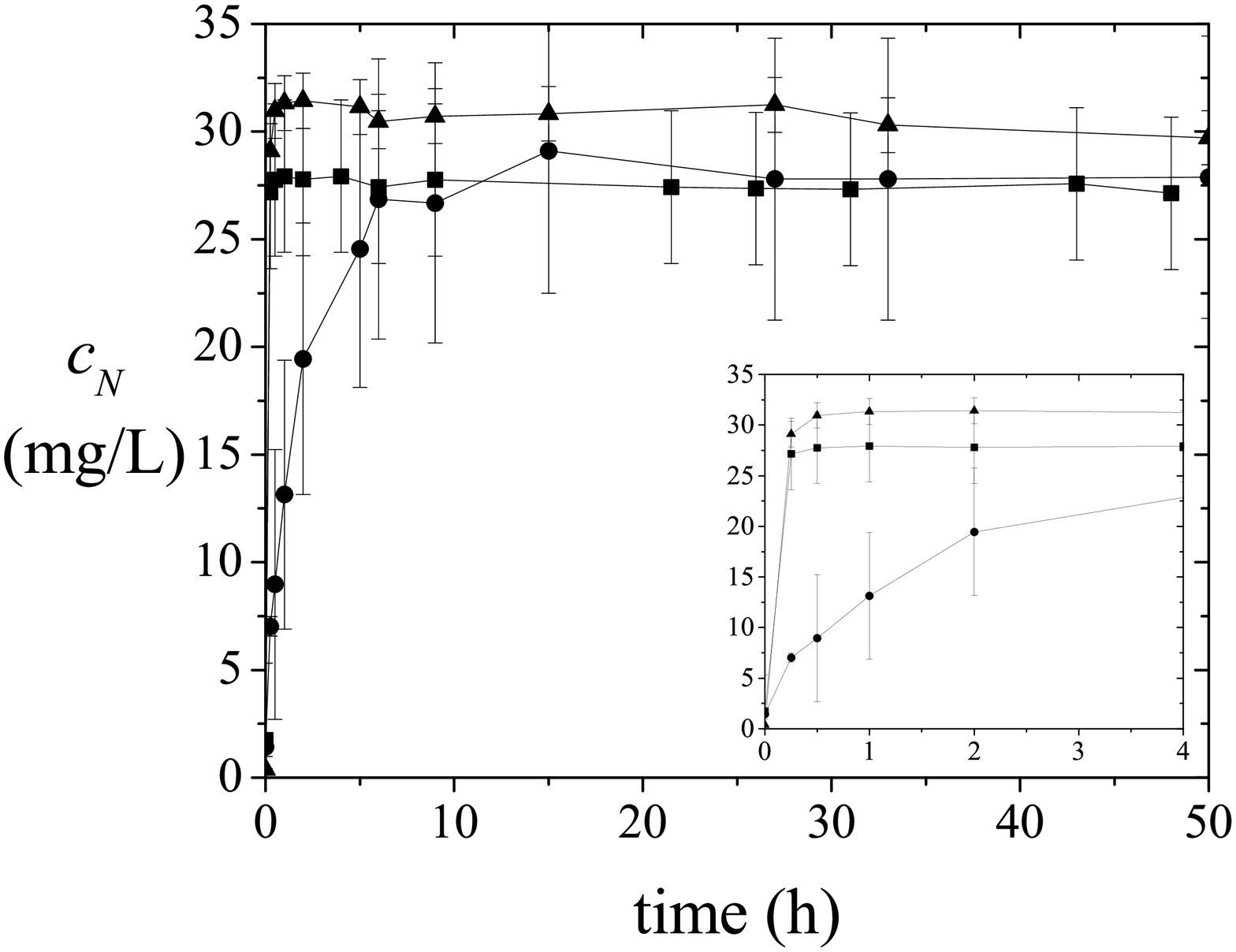}
\end{center}
\caption{Variation of the concentration of \ce{NO3-}, $c_{N}$ with the time for an initial concentration of 40 mg/L of \ce{F-} and different initial concentrations of \ce{Cl-} in mg/L: {\huge$\fontdimen16\textfont2=2.5pt _\bullet$}, 800; $\filledmedsquare$, 200; $\filledmedtriangleup$, 0. Parameter values are as in Fig.~\ref{Re2-f8}.}
\label{Re2-f10}
\end{figure}

There was no effect of \ce{Cl-} on the uptake of \ce{F-} (Fig.~\ref{Re2-f8}). This was reflected in the concentration of \ce{Cl-}, which shows that there was no adsorption of this ion onto the adsorbent (Fig.~\ref{Re2-f9}). However, when only \ce{Cl-} was present in the system, there was about 20\% uptake (Fig.~\ref{Re2-f4}). This clearly indicates that the adsorption of \ce{Cl-} is mostly to neutralize the charge present on the surface of the adsorbent i.e. the ion that is present in the double layer (Fig.~\ref{MM2-f1.1}) rather than forming a new complex species on the adsorbent. As mentioned earlier, there was leaching of \ce{NO3-} into the solution from the AA. When the concentrations of other ions such as \ce{F-} and \ce{Cl-} were increased, the leaching of \ce{NO3-} into the solution was hindered, even though the value attained at long times was almost the same (Fig.~\ref{Re2-f10}). This delay may be caused by a decrease in the effective diffusion coefficient of \ce{NO3-}, which is affected because of the higher concentration of \ce{Cl-}.  Therefore, there is a relatively higher influx of \ce{Cl-} compared to the outflow of \ce{NO3-} from the pellet.

\begin{figure}[ht!]
\begin{center}
\includegraphics[width=0.4\textheight]{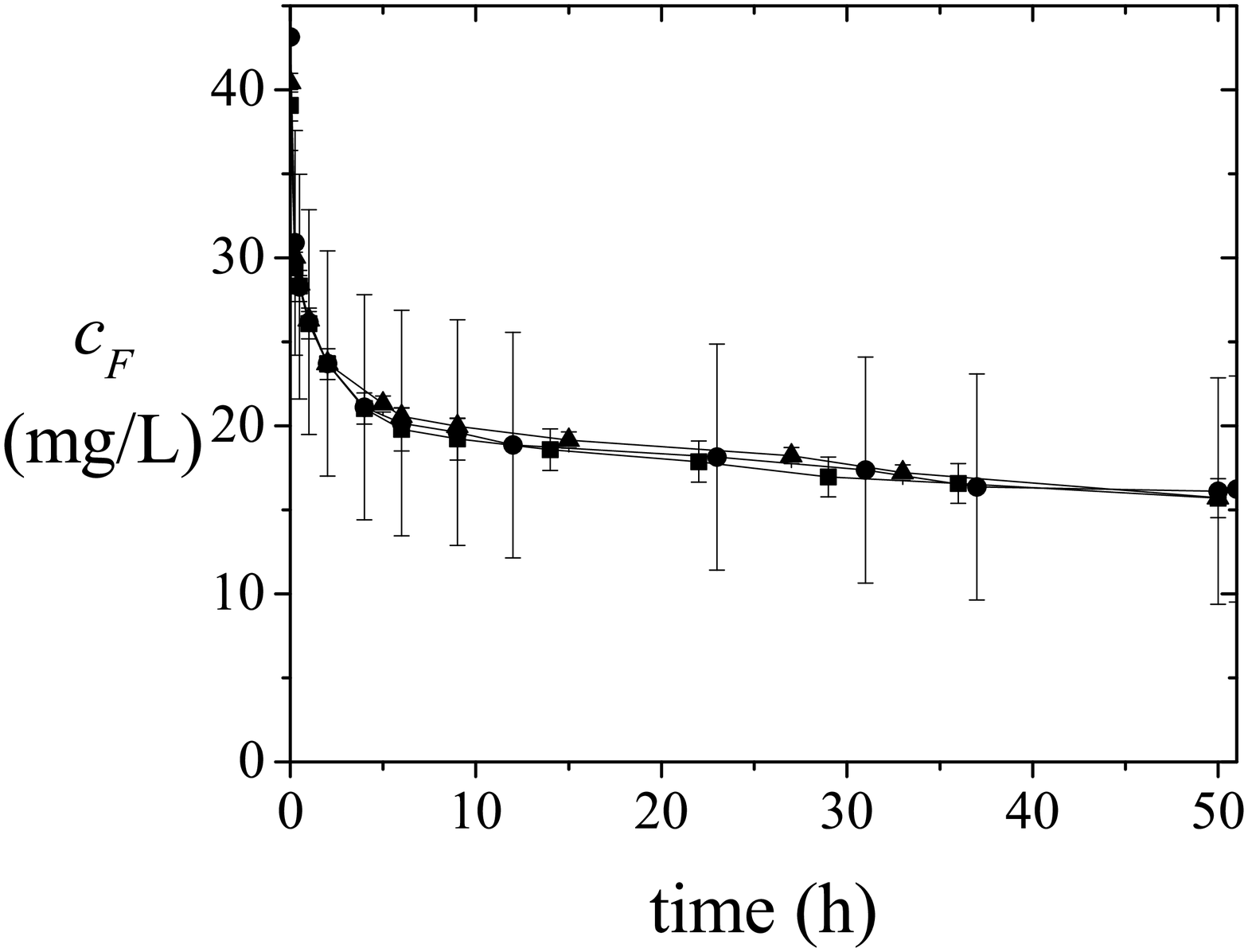}
\end{center}
\caption{Variation of the concentration of \ce{F-}, $c_{F}$ with the time for an initial concentration of 40 mg/L of \ce{F-} and different initial concentrations of \ce{NO3-} in mg/L: {\huge$\fontdimen16\textfont2=2.5pt _\bullet$}, 800; $\filledmedsquare$, 200; $\filledmedtriangleup$, 0. Parameter values are as in Fig.~\ref{Re2-f8}.}
\label{Re2-f11}
\end{figure}

\begin{figure}[ht!]
\begin{center}
\includegraphics[width=0.4\textheight]{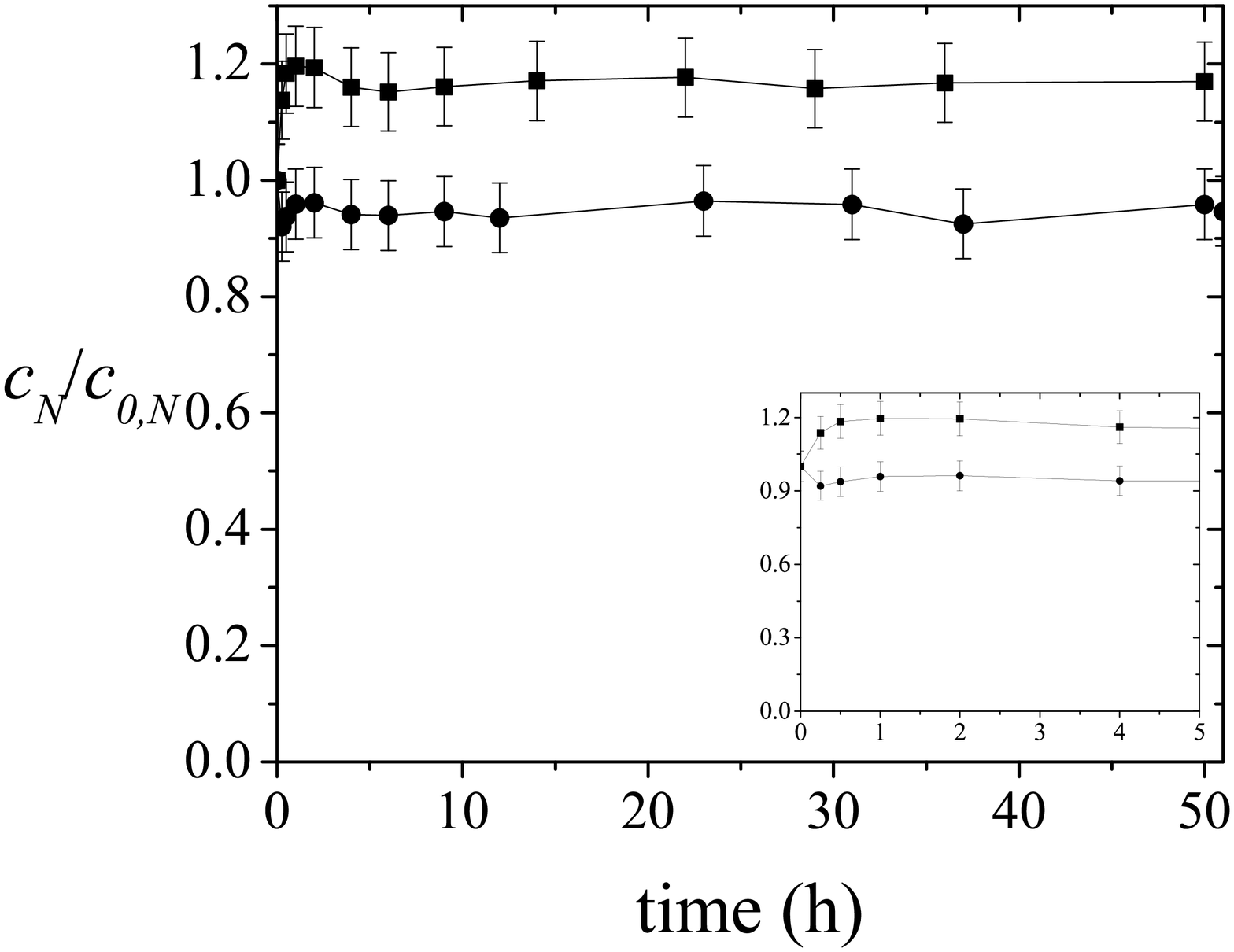}
\end{center}
\caption{Variation of the concentration of \ce{NO3-}, $c_{N}$ with the time for an initial concentration of 40 mg/L of \ce{F-} and different initial concentrations of \ce{NO3-} in mg/L: {\huge$\fontdimen16\textfont2=2.5pt _\bullet$}, 800; $\filledmedsquare$, 200. Parameter values are as in Fig.~\ref{Re2-f8}.}
\label{Re2-f12}
\end{figure}

Similarly, \ce{NO3-} had a negligible effect on the adsorption of \ce{F-} (Fig.~\ref{Re2-f11}). For an initial \ce{NO3-} concentration of 200 mg/L, there was an increase in the \ce{NO3-} concentration because of the release of the nitrate from the adsorbent. For a higher initial concentration, there was no leaching (Fig.~\ref{Re2-f12}). This may be caused by a lower concentration gradient between the \ce{NO3-} in the pellets and in the solution, which is responsible for the diffusion of \ce{NO3-} into the solution. 

\begin{figure}[ht!]
\begin{center}
\includegraphics[width=0.375\textheight]{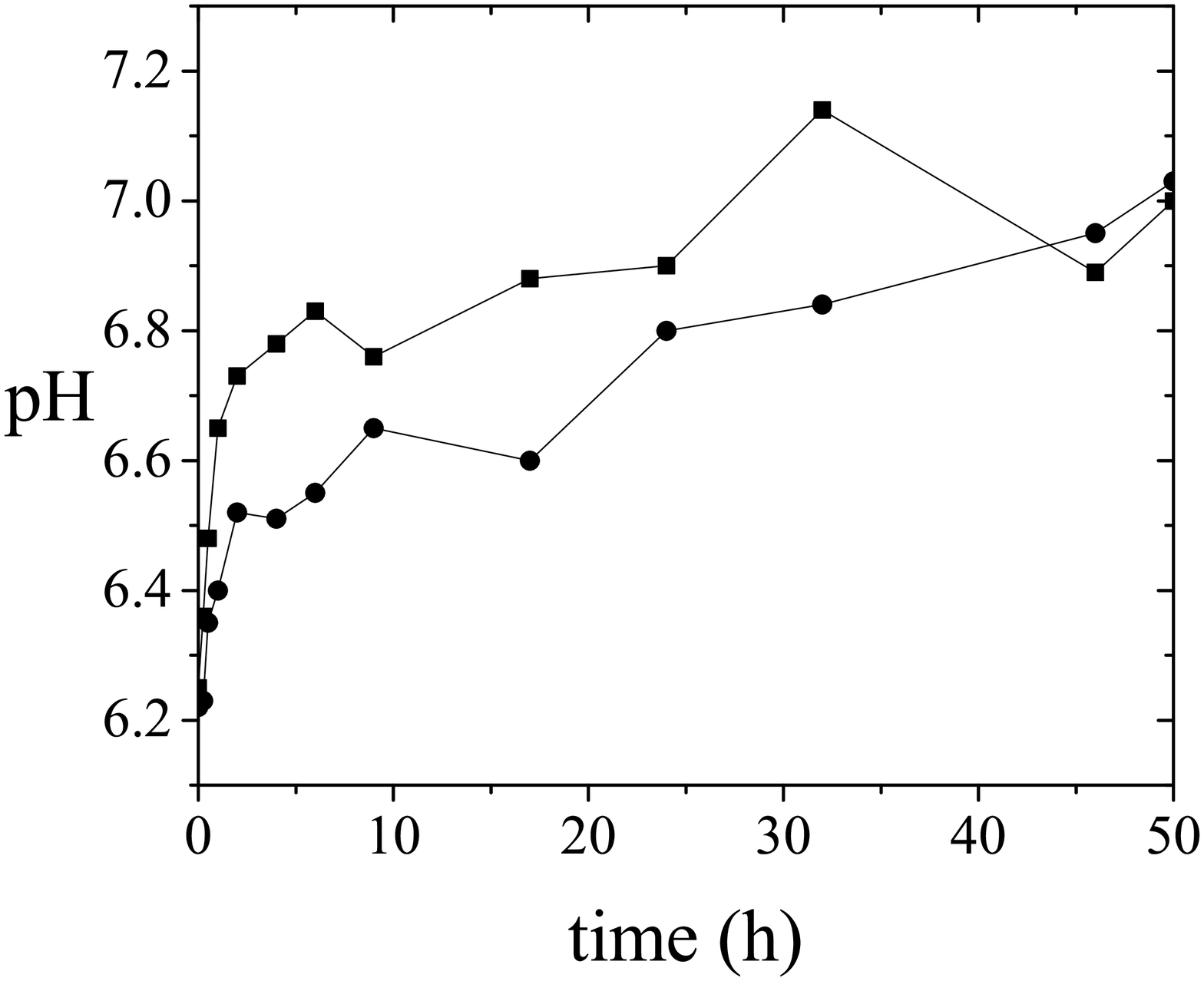}
\end{center}
\caption{Variation of the pH with time for different concentrations of \ce{Na+} and \ce{F-} in mg/L: {\huge$\fontdimen16\textfont2=2.5pt _\bullet$}, $c_{Na} = 14.6$ mg/L, $c_{F} = 9.2$ mg/L; $\filledmedsquare$, $c_{Na} = 22.6$ mg/L, $c_{F} = 14.4$ mg/L. Parameter values are as in Fig.~\ref{Re2-f8}.}
\label{Re2-f13}
\end{figure}

\begin{figure}[htbp]
\begin{center}
\includegraphics[width=0.4\textheight]{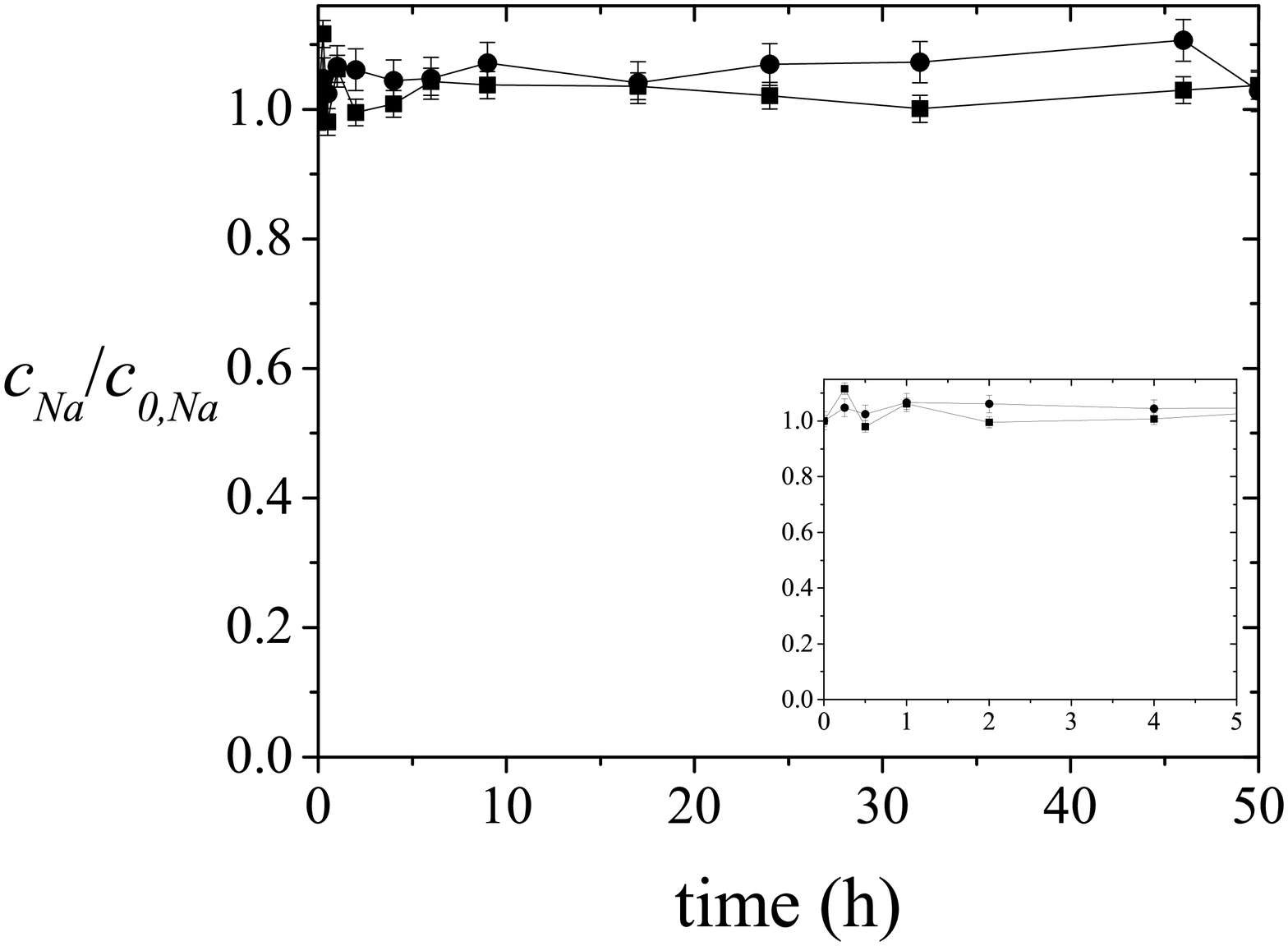}
\end{center}
\caption{Variation of the dimensionless concentration of \ce{Na+}, $c_{Na}$ with time for different concentrations of \ce{Na+} and \ce{F-}: {\huge$\fontdimen16\textfont2=2.5pt _\bullet$}, $c_{Na} = 14.6$ mg/L, $c_{F} = 9.2$ mg/L; $\filledmedsquare$, $c_{Na} = 22.6$ mg/L, $c_{F} = 14.4$ mg/L. Parameter values are as in Fig.~\ref{Re2-f8}.}
\label{Re2-f14}
\end{figure}

The adsorption of positive ions onto AA is mainly assumed to occur when the surface is negatively charged, i.e. when the pH of the solution is $>$ pH$_{\ce{zpc}}$. In the presence of \ce{F-}, there was an increase in the pH of the solution upon adsorption of \ce{F-} (Fig.~\ref{Re2-f13}), but not $>$ pH$_{\ce{zpc}}$ of the adsorbent i.e. $7.8$. The pH$_{\ce{zpc}}$ was obtained from the equilibrium constants $K_1$ and $K_2$ for the alumina using the equation \citep{Charmas95} \begin{equation}
\ce{pH_{zpc}} = \frac{1}{2}(\ce{pK1} + \ce{pK2})
\end{equation} The values of $K_1$ and $K_2$ are determined in the next section. Therefore there were not many negative sites on the adsorbent for the adsorption or attraction of the \ce{Na+} ions. This was observed in the negligible adsorption of \ce{Na+} on AA (Fig.~\ref{Re2-f14}). In these experiments, the \ce{Na+} ion was added in the form of \ce{NaF}. Hence there was a simultaneous increase in concentration of both these ions. An increment of \ce{Na+} with \ce{F-} maintained constant is possible by the addition of \ce{NaOH}. However, this leads to an increase in the pH of the solution, thereby changing the initial condition for the adsorption of \ce{F-}. With \ce{NaF}, there is negligible or no change in the pH of the solution.

\subsection{Effect of soaking on the surface of the adsorbent}
\begin{figure}[ht!]
\begin{center}
\includegraphics[width=0.35\textheight]{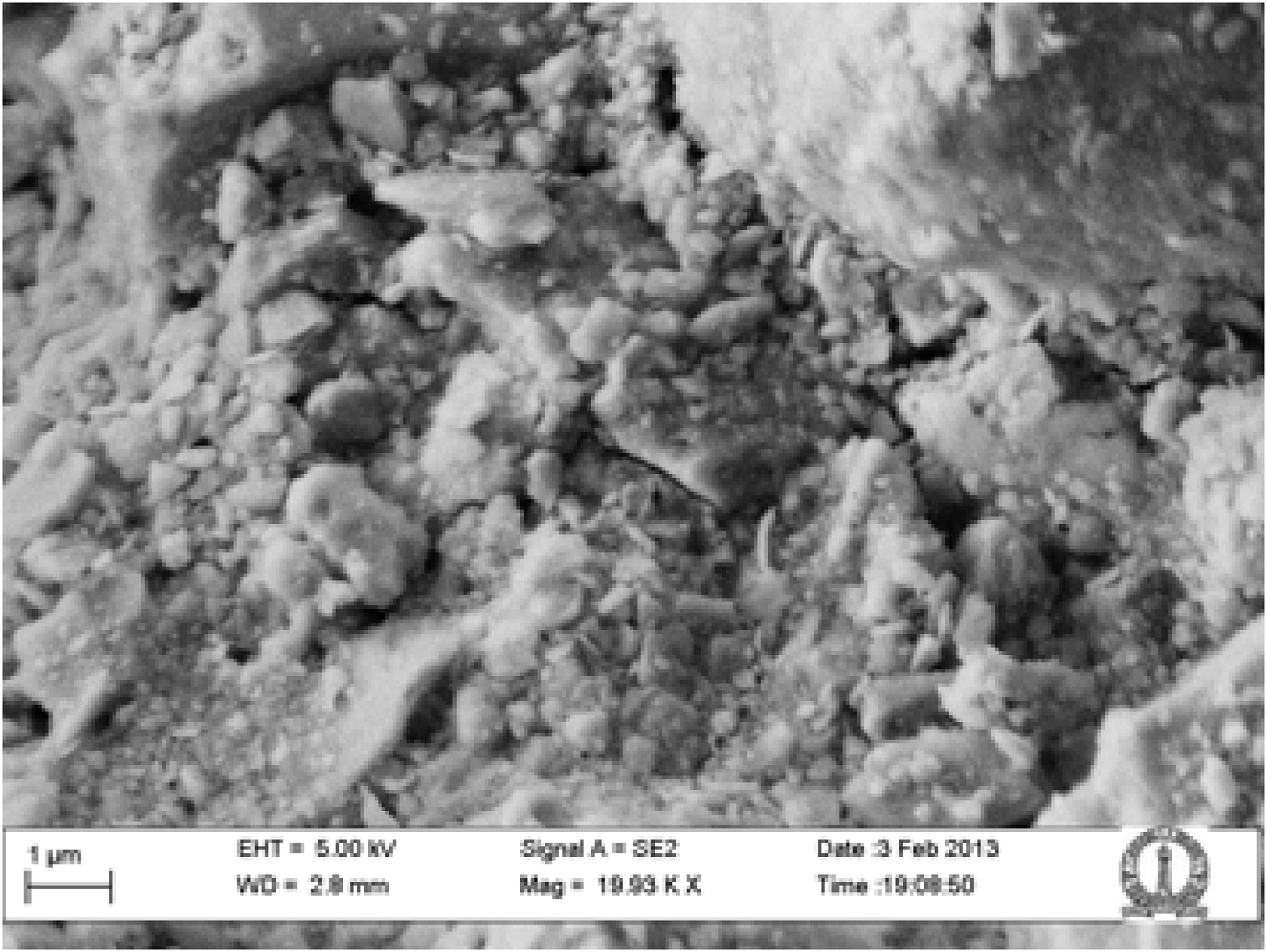}
\end{center}
\caption{Scanning electron microscope image of fresh AA. Magnification $\approx$ 20 KX.}
\label{Re2-f14a}
\end{figure}

\begin{figure}[ht!]
\begin{center}
\includegraphics[width=0.35\textheight]{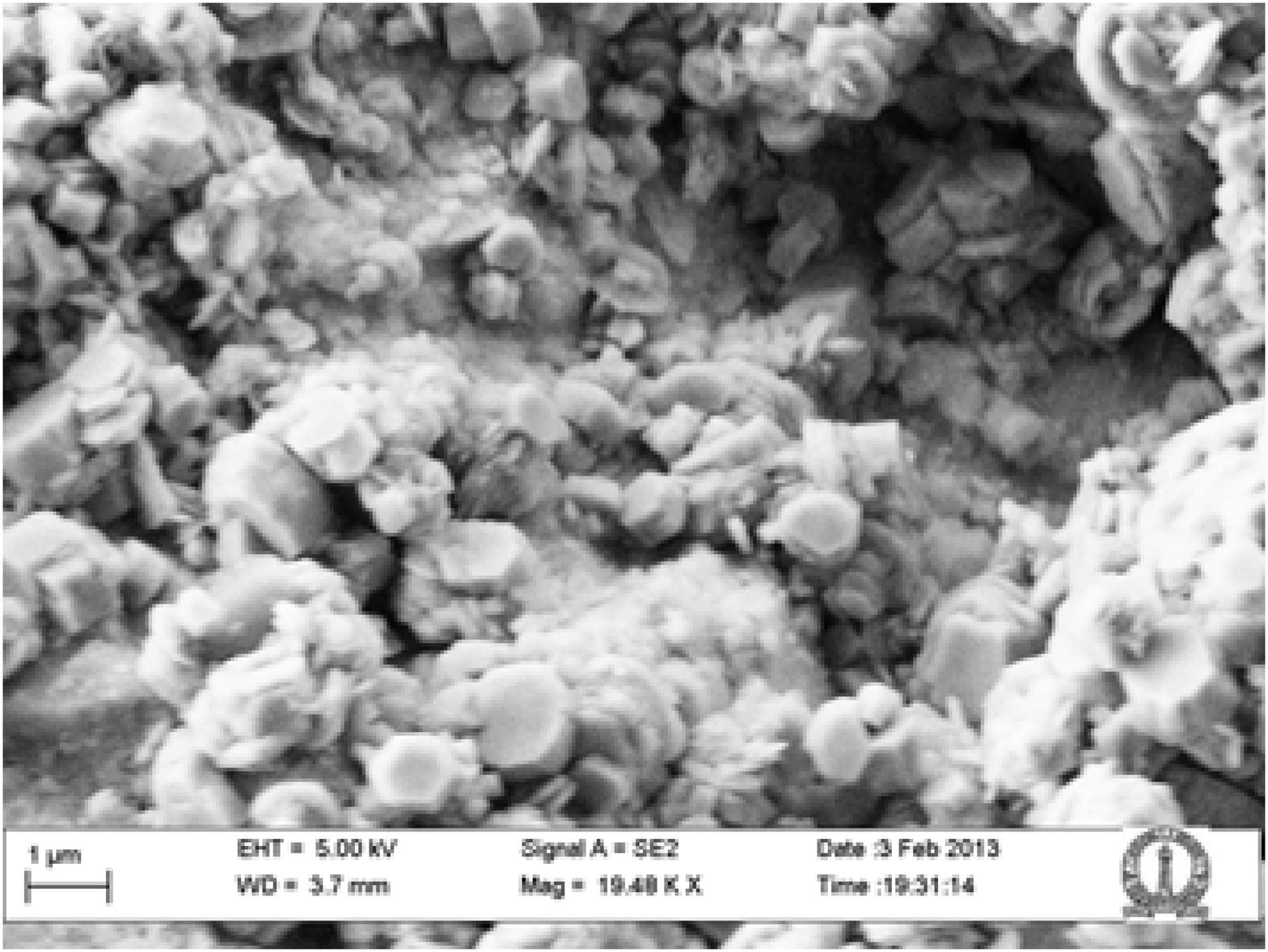}
\end{center}
\caption{Scanning electron microscope image of AA soaked for 24 h. Magnification $\approx$ 20 KX.}
\label{Re2-f14b}
\end{figure}

\begin{figure}[ht!]
\begin{center}
\includegraphics[width=0.35\textheight]{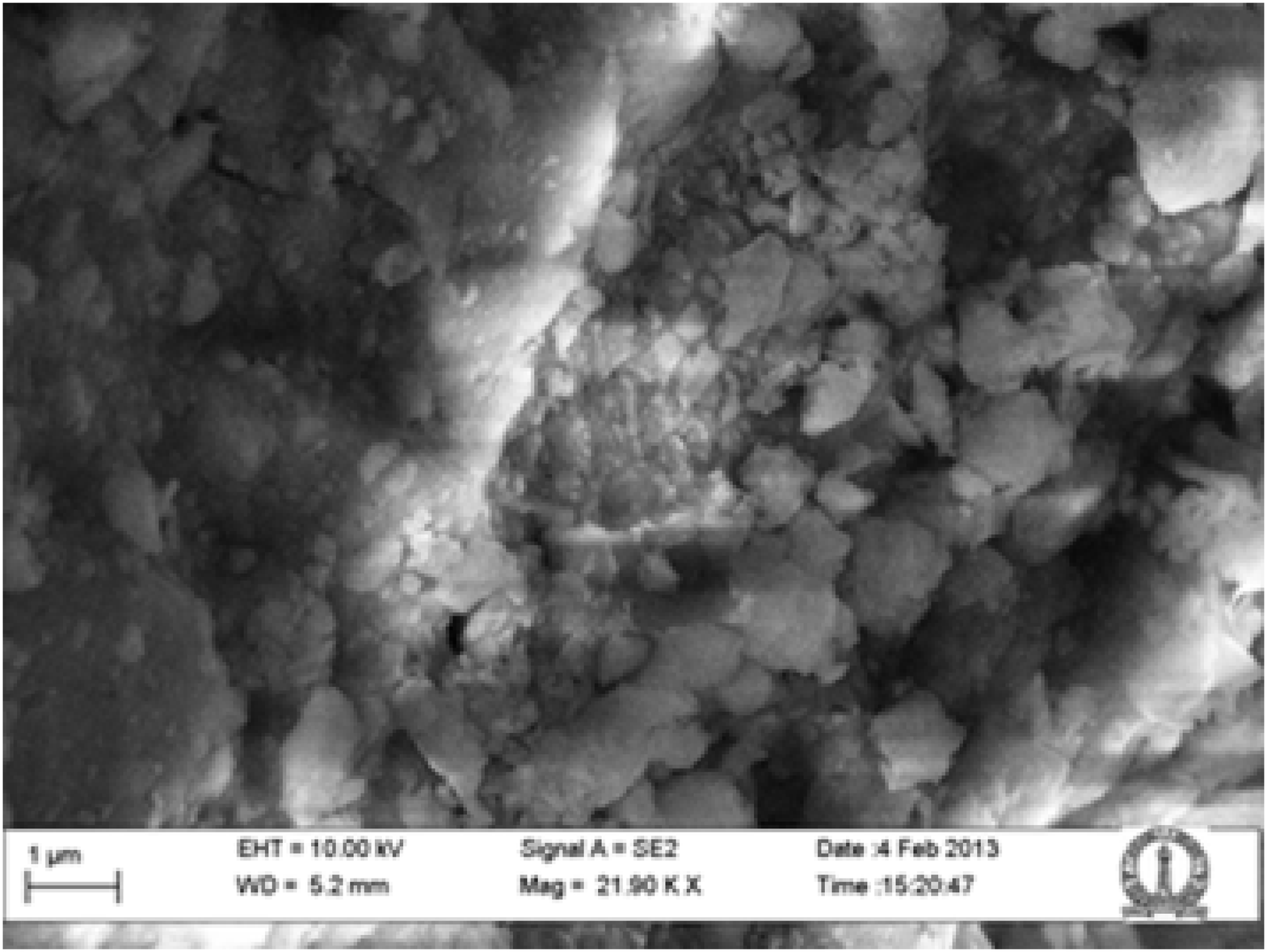}
\end{center}
\caption{Scanning electron microscope image of AA after adsorption of \ce{F-}. Magnification $\approx$ 20 KX.}
\label{Re2-f14c}
\end{figure}

Soaking of activated alumina is observed to have profound effect on the surface and reactivity of the adsorbent. \cite{Lalitha10} observed that soaking of AA in deionized water for 24 h had increased the uptake of \ce{F-} compared to fresh AA or AA soaked for 48 h. Scanning electron microscope (SEM) images show that the fresh AA had a rough surface with many small particles on it (Fig.~\ref{Re2-f14a})  but when the AA was soaked for 24 h, the surface was covered with hexagonal rings (Fig.~\ref{Re2-f14b}).  This indicates that there is some change in the crystalline structure of alumina and it might be the reason for an increased uptake of \ce{F-}. Upon adsorption of \ce{F-}, there is a change in structure to a smooth surface with mostly large particles compared to fresh AA (Fig.~\ref{Re2-f14c}).  

\subsection{Prediction of the concentrations of different ions}
\subsubsection{Estimation of the equilibrium constants}
First consider the equilibrium titration data, obtained as discussed in section~\ref{MM2-SS}. To predict the variation of the pH with the volume of acid added, estimates for the equilibrium constants $K_i$, and the total number of sites on the adsorbent $c_T$ are needed. As mentioned earlier, \ce{NO3-} leaches from AA. In order to account for \ce{NO3-} in (\ref{Th-r8.1}), along with $K_1, K_2$, and $c_T$, the initial concentration of nitrate present when the adsorbent is added to solution i.e. $c_{imp}$ was also taken as a parameter.

Initial guesses for $K_1, K_2$, and $c_T$ were obtained from the results of \cite{Lalitha10}, and an estimate for $c_{imp}$ was obtained from the experiment conducted by soaking the adsorbent in DI water before the start of the experiment. The final values of the parameters were obtained using the MATLAB routine Lsqcurvefit. However, there was a large variation in the values of $K_1$ and $K_2$ for different initial guesses (Table~\ref{Re2-t1}).\begin{table}[ht!]
\begin{center}
\caption{Estimated parameters for the titration data using the MATLAB routine Lsqcurvefit for three different initial guess values.}
\begin{tabular}{ccccccc}
\\\hline \multirow{2}*{Parameter} & \multicolumn{3}{c}{initial guess} & \multicolumn{3}{c}{final value}\\
&1&2&3&1&2&3\\\hline
$K_1*10^{-6}$ & $0.1$ & $1.0$&$100$& $5.74$&$1.66$&$100$\\
$K_2*10^{9}$ & $3.16$ & $31.6$& $1.0$&$83.2$&$0.635$&$1.03$\\
$c_T$ & \multirow{2}*{$0.115$} & \multirow{2}*{$0.12$}& \multirow{2}*{$0.042$}&\multirow{2}*{$0.236$}&\multirow{2}*{$0.277$}&\multirow{2}*{$0.24$}\\
 (mmol/g)&&&&&&\\
$c_{imp}$& \multirow{2}*{$0.2$} & \multirow{2}*{$0.018$}&\multirow{2}*{$0.02$}&\multirow{2}*{$0.018$}&\multirow{2}*{$0.017$}&\multirow{2}*{$0.02$} \\
 (mol/L)&&&&&&\\\hline
\label{Re2-t1}
\end{tabular}
\end{center}
\end{table} As the values of $c_T$ and $c_{imp}$ were approximately constant for different initial guesses, average values were used for these parameters. 

\begin{figure}[ht!]
\begin{center}
\includegraphics[width=0.4\textheight]{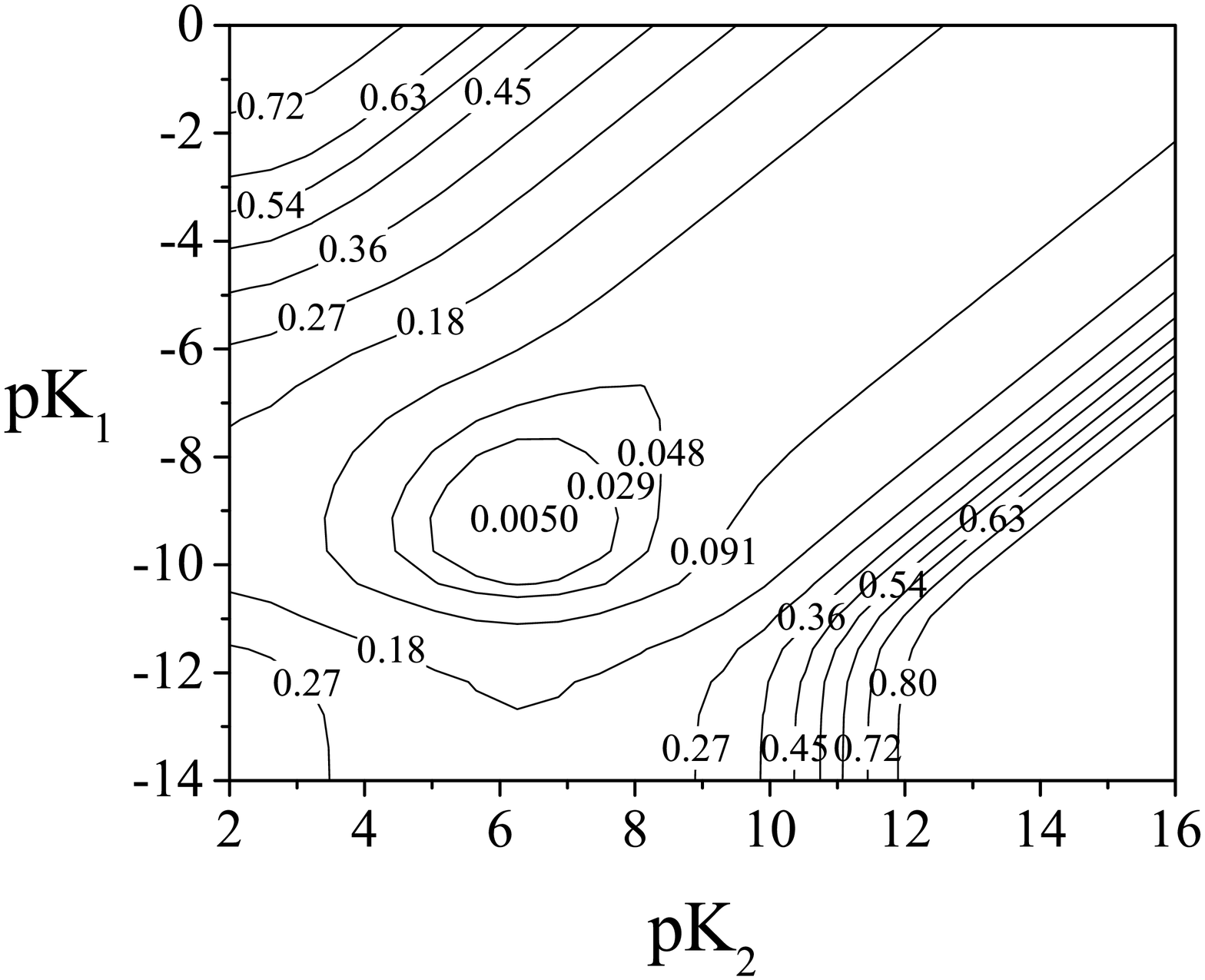}
\end{center}
\caption{Contour map showing the variation of the objective function $f_o$ with change in the values of pK$_1 = -$log$_{10} K_1$ and pK$_2= -$log$_{10} K_2$.}
\label{Re2-f14.1}
\end{figure}

\begin{figure}[ht!]
\begin{center}
\includegraphics[width=0.4\textheight]{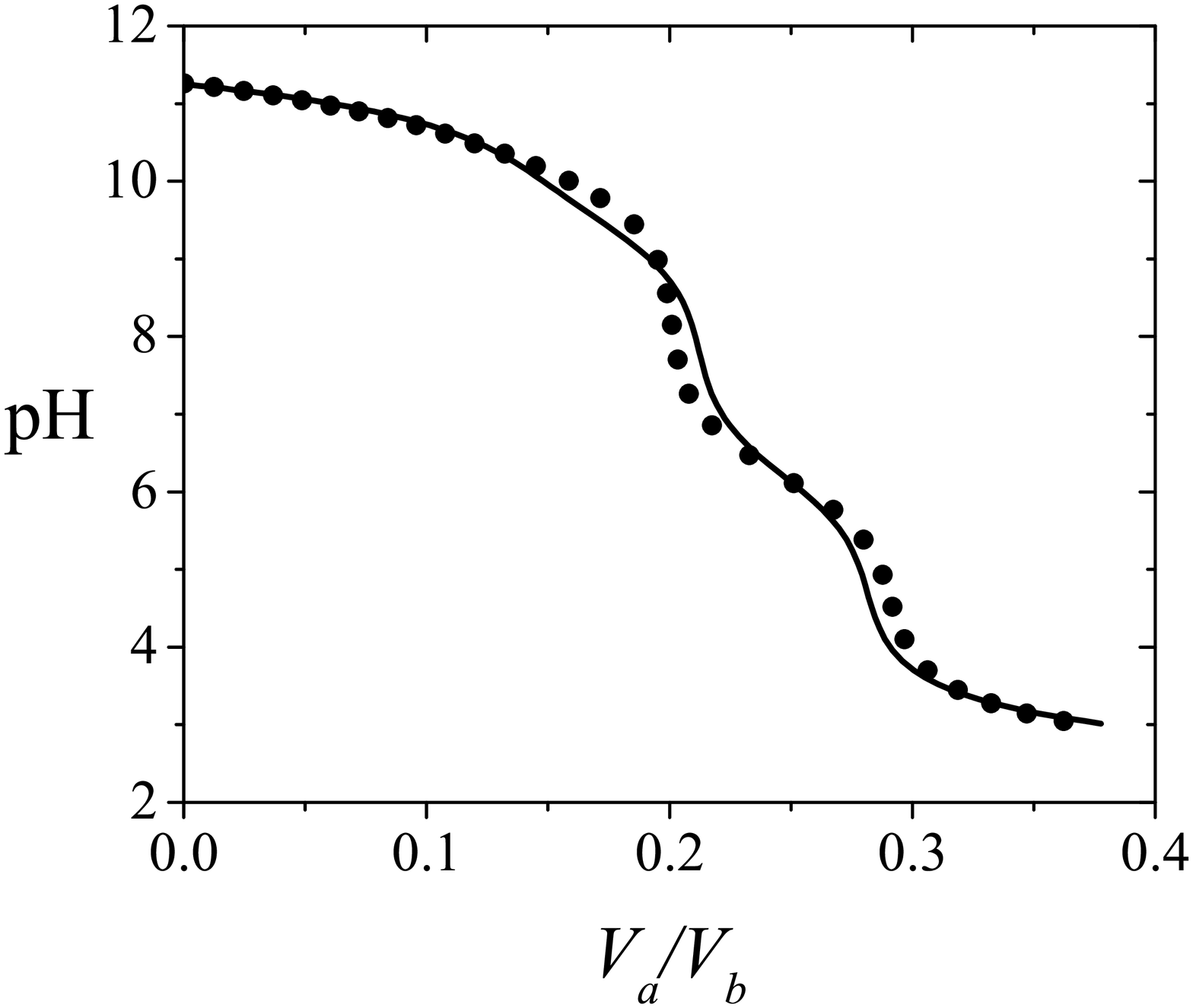}
\end{center}
\caption{Titration of a solution containing 0.02 N \ce{NaOH} and 0.25 g of AA with 0.02 N \ce{HCl}: {\huge$\fontdimen16\textfont2=2.5pt _\bullet$}, data; \protect\rule[0.5ex]{1cm}{1pt}, model. Here $V_b = 50$ mL is the volume of the base taken and $V_a$ is the volume of the acid added.}
\label{Re2-f15}
\end{figure}

\begin{figure}[ht!]
\begin{center}
\includegraphics[width=0.4\textheight]{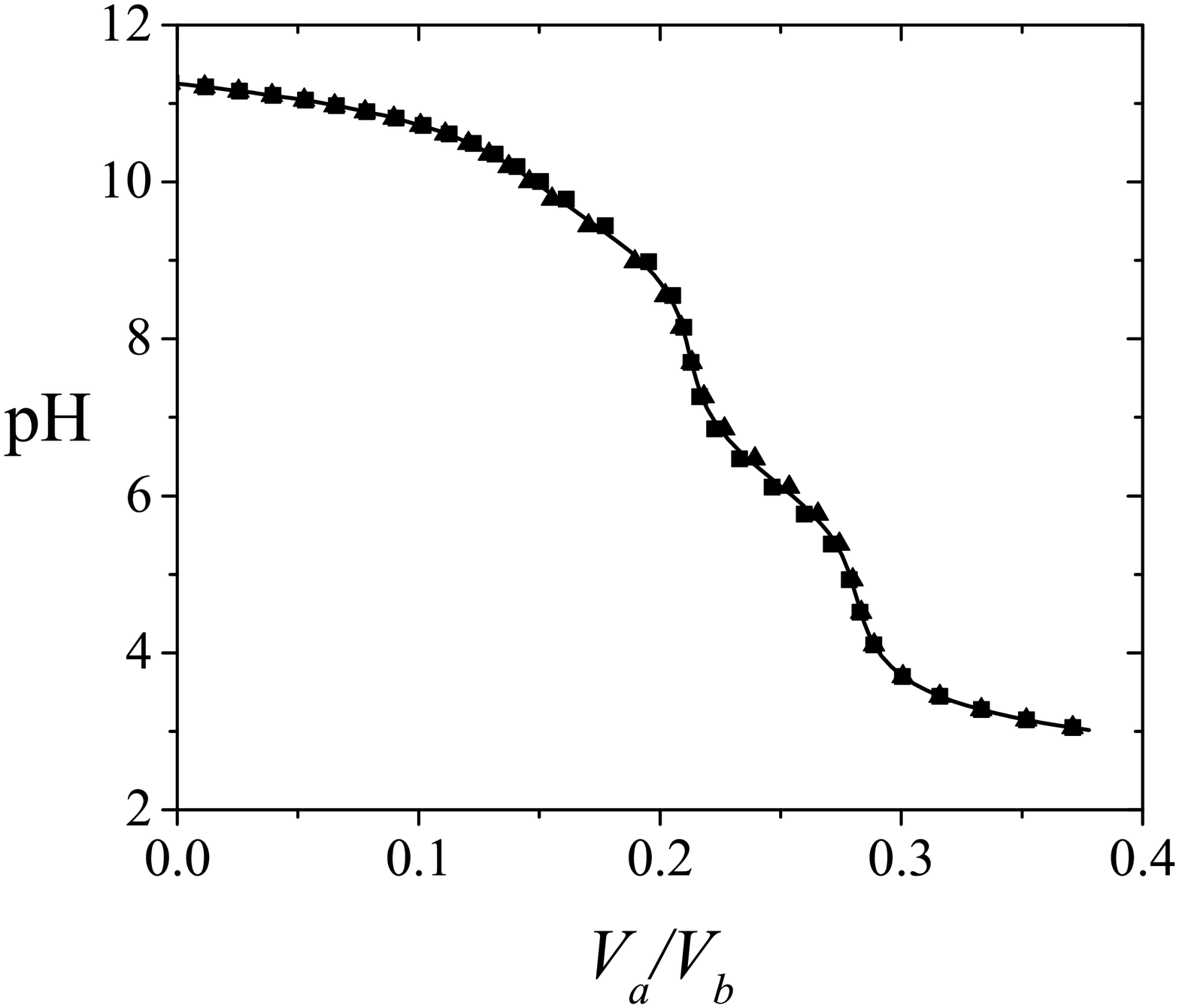}
\end{center}
\caption{Simulated titration curves for titration of a solution containing 0.02 N \ce{NaOH} and 0.25 g of AA with 0.02 N \ce{HCl}: \protect\rule[0.5ex]{1cm}{1pt}, $K_1$ and $K_2$ corresponding to the minimum of $f_o$ ($K_1 = 1.58 * 10^6 \equiv K_{1*}$, $K_2 = 3.98 * 10^{-10} \equiv K_{2*}$); $\filledmedtriangleup$, $K_2 = 0.8 K_{2*}$, $K_1 = 0.8 K_{1*}$; $\filledmedsquare$, $K_2 = 1.2 K_{2*}$, $K_2 = 1.2 K_{1*}$. Here $V_b = 50$ mL is the volume of the base taken and $V_a$ is the volume of the acid added.}
\label{Re2-f15.1}
\end{figure}

The constants are obtained by minimizing the objective function \begin{equation} \label{Re2-r1}
f_o = \sum_i (F(p,xdata_i)-ydata_i)^2
\end{equation} where $xdata_i$ and $ydata_i$ are the pH and the ratio of the volume of the acid added $V_a$ to the volume of base taken $V_b$, respectively, corresponding to the $i^{\text{th}}$ data point, and $F(p,xdata_i)$ is the predicted value of $V_a/V_b$, obtained using (\ref{Th-r8.1}). Here $p$ denotes the parameters that have to be estimated. Keeping $c_T$ and $c_{imp}$ fixed, contours of constant $f_o$ are plotted in the pK$_1$ - pK$_2$ plane, where pK$_i \equiv -$ log$_{10} K_i$. The minimum value of $f_o$ occurs at $K_1 = 1.58*10^{6}$ and $K_2 = 3.98*10^{-10}$. The poor performance of Lsqcurvefit may be because of a shallow region near the minimum (Fig.~\ref{Re2-f14.1}). A good fit was obtained with $f_o = 0.0036$ (Fig.~\ref{Re2-f15}). The amount of impurity in the adsorbent in the form of \ce{NO3-} i.e. $c_{imp}/\rho_b = 3.8$ mmol/g is more than the total number of sites ($c_T = 0.28$ mmol/g) on the adsorbent (Table~\ref{Re2-t1}). Thus \ce{NO3-} is probably not adsorbed, but remains in the liquid in the pores of the adsorbent. This is also consistent with our assumption that \ce{NO3-} and \ce{Cl-} are the ions which are present only in the double layer adjacent to the charged surface as a means to counter the charge of the surface. The obtained pK values i.e. 6.2 and 9.4, correspond to the range of values reported in literature \citep{Davis78,Ryazanov03,Ryazanov04}. Sensitivity analysis for the obtained equilibrium constants $K_1$ and $K_2$ was done by varying the values by 20\%. The simulated curves were similar to the values corresponding to the minimum $f_o$ (Fig.~\ref{Re2-f15.1}).

\begin{figure}[ht!]
\begin{center}
\includegraphics[width=0.4\textheight]{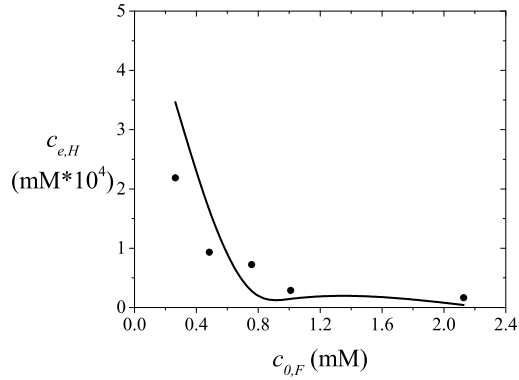}
\end{center}
\caption{Variation of the equilibrium concentration of \ce{H+} ($c_{e,H}$) with the initial concentration of \ce{F-} ($c_{0,F}$): {\huge$\fontdimen16\textfont2=2.5pt _\bullet$}, data; \protect\rule[0.5ex]{1cm}{1pt}, model predictions.}
\label{Re2-f16}
\end{figure}

\begin{figure}[ht!]
\begin{center}
\includegraphics[width=0.35\textheight]{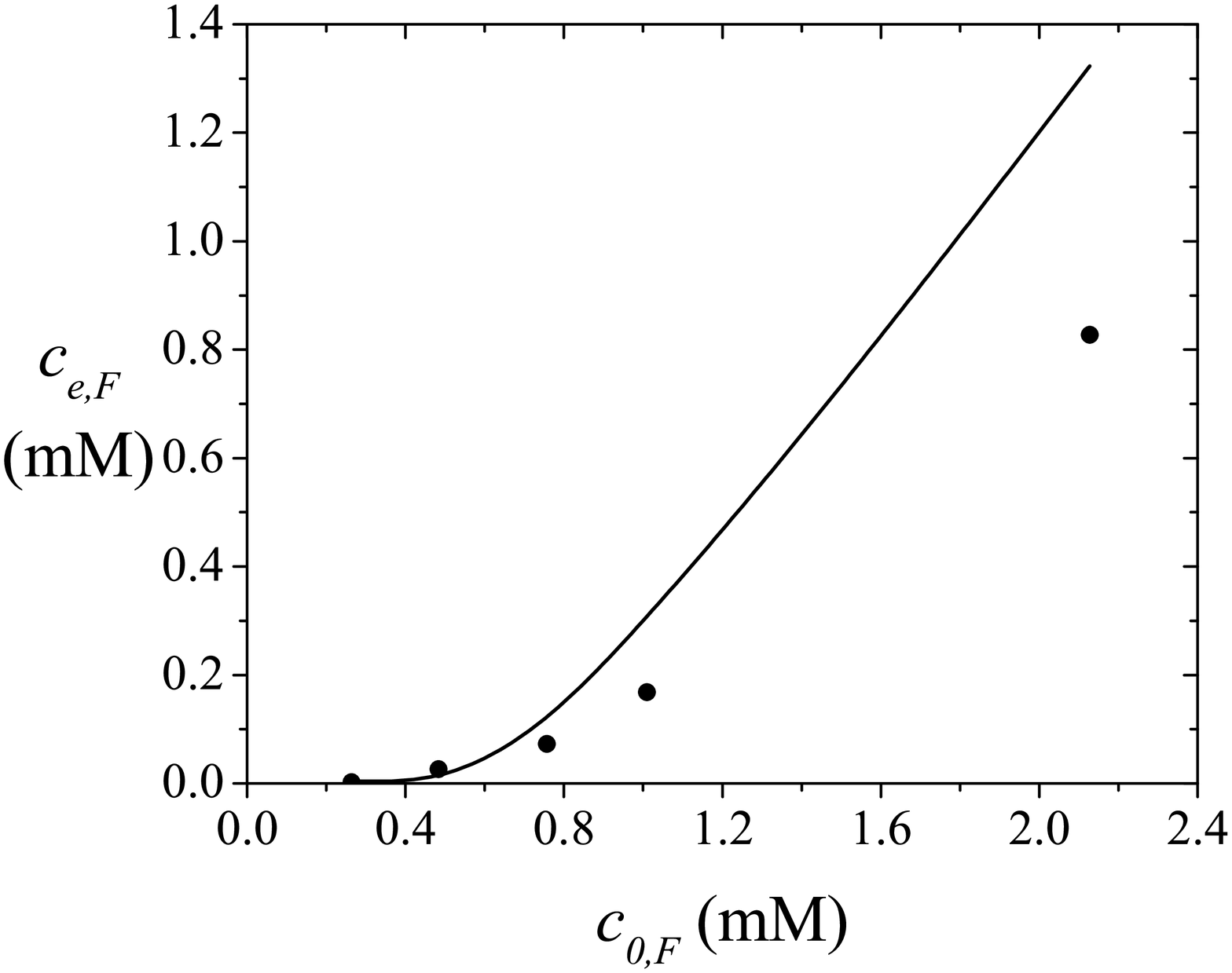}
\end{center}
\caption{Variation of the equilibrium concentration of \ce{F-} ($c_{e,F}$) with the initial concentration of \ce{F-} ($c_{0,F}$): {\huge$\fontdimen16\textfont2=2.5pt _\bullet$}, data; \protect\rule[0.5ex]{1cm}{1pt}, model predictions.}
\label{Re2-f17}
\end{figure}

\begin{figure}[ht!]
\begin{center}
\includegraphics[width=0.35\textheight]{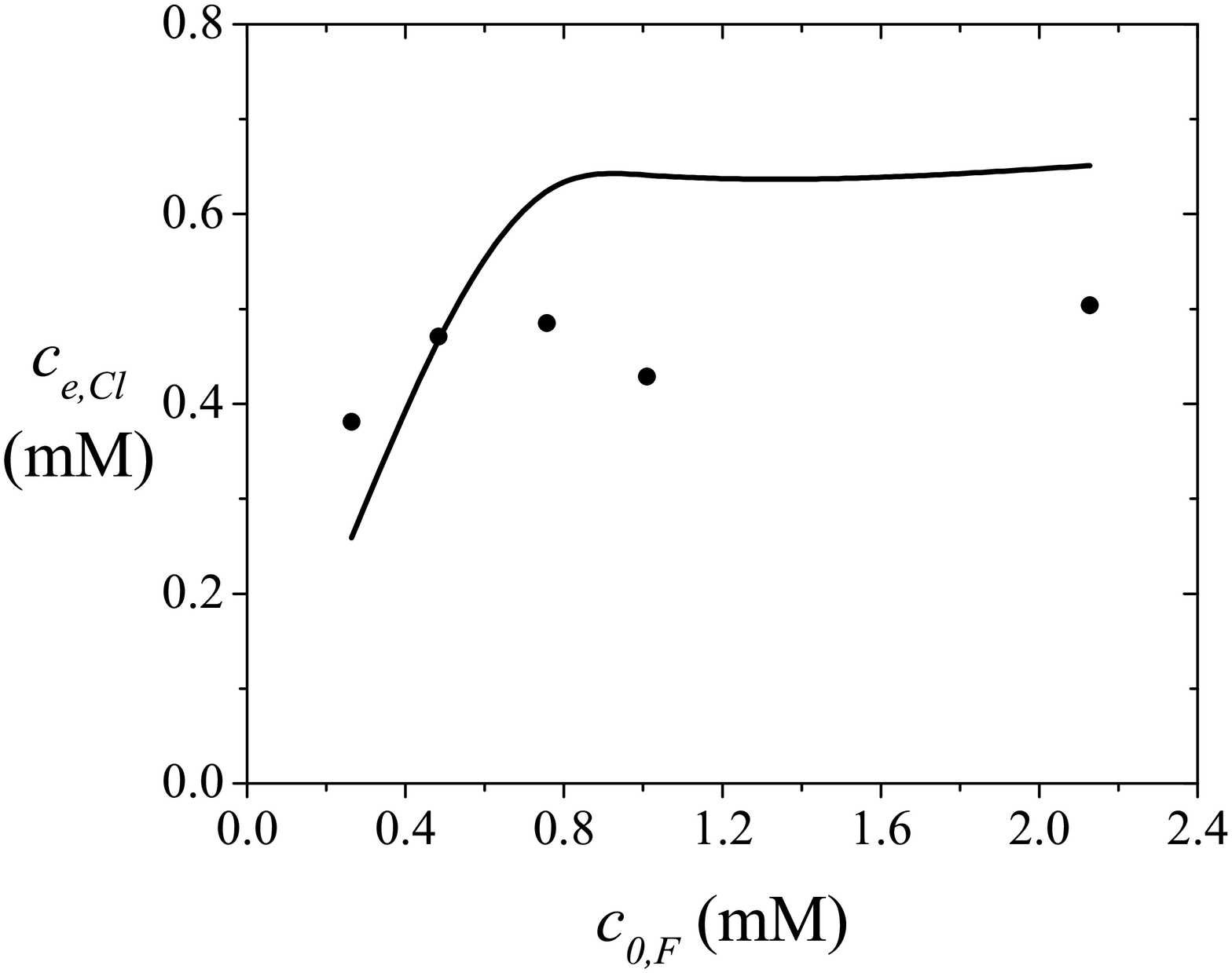}
\end{center}
\caption{Variation of the equilibrium concentration of \ce{Cl- + NO3-} ($c_{e,Cl}$) with the initial concentration of \ce{F-} ($c_{0,F}$): {\huge$\fontdimen16\textfont2=2.5pt _\bullet$}, data; \protect\rule[0.5ex]{1cm}{1pt}, model predictions.}
\label{Re2-f17.5}
\end{figure}

In order to estimate the equilibrium constants for the adsorption of ions such as \ce{F-} and \ce{HCO3-}, batch adsorption data and equilibrium values are required for a series of initial concentrations. Using the equilibrium data for \ce{F-}, \ce{Cl-} and \ce{H+} (Fig.~\ref{Re2-f16}), the equilibrium constant $K_3$ (corresponding to (\ref{Th-r4c})) was obtained by minimizing the objective function. During the adsorption experiment, it was observed that even after soaking AA in DI water there was still some amount of impurity present in the adsorbent. So, in order to account for this a parameter $q_{1i}$ (see (\ref{Th-r9b})) is also taken into account along with $K_3$. The objective function (\ref{Re2-r1}) was minimized using the MATLAB routine Lsqcurvefit and the parameters obtained were $K_3 = 1.66*10^2$ and $q_{1i} = 0.13$ mmol/g. The model ((\ref{Th-r9a}) - (\ref{Th-r11})) fits the data for \ce{H+} fairly well (Fig.~\ref{Re2-f16}),  but overestimates the data for \ce{F-} and \ce{Cl-} at higher concentrations (Figs.~\ref{Re2-f17} and \ref{Re2-f17.5}).   

\subsubsection{Estimation of the rate constants for the reactions (\ref{Th-r4a}) - (\ref{Th-r4c})}
The rate expressions for these reactions (\ref{Th-r4a}) - (\ref{Th-r4c}) are given as 
\begin{align}
r_1 = k_{f1}(c_1 q_2 - q_1/K_1) \label{Re2-r3} \\
r_2 = k_{f2}(q_2 - q_3 c_1/K_1) \label{Re2-r4}\\
r_3 = k_{f3}(q_1 c_2 - q_4/K_3) \label{Re2-r5}
\end{align} where $q_1, q_2, \text{and} q_3$ correspond to the adsorbate concentrations of \ce{\bond{3}AlOH2+Cl-}, \ce{\bond{3}AlOH}, and \ce{\bond{3}AlO^-Na+}, respectively. For the differential adsorber, the variation of the concentration of the ions with time can be obtained by integrating (\ref{Th-r56}) and (\ref{Th-r57}), and the mass balances for the adsorbates simultaneously. In order to solve these equations the values of the diffusivities $D_i$, the film thickness in bulk phase $\delta$ and in the pellet phase $\delta '$, the rate constants for the reactions ${k_{fi}}$, and the surface concentrations ${c_{j,s}}$ are required. The values of the diffusion coefficients were obtained from \cite{Cussler09} (Table~\ref{Re2-t2}). \begin{table}[ht!]
\begin{center}
\caption{Diffusion coefficients at infinite dilution used for different ions.}
\begin{tabular}{cc}
\\\hline Ion & $D_{iw}$ (m$^2$/s)\\\hline
\ce{H+} & $9.31 * 10^{-9}$ \\
\ce{Na+} & $1.33 * 10^{-9}$ \\
\ce{F-} & $1.47 * 10^{-9}$ \\
\ce{Cl-} & $2.03 * 10^{-9}$ \\
\ce{OH-} & $5.28 * 10^{-9}$ \\\hline
\label{Re2-t2}
\end{tabular}
\end{center}
\end{table} The film thickness in the bulk phase was obtained from the correlation \citep{Wakao78,Ruthven84} \begin{equation} \label{Re2-r2}
\delta = \dfrac{d_p}{2 + 1.1 \: \text{Sc}^{(1/3)} \: \text{Re}^{0.6}} \:, \qquad 3 < \text{Re} < 10^4
\end{equation} and in the pellet phase it was taken as $\delta ' = d_p/10$ \citep{Liaw79,Seader06}. Using (\ref{Th-r46a}) and (\ref{Th-r46b}), the the surface concentrations can be calculated for known concentrations in the bulk and the particle phases. This was obtained using a modified Gauss elimination method, which is discussed in \ref{app4}. 

\begin{figure}[htbp]
\begin{center}
\includegraphics[width=0.65\textheight]{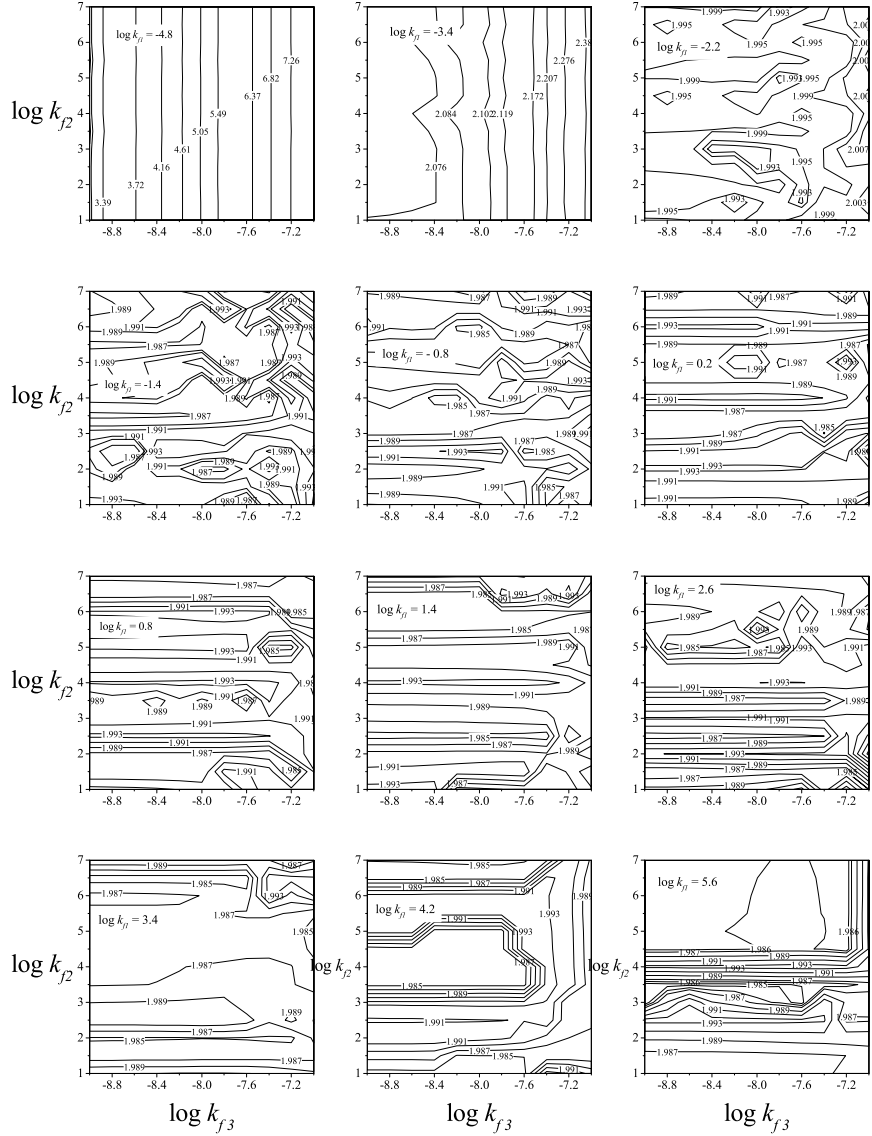}
\end{center}
\caption{Contour map showing the variation of the objective function $f_o$ with of $k_{f1}$, $k_{f2}$, and $k_{f3}$ when diffusional resistance was considered. Each panel corresponds to a fixed value of $k_{f1}$.}
\label{Re2-f24}
\end{figure}

\begin{figure}[ht!]
\begin{center}
\includegraphics[width=0.4\textheight]{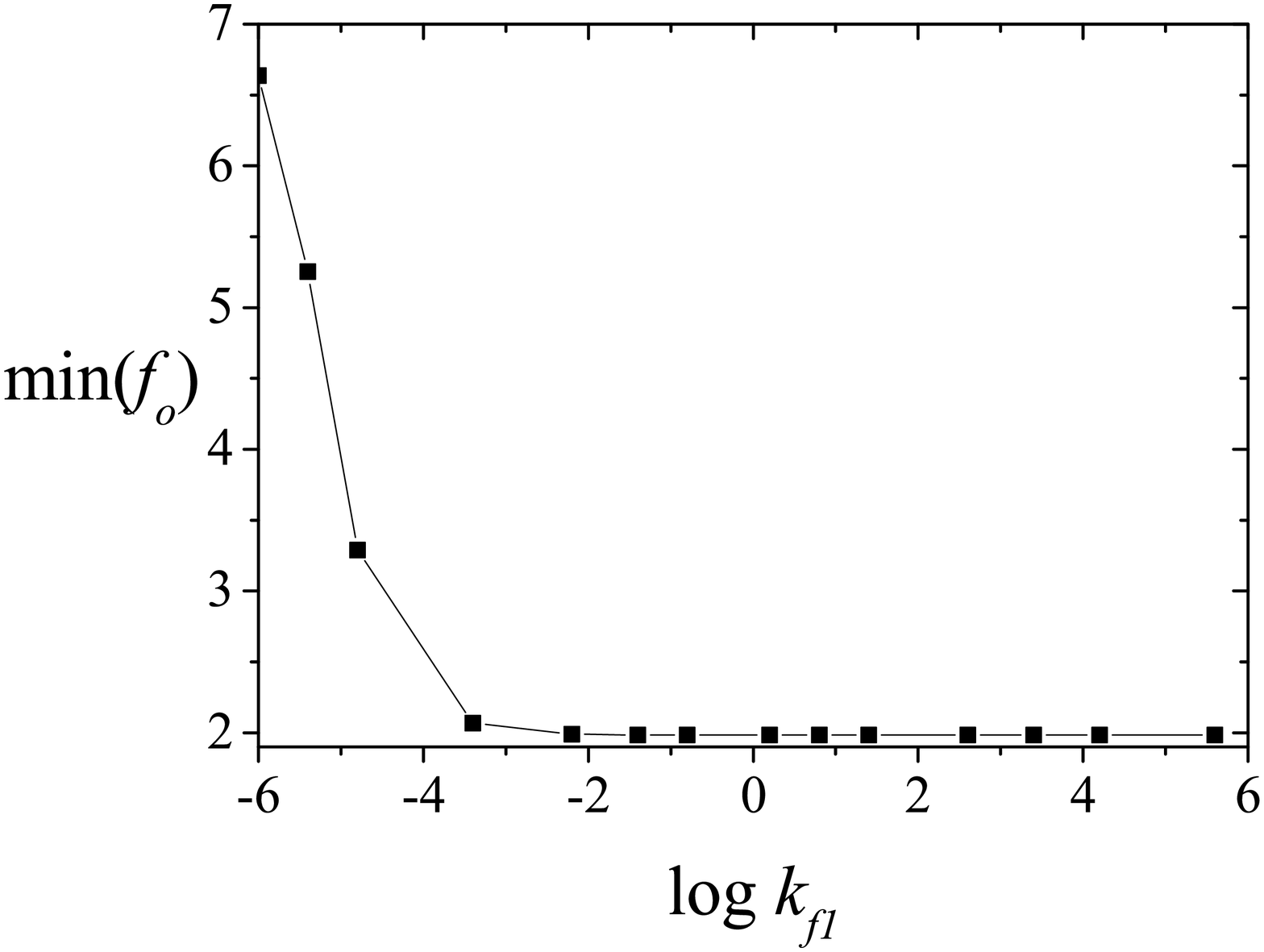}
\end{center}
\caption{Variation with $k_{f1}$ of the minimum value of the objective function min$(f_o)$ obtained for different values of $k_{f2}$ and $k_{f3}$.}
\label{Re2-f25}
\end{figure}

\begin{figure}[ht!]
\begin{center}
\includegraphics[width=0.425\textheight]{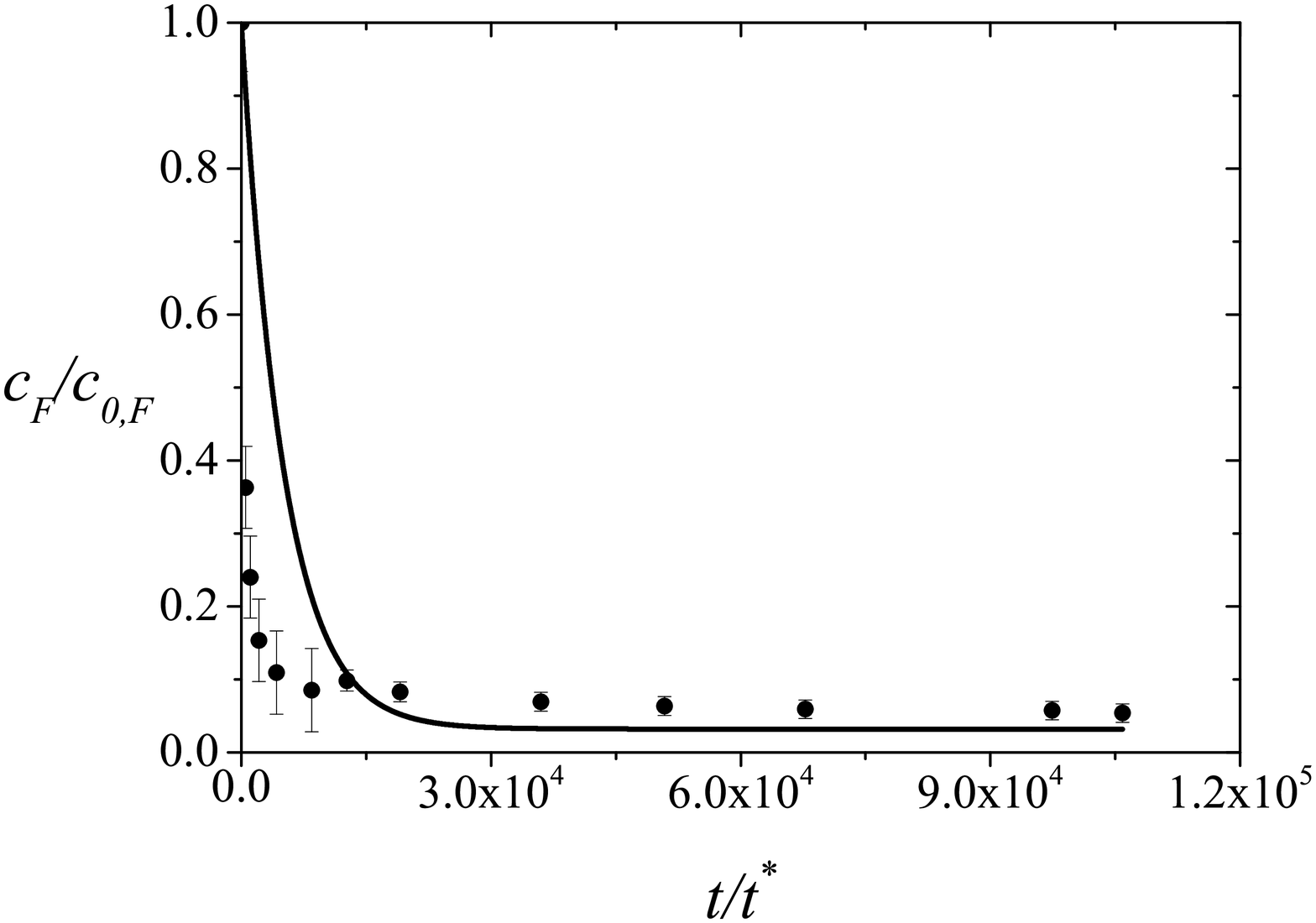}
\end{center}
\caption{Variation of the dimensionless concentration of \ce{F-} ($c_{F}/c_{0,F}$) with dimensionless time of operation ($t/t^*$): {\huge$\fontdimen16\textfont2=2.5pt _\bullet$}, data; \protect\rule[0.5ex]{1cm}{1pt}, model predictions. Here $c_{0,F} = 0.48$ mM is the initial concentrations of \ce{F-} and $t^* = 1.7$ s is the empty bed contact time.}
\label{Re2-f26}
\end{figure}

\begin{figure}[ht!]
\begin{center}
\includegraphics[width=0.425\textheight]{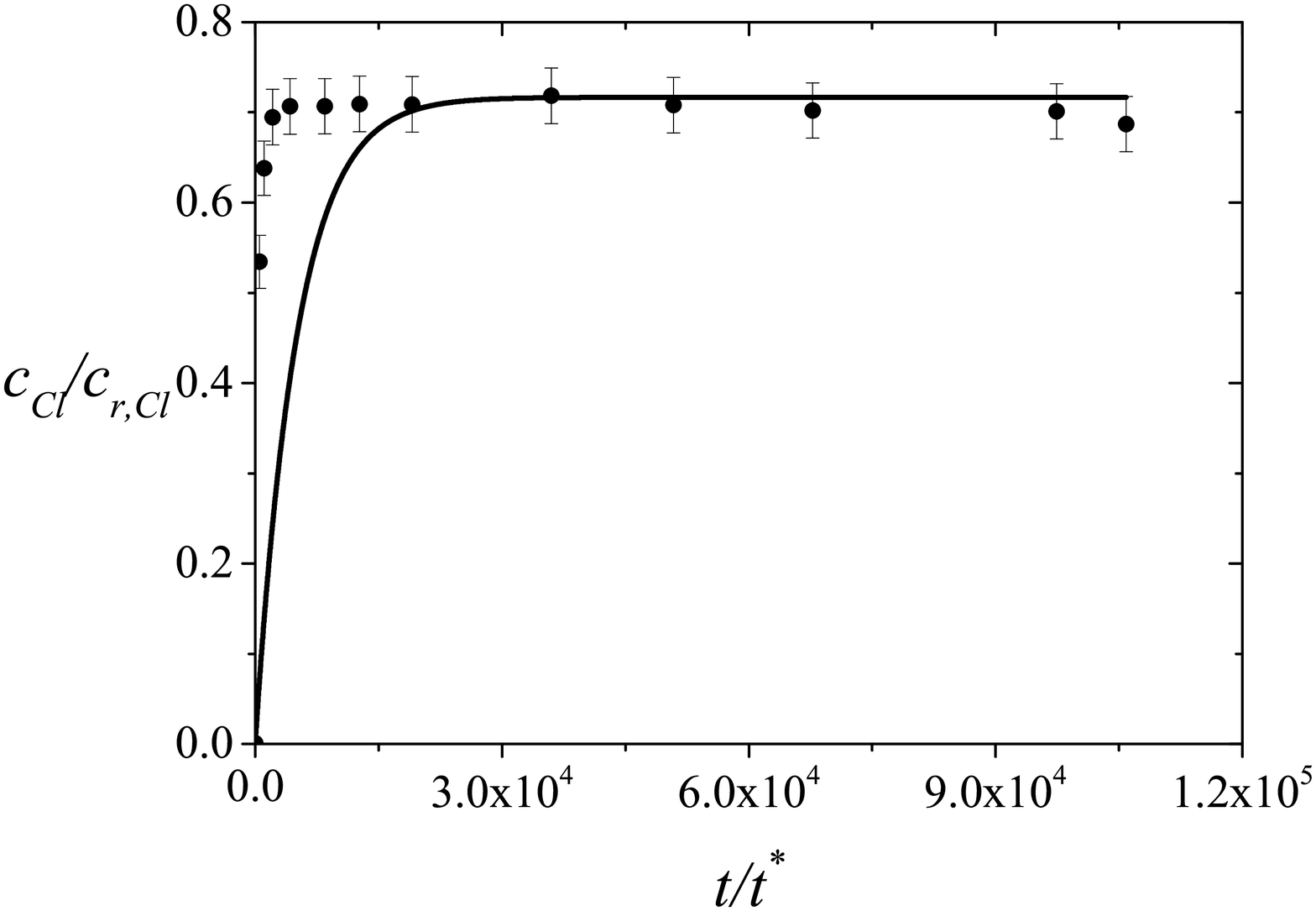}
\end{center}
\caption{Variation of the dimensionless concentration of \ce{Cl- + NO3-} ($c_{Cl}/c_{r,Cl}$) with dimensionless time of operation ($t/t^*$): {\huge$\fontdimen16\textfont2=2.5pt _\bullet$}, data; \protect\rule[0.5ex]{1cm}{1pt}, model predictions. Here $c_{0,F} = 0.48$ mM is the initial concentration of             \ce{F-}, $c_{r,Cl} = \rho_b q_{1i} = 0.655$ mM is the maximum amount of impurity present on the adsorbent, and $t^* = 1.7$ s is the empty bed contact time.}
\label{Re2-f27}
\end{figure}

\begin{figure}[htpb]
\begin{center}
\includegraphics[width=0.425\textheight]{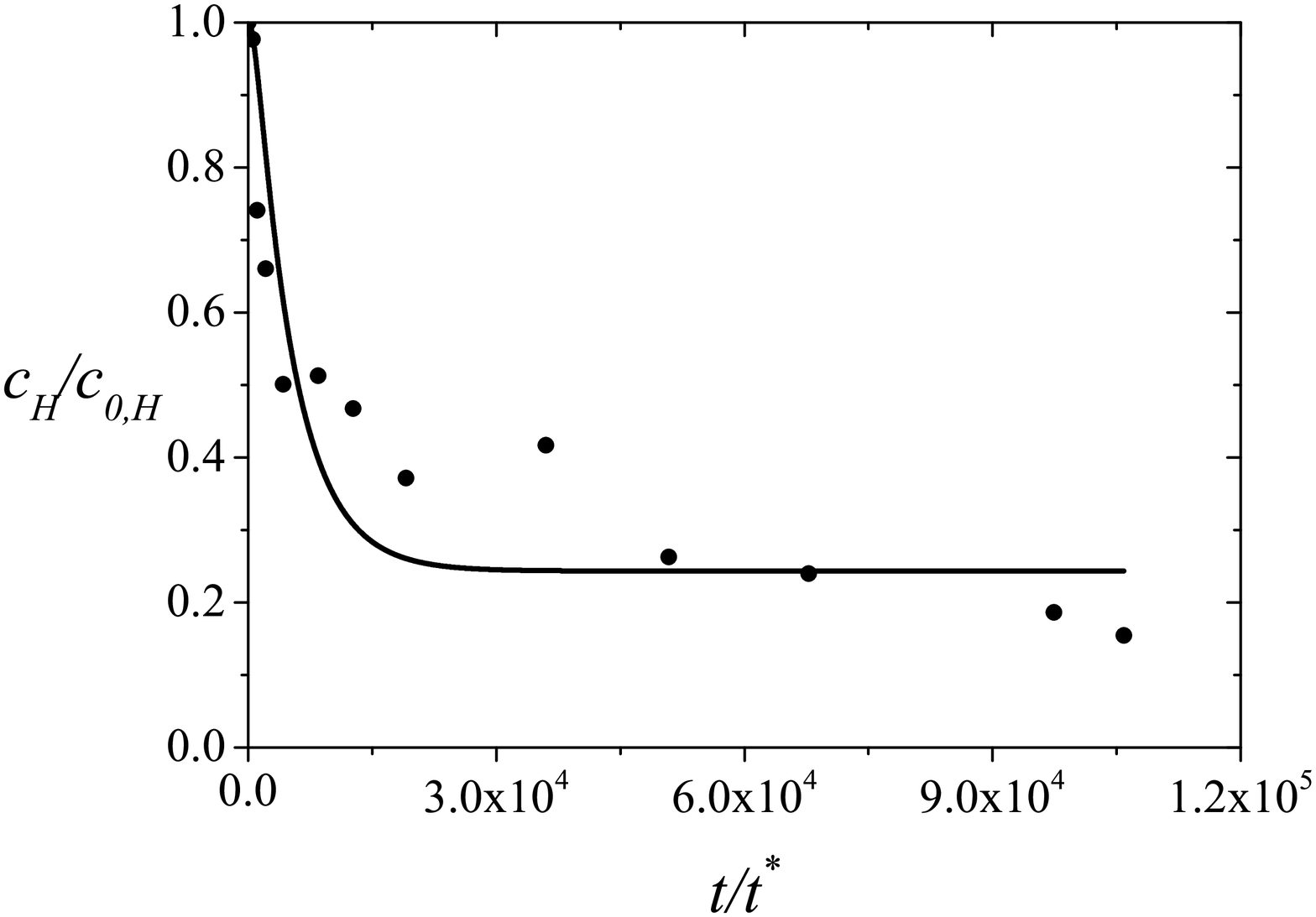}
\end{center}
\caption{Variation of the dimensionless concentration of \ce{H+} ($c_{H}/c_{0,H}$) with dimensionless time of operation ($t/t^*$): {\huge$\fontdimen16\textfont2=2.5pt _\bullet$}, data; \protect\rule[0.5ex]{1cm}{1pt}, model predictions. Here $c_{0,F} = 0.48$ mM, $c_{0,H} = 6.02 * 10^{-4}$ mM are the initial concentration of \ce{F-} and \ce{H+}, respectively, and $t^* = 1.7$ s is the empty bed contact time.}
\label{Re2-f28}
\end{figure}

As there are no predetermined values for the rate constants, an initial estimate was obtained for the special case of negligible diffusional resistance, and the rate constants were spanned over a range of values to obtain a minimum for the objective function. The minimum was obtained using an inhouse code . In the developed routine, the differential equations are solved using the MATLAB inbuilt routine ODE15s, which works well for a system of stiff equations. For an initial condition $c_{0,F} = 0.48$ mM and $c_{0,H} = 6.02 * 10^{-4}$ mM, the above equations were solved. Values of the objective function $f_o$, defined by an equation similar to (\ref{Re2-r1}), were plotted with respect to different values of the rate constants. Here, the $xdata_i$ and $ydata_i$ correspond to the dimensionless time and the experimental concentrations of all the ions at $i$, and $F(p,xdata_i)$ is the simulated dimensionless concentrations with parameters $p$. For each value of $k_{f1}$, contours of constant $f_o$ in the log$k_{f2}$ - log$k_{f3}$ plane are shown in Fig.~\ref{Re2-f24}. It was found that there was a minimum value of $f_o$ for each value of $k_{f1}$. This is denoted by min$(f_o)$, and its variation with $k_{f1}$ is shown in Fig.~\ref{Re2-f25}.  The minimum value of $f_o$ was obtained for the rate constants $k_{f1} = 1.59*10^{4}$ m$^3$mol$^{-1}$s$^{-1}$, $k_{f2} = 1.0*10^{-9}$ s$^{-1}$, and $k_{f3} = 3.16*10^{4}$ m$^3$mol$^{-1}$s$^{-1}$. Reasonably good fits were obtained for all the ions (Figs.~\ref{Re2-f26} - \ref{Re2-f28}).   

\begin{figure}[htpb]
\begin{center}
\includegraphics[width=0.425\textheight]{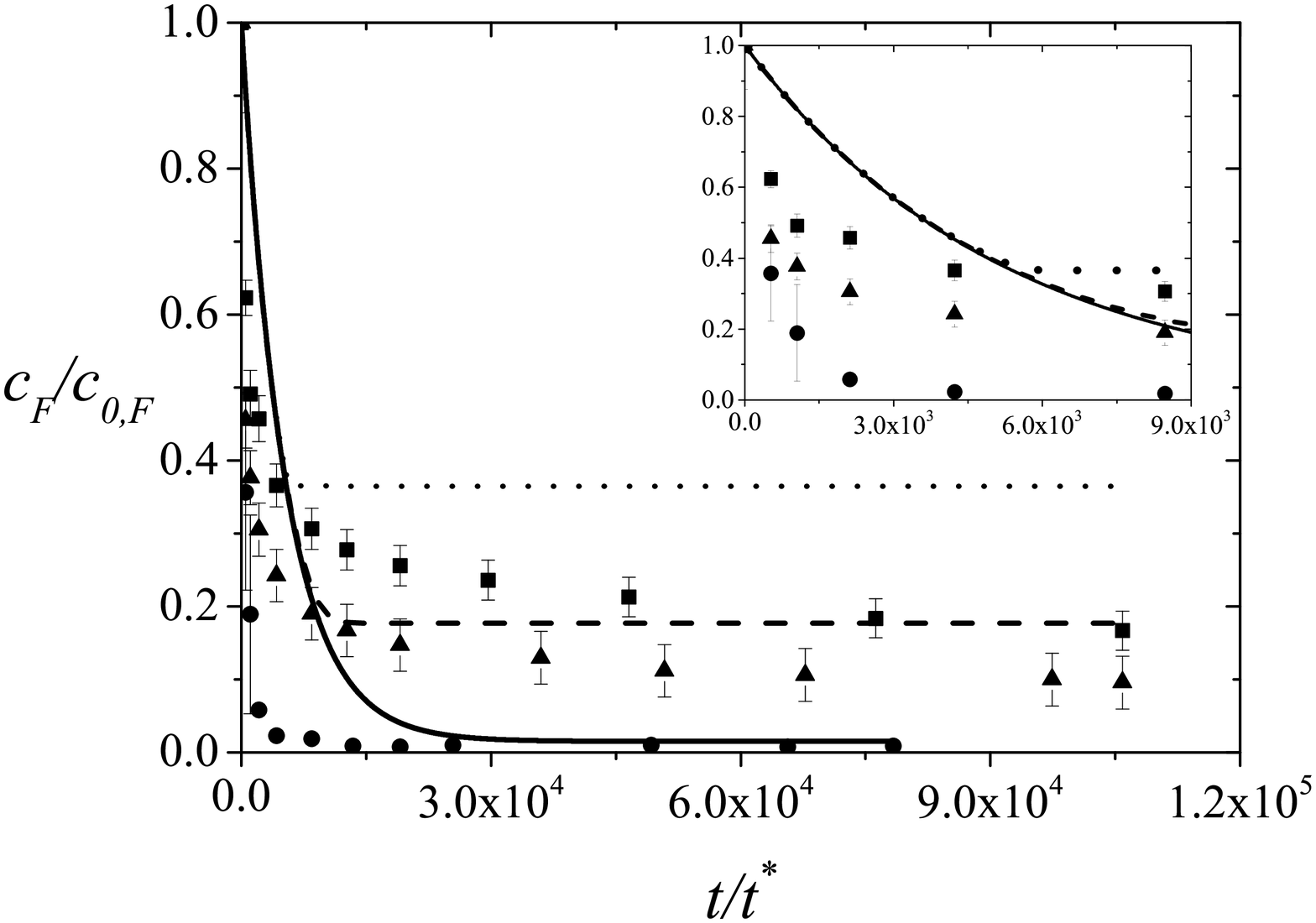}
\end{center}
\caption{Variation of the dimensionless concentration of \ce{F-} ($c_{F}/c_{0,F}$) with dimensionless time of operation ($t/t^*$): {\huge$\fontdimen16\textfont2=2.5pt _\bullet$}, \protect\rule[0.5ex]{1cm}{1pt}, $c_{0,F} = 0.26$ mM; $\filledmedtriangleup$, \hdashrule[0.5ex][c]{1cm}{1pt}{2pt}, $c_{0,F} = 0.79$ mM; $\filledmedsquare$, \hdashrule[0.5ex][c]{1cm}{1pt}{0.5pt 2pt}, $c_{0,F} = 1.05$ mM; symbols, data; curves, predictions. Here $c_{0,F}$ is the initial concentration of \ce{F-} and $t^* = 1.7$ s is the empty bed contact time.}
\label{Re2-f29}
\end{figure}

\begin{figure}[htpb]
\begin{center}
\includegraphics[width=0.42\textheight]{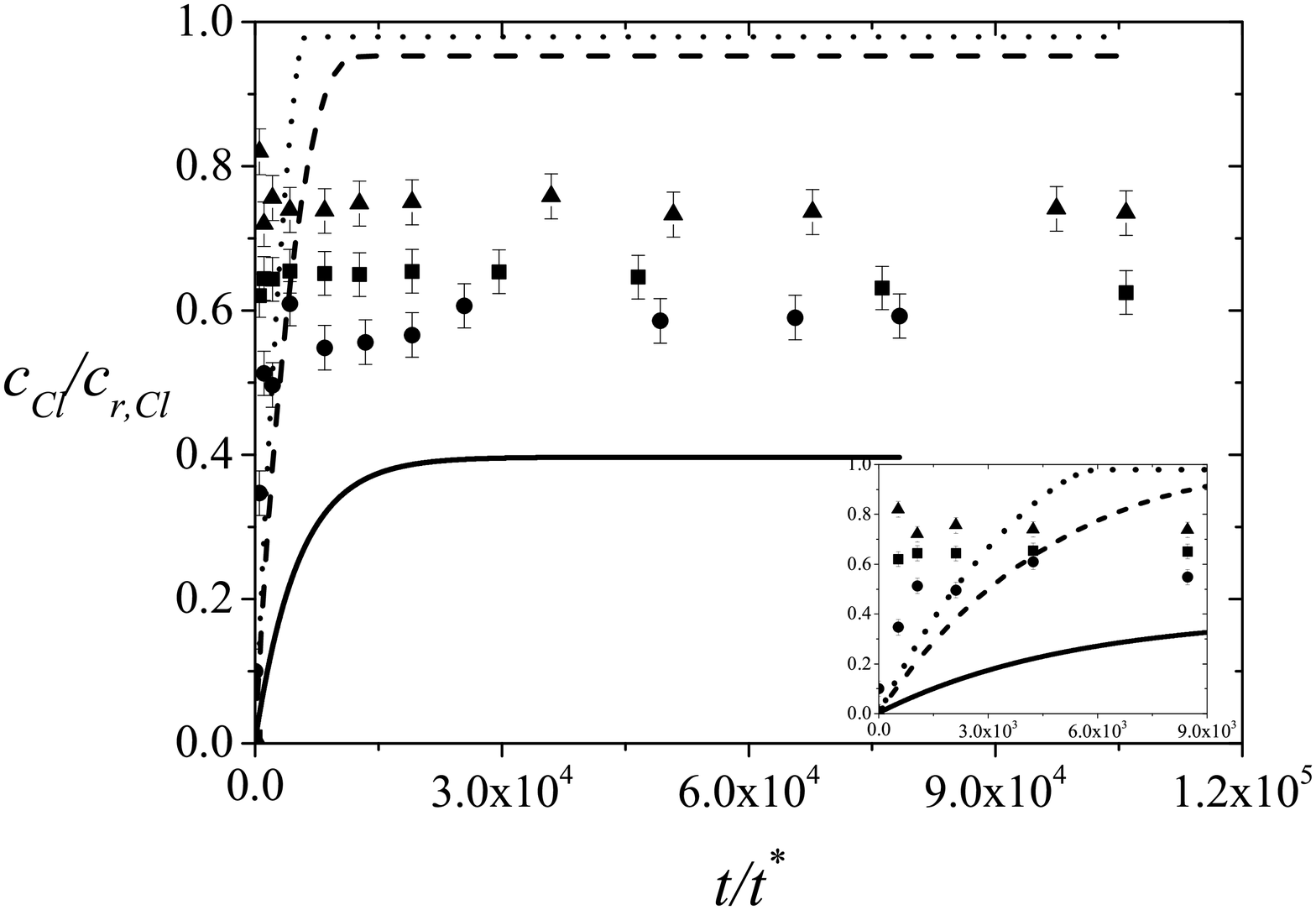}
\end{center}
\caption{Variation of the dimensionless concentration of \ce{Cl- + NO3-} ($c_{Cl}/c_{r,Cl}$) with dimensionless time of operation ($t/t^*$): {\huge$\fontdimen16\textfont2=2.5pt _\bullet$}, \protect\rule[0.5ex]{1cm}{1pt}, $c_{0,F} = 0.26$ mM; $\filledmedtriangleup$, \hdashrule[0.5ex][c]{1cm}{1pt}{2pt}, $c_{0,F} = 0.79$ mM; $\filledmedsquare$, \hdashrule[0.5ex][c]{1cm}{1pt}{0.5pt 2pt}, $c_{0,F} = 1.05$ mM; symbols, data; curves, predictions. Here $c_{r,Cl} = \rho_b q_{1i} = 0.655$ mM and $t^* = 1.7$ s is the empty bed contact time.}
\label{Re2-f30}
\end{figure}

\begin{figure}[htpb]
\begin{center}
\includegraphics[width=0.42\textheight]{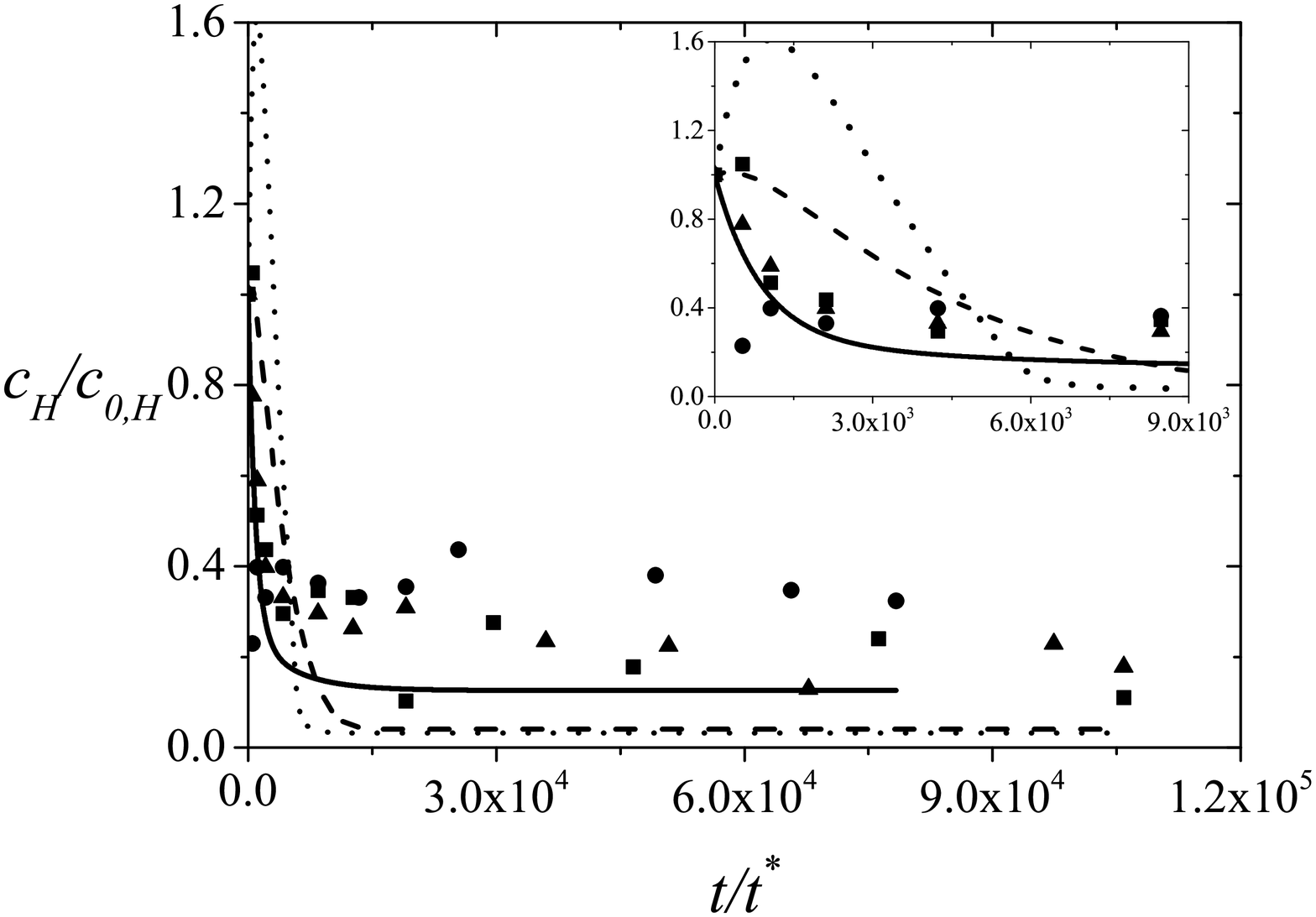}
\end{center}
\caption{Variation of the dimensionless concentration of \ce{H+} ($c_{H}/c_{0,H}$) with dimensionless time of operation ($t/t^*$): {\huge$\fontdimen16\textfont2=2.5pt _\bullet$}, \protect\rule[0.5ex]{1cm}{1pt}, $c_{0,F} = 0.26$ mM; $\filledmedtriangleup$, \hdashrule[0.5ex][c]{1cm}{1pt}{2pt}, $c_{0,F} = 0.79$ mM; $\filledmedsquare$, \hdashrule[0.5ex][c]{1cm}{1pt}{0.5pt 2pt}, $c_{0,F} = 1.05$ mM; symbols, data; curves, predictions. Here $c_{0,H}$ is the initial concentration of \ce{H+} and $t^* = 1.7$ s is the empty bed contact time.}
\label{Re2-f31}
\end{figure}

The predicted rate constants were used to simulate the concentration profiles for initial \ce{F-} concentrations of 0.26 mM, 0.79 mM, and 1.05 mM, but the profiles did not fit the data well (Figs.~\ref{Re2-f29} - \ref{Re2-f31}).    In Fig.~\ref{Re2-f30}, $c_{Cl}$ increases with time because \ce{NO3-} is released into the solution and \ce{Cl-} and \ce{NO3-} have been treated as interchangeable in the present work. There is an inherent error in the predictions at long times, as the theoretical equilibrium values for higher initial concentrations of \ce{F-} did not match the data (Figs.~\ref{Re2-f16} - \ref{Re2-f17.5}). The assumption of neutralizing the charge on the surface of pellet with a counter ion from the solution may be affecting the equilibrium and kinetics of ion adsorption.

\clearpage 
\section{Conclusions}
The adsorption of \ce{F-} on activated alumina is observed to be highly influenced by the presence of many ions like \ce{HCO3-}, \ce{SO4^{2-}}. Here single component adsorption of the ions was performed. There was adsorption of \ce{HCO3-} and \ce{SO4^{2-}}, but a negligible adsorption of \ce{Cl-}, \ce{NO3-}, and \ce{Na+} onto AA. When a two-component adsorption was performed with \ce{F-}, there was no influence of \ce{Cl-}, \ce{NO3-}, and \ce{Na+} on the uptake of \ce{F-}.

Equilibrium constants for the reactions (\ref{Th-r4a}) and (\ref{Th-r4b}) were estimated by fitting model predictions to data obtained by titration. A good fit was obtained for the variation of the pH with the volume of acid added.  The present model predicts the concentration profiles of the ions including \ce{H+}, whereas existing models specify the concentration of \ce{H+} and predict the equilibrium concentrations of the ion that is adsorbed. Here the formation of \ce{\bond{3}AlF} sites is dependent on all the sites and species in solution. For the reaction (\ref{Th-r4c}) i.e. the adsorption of \ce{F-}, a reasonable fit was obtained for the variation of equilibrium values of \ce{F-}, \ce{H+}, and \ce{Cl-} with the initial concentration of \ce{F-} ($c_{0,F}$) in the range 0.3 - 0.8 mM. However, for $c_{0,F}$ in the range of 0.8 - 2.0 mM, the agreement between the model predictions and the data was not good.

A model describing the kinetics of adsorption in a differential bed adsorber was developed, accounting for the various ions in solution. For $c_{0,F} = 0.48$ mM, rate constants for the reactions (\ref{Th-r4a}) - (\ref{Th-r4c}) were estimated by fitting predictions to data. For other values of $c_{0,F}$, the performance of the model was not good.

 \appendix


\section{Derivation of (\ref{Th-r47a}) and (\ref{Th-r47b})} \label{app5}
The derivation of (\ref{Th-r47a}) is discussed in this appendix. A similar approach can be used to derive (\ref{Th-r47b}). The flux terms used in the mass balance equations with the assumption of a linear driving force depend on the $n$ species present in the system. But the dependence on the $n$ species can be reduced to $n-2$ species with the use of charge neutrality condition and the assumption that the water dissociation reaction attains equilibrium very rapidly. The general flux term is given by \begin{equation} \label{Ap5-r1}
N_i = \sum_{j=1}^n \mathfrak{D}_{ij} \dfrac{\Delta c_j}{\delta}, \quad i =1,n
\end{equation} where $\mathfrak{D}_{ij} = D_i \delta_{ij} - \frac{t_i}{z_i}z_j D_j$, $\Delta c_j$ is the difference in the concentration of species in the phases separated by a film of thickness $\delta$. Let $i=1$ denote \ce{H+}, $i=n-1$ denote \ce{OH-}, and $i=n$ be the species eliminated by the charge neutrality condition. 

Using the assumption that the solution is electrically neutral, we have \begin{equation} \label{Ap5-r2}
\sum_{i=1}^n z_i c_i = 0
\end{equation} and substituting (\ref{Ap5-r2}) in (\ref{Ap5-r1}), we obtain \begin{equation} \label{Ap5-r3}
N_i = - \sum_{j=1}^{n-1} \xi_{nj} \mathfrak{D}_{in} \dfrac{\Delta c_j}{\delta} + \sum_{j=1}^{n-1} \mathfrak{D}_{ij} \dfrac{\Delta c_j}{\delta}
\end{equation} where $\xi_{ij} = z_j/z_i$. If $\mathbb{D}_{ij} = \dfrac{\mathfrak{D}_{ij} - \xi_{ij} \mathfrak{D}_{in}}{\delta}$, then (\ref{Ap5-r3}) can be rewritten as \begin{equation} \label{Ap5-r4}
N_i = \sum_{j=1}^{n-1} \mathbb{D}_{ij} \Delta c_j
\end{equation}

Now using the condition that the water reaction is very fast, we have \begin{equation} \label{Ap5-r4.1}
c_{n-1} = \dfrac{K_w}{c_1}
\end{equation} Equations (\ref{Ap5-r4}) and (\ref{Ap5-r4.1}) imply that \begin{equation} \label{Ap5-r5}
N_i = \mathbb{D}_{i,n-1} K_w \left(\dfrac{1}{c_{1,b}} - \dfrac{1}{c_{1,s}} \right) + \sum_{j=1}^{n-2} \mathbb{D}_{ij} \Delta c_j
\end{equation} \begin{equation} \nonumber
N_i = \mathbb{D}_{i,n-1} K_w \dfrac{c_{1,s} - c_{1,b}}{c_{1,b} c_{1,s}} + \sum_{j=1}^{n-2} \mathbb{D}_{ij} \Delta c_j
\end{equation} Let $\phi = \dfrac{K_w}{c_{1,b}}$, then \begin{equation} \nonumber
N_i = - \dfrac{\phi}{c_{1,s}} \mathbb{D}_{i,n-1} \Delta c_1 + \sum_{j=1}^{n-2} \mathbb{D}_{ij} \Delta c_j
\end{equation} \begin{equation} \label{Ap5-r6}
N_i = (\mathbb{D}_{i1} - \dfrac{\phi}{c_{1,s}} \mathbb{D}_{i,n-1}) \Delta c_1 + \sum_{j=2}^{n-2} \mathbb{D}_{ij} \Delta c_j, \quad i=1,n
\end{equation}

For $i =2,n-2$ and $i=n$, $N_i$ can be written in the simplified form (\ref{Th-r47a}), with $\lambda_{ij} = \mathbb{D}_{ij}$. For $i=1$ (\ref{Th-r33}) implies that \begin{equation} \nonumber
\begin{split}
N_1 - N_{n-1} &= (\mathbb{D}_{11} - \dfrac{\phi}{c_{1,s}} \mathbb{D}_{1,n-1}) \Delta c_1 + \sum_{j=2}^{n-2} \mathbb{D}_{1j} \Delta c_j \\
&- (\mathbb{D}_{n-1,1} - \dfrac{\phi}{c_{1,s}} \mathbb{D}_{n-1,n-1}) \Delta c_1 - \sum_{j=2}^{n-2} \mathbb{D}_{n-1,j} \Delta c_j
\end{split}
\end{equation} If $\lambda_{1j} = \mathbb{D}_{1j} - \mathbb{D}_{n-1,j}$, then \begin{equation} \label{Ap5-r7}
N_1 - N_{n-1} = - \dfrac{\phi}{c_{1,s}} \lambda_{1,n-1} \Delta c_1 + \sum_{j=1}^{n-2} \lambda_{1j} \Delta c_j
\end{equation}

Equation (\ref{Ap5-r7}) can be written in terms of an effective mass transfer coefficient as \begin{equation} \label{Ap5-r8}
N_i = \sum_{j=1}^{n-2} k_{b,ij} \Delta c_j
\end{equation} where $k_{b,ij}$ is given by (\ref{Th-r48}). Using a similar approach, expressions for $k_{p,ij}$, which are shown in (\ref{Th-r48}), can be derived.

\section{Modified Gauss elimination to calculate the surface concentration} \label{app4}
Calculation of the surface concentration is dependent on the concentrations of species in the bulk and the particle. As there is no accumulation of the mass in the films at the interfaces, there is a continuity in the flux from the bulk to particle. So, we obtain (\ref{Th-r46a}) and (\ref{Th-r46b}), and based on the derivation given in Appendix~\ref{app5}, they can be modified as \begin{equation} \label{Ap4-r1}
\sum_{j=1}^{n-2} k_{b,ij} (c_{j,b} - c_{j,s}) = \sum_{j=1}^{n-2} k_{p,ij} (c_{j,s} - \bar{c}_{j,p})
\end{equation} Upon rearranging the terms we obtain \begin{equation} \label{Ap4-r2}
\sum_{j=1}^{n-2} c_{j,s} (k_{b,ij} + k_{p,ij}) = \sum_{j=1}^{n-2} (k_{b,ij} c_{j,b} + k_{p,ij} \bar{c}_{j,p})
\end{equation} Equation (\ref{Ap4-r2}) can be represented in a matrix form by introducing effective mass transfer coefficients matrices $[k_b]$ and $[k_p]$ of size $[n-2 \;\text{x}\; n-2]$ and concentration vectors $c_s, c_p, c_b$ of size $n-2$. The resulting matrix form will be \begin{equation} \label{Ap4-r3}
[k_b + k_p] c_s = [k_b] c_b + [k_p] c_p
\end{equation} 

Equation (\ref{Ap4-r3}) is of the form $A X = B$ and can be solved using Gauss elimination to obtain the unknown vector $X$ if $A$ and $B$ are independent of the components of $X$. In the present case, the matrices are dependent on $c_{1,s}$. So, (\ref{Ap4-r3}) is modified to solve for the vector $c_s$. 

Using (\ref{Th-r48}) for any $i$, from (\ref{Ap4-r3}) we can obtain \begin{equation} \label{Ap4-r4}
\begin{split}
- (\phi \lambda_{i,n-1} + \phi' \lambda_{i,n-1}') &+ \sum_{j=1}^{n-2} (\lambda_{ij} + \lambda_{ij}') c_{j,s} = - (\phi \lambda_{i,n-1} c_{1,b} + \phi' \lambda_{i,n-1}' \bar{c}_{1,p}) \dfrac{1}{c_{1,s}} \\
&+ \sum_{j=1}^{n-2} (\lambda_{ij} c_{j,b} +  \lambda_{ij}' \bar{c}_{1,p})   
\end{split}
\end{equation} Upon rearranging and multiplying both sides by $c_{1,s}$ we obtain \begin{equation} \label{Ap4-r5}
\begin{split}
(\lambda_{i1} + \lambda_{i1}') c_{1,s}^2 &+ \sum_{j=2}^{n-2} (\lambda_{ij} + \lambda_{ij}') c_{1,s} c_{j,s} + \left[ - \sum_{j=1}^{n-2} (\lambda_{ij} c_{j,b} + \lambda_{ij}' \bar{c}_{j,p}) - \phi \lambda_{i,n-1} - \phi' \lambda_{i,n-1}' \right] c_{1,s} \\ 
&+ \phi \lambda_{i,n-1} c_{1,b} + \phi' \lambda_{i,n-1}' \bar{c}_{1,p} = 0
\end{split}
\end{equation} 
If \begin{align}
\tau_{i,j-1} &= \lambda_{ij} + \lambda_{ij}'; \quad j =2,n-2; \quad i =1,n-2\,\&\,n \\
\tau_{i,n-2} &= \lambda_{i1} + \lambda_{i1}'; \quad i =1,n-2\,\&\,n  \\
\tau_{i,n-1} &= - \sum_{j=1}^{n-2} (\lambda_{ij} c_{j,b} + \lambda_{ij}' \bar{c}_{j,p}) - \phi \lambda_{i,n-1} - \phi' \lambda_{i,n-1}'; \quad i =1,n-2\,\&\,n \\ \tau_{v,i} &= - (\phi \lambda_{i,n-1} c_{1,b} + \phi' \lambda_{i,n-1}' \bar{c}_{1,p}); \quad i =1,n-2\,\&\,n
\end{align} then (\ref{Ap4-r5}) becomes \begin{equation} \label{Ap4-r6}
\sum_{j=2}^{n-2} (\tau_{i,j-1} c_{j,s} c_{1,s}) + \tau_{i,n-2} c_{1,s}^2 + \tau_{i,n-1} c_{1,s} = \tau_{v,i}
\end{equation}

Now $[\tau]$ is a matrix of dimension $[n-2\;\text{x}\;n-1]$ and $\tau_v$ is a vector of dimension $n-2$. Using (\ref{Ap4-r6}), a new unknown vector $P$ can be defined with elements being $c_{j,s}c_{1,s}(j=2,n-2), c_{1,s}^2$, and $c_{1,s}$. Therefore, (\ref{Ap4-r6}) takes the form $\tau P = \tau_v$. As $\tau$ is a rectangular matrix, the last row will have two elements when Gaussian elimination is used. Upon multiplication of the matrices and vectors after elimination, the last row becomes a quadratic in $c_{1,s}$. With the selection of a suitable root for $c_{1,s}$ which satisfies $c_{1,s} \geq 0$, the values $c_{j,s}, j=2,n-2$ can be computed.

\section*{Acknowledgements}
The authors would like to thank Dr. Jean Riotte for his help with the titration experiments and Indo-French Cell for Water Sciences, IISc-IRD Joint International Laboratory, IISc, Bengaluru for providing the equipment.

\section*{References}

\begin{thebibliography}{49}
\expandafter\ifx\csname natexlab\endcsname\relax\def\natexlab#1{#1}\fi
\expandafter\ifx\csname url\endcsname\relax
  \def\url#1{\texttt{#1}}\fi
\expandafter\ifx\csname urlprefix\endcsname\relax\def\urlprefix{URL }\fi

\bibitem[{Al-Abadleh and Grassian(2003)}]{Al03}
Al-Abadleh, H.~A., Grassian, V., 2003. {FT-IR} study of water adsorption on
  aluminum oxide surfaces. Langmuir 19, 341--347.

\bibitem[{Appel et~al.(2013)Appel, Rhue, Kabengi, and Harris}]{Appel13}
Appel, C., Rhue, D., Kabengi, N., Harris, W., 2013. Calorimetric investigation
  of the nature of sulfate and phosphate sorption on amorphous aluminum
  hydroxide. Soil Sci. 178, 180--188.

\bibitem[{Bachet et~al.(2014)Bachet, Jauberty, De~Windt, Tevissen,
  De~Dieuleveult, and Schneider}]{Bachet14}
Bachet, M., Jauberty, L., De~Windt, L., Tevissen, E., De~Dieuleveult, C.,
  Schneider, H., 2014. Comparison of mass transfer coefficient approach and
  {N}ernst--{P}lanck formulation in the reactive transport modeling of \ce{Co},
  \ce{Ni}, and \ce{Ag} removal by mixed-bed ion-exchange resins. Ind. Eng.
  Chem. Res. 53, 11096--11106.

\bibitem[{Benjamin(2002)}]{Benjamin02}
Benjamin, M.~M., 2002. Water {C}hemistry. McGraw-Hill, New York.

\bibitem[{Bird et~al.(2002)Bird, Stewart, and Lightfoot}]{Bird02}
Bird, R.~B., Stewart, W.~E., Lightfoot, E.~N., 2002. Transport phenomena. John
  Wiley \& Sons.

\bibitem[{Bockris et~al.(2006)Bockris, Reddy, and Gamboa-Aldeco}]{Bockris06}
Bockris, J., Reddy, A., Gamboa-Aldeco, M., 2006. Modern {E}lectrochemistry
  2{A}: Fundamentals of {E}lectrodics. Springer International edition.

\bibitem[{Carrier et~al.(2007)Carrier, Marceau, Lambert, and Che}]{Carrier07}
Carrier, X., Marceau, E., Lambert, J.-F., Che, M., 2007. Transformations of
  $\gamma$-alumina in aqueous suspensions: 1. alumina chemical weathering
  studied as a function of p{H}. J. Colloid Interf. Sci. 308, 429--437.

\bibitem[{Charmas et~al.(1995)Charmas, Piasecki, and Rudzinski}]{Charmas95}
Charmas, R., Piasecki, W., Rudzinski, W., 1995. Four layer complexation model
  for ion adsorption at electrolyte/oxide interface: Theoretical foundations.
  Langmuir 11, 3199--3210.

\bibitem[{Chatterjee and De(2014)}]{Chatterjee14}
Chatterjee, S., De, S., 2014. Adsorptive removal of fluoride by activated
  alumina doped cellulose acetate phthalate {(CAP)} mixed matrix membrane. Sep.
  Purif. Technol. 125, 223--238.

\bibitem[{Curtiss and Bird(1999)}]{Curtiss99}
Curtiss, C., Bird, R.~B., 1999. Multicomponent diffusion. Ind. Eng. Chem. Res.
  38, 2515--2522.

\bibitem[{Cussler(2009)}]{Cussler09}
Cussler, E.~L., 2009. Diffusion: {M}ass {T}ransfer in {F}luid {S}ystems.
  Cambridge University Press.

\bibitem[{Davis et~al.(1978)Davis, James, and Leckie}]{Davis78}
Davis, J.~A., James, R.~O., Leckie, J.~O., 1978. Surface ionization and
  complexation at the oxide/water interface: {I.} {C}omputation of electrical
  double layer properties in simple electrolytes. J. Colloid Interface Sci. 63,
  480--499.

\bibitem[{Dou et~al.(2011)Dou, Zhang, Wang, Wang, and Wang}]{Dou11}
Dou, X., Zhang, Y., Wang, H., Wang, T., Wang, Y., 2011. Performance of granular
  zirconium--iron oxide in the removal of fluoride from drinking water. Water
  Res. 45, 3571--3578.

\bibitem[{Eaton et~al.(2005)Eaton, Clesceri, Rice, and Greenberg}]{Standards}
Eaton, A.~D., Clesceri, L.~S., Rice, E.~W., Greenberg, A.~E., 2005. Standard
  {M}ethods for the {E}xamination of {W}ater and {W}astewater. American Public
  Health Association (APHA), Washington, DC, USA.

\bibitem[{Fletcher et~al.(2006)Fletcher, Smith, and Pivonka}]{Fletcher06}
Fletcher, H., Smith, D., Pivonka, P., 2006. Modeling the sorption of fluoride
  onto alumina. J. Environ. Eng.-ASCE 132, 229--246.

\bibitem[{Frey(1986)}]{Frey86}
Frey, D.~D., 1986. Prediction of liquid-phase mass-transfer coefficients in
  multicomponent ion exchange: comparison of matrix, film-model, and
  effective-diffusivity methods. Chem. Eng. Commun. 47, 273--293.

\bibitem[{George et~al.(2010)George, Pandit, and Gupta}]{George10}
George, S., Pandit, P., Gupta, A., 2010. Residual aluminium in water
  defluoridated using activated alumina adsorption--modeling and simulation
  studies. Water Res. 44, 3055--3064.

\bibitem[{Gleuckauf(1955)}]{Gleuckauf55}
Gleuckauf, E., 1955. Theory of chromatography part 10: Formula for diffusion
  into spheres and their applications in chromatography. T. Faraday Soc. 51,
  1540--1551.

\bibitem[{Goswami and Purkait(2012)}]{Goswami12}
Goswami, A., Purkait, M.~K., 2012. The defluoridation of water by acidic
  alumina. Chem. Eng. Res. Des. 90, 2316 -- 2324.

\bibitem[{Gu et~al.(1997)Gu, Wang, and Liaw}]{Gu97}
Gu, W., Wang, C., Liaw, B., 1997. Numerical modeling of coupled electrochemical
  and transport processes in lead-acid batteries. J. Electrochem. Soc. 144,
  2053--2061.

\bibitem[{Hao and Huang(1986)}]{Hao86}
Hao, O.~J., Huang, C., 1986. Adsorption characteristics of fluoride onto
  hydrous alumina. J. Environ. Eng. ASCE 112, 1054--1069.

\bibitem[{Islam and Patel(2007)}]{Islam07}
Islam, M., Patel, R., 2007. Evaluation of removal efficiency of fluoride from
  aqueous solution using quick lime. J. Hazard. Mater. 143, 303--310.

\bibitem[{Jia and Foutch(2004)}]{Jia04}
Jia, Y., Foutch, G.~L., 2004. True multi-component mixed-bed ion-exchange
  modeling. React. Funct. Polym. 60, 121--135.

\bibitem[{Jin et~al.(2010)Jin, Qian, Lu, Yang, and Bi}]{Jin10}
Jin, X., Qian, Z., Lu, B., Yang, W., Bi, S., 2010. Density functional theory
  study on aqueous aluminum-fluoride complexes: Exploration of the intrinsic
  relationship between water-exchange rate constants and structural parameters
  for monomer aluminum complexes. Environ. Sci. Technol. 45, 288--293.

\bibitem[{Kanwar(2010)}]{Lalitha10}
Kanwar, L., 2010. Defluoridation of drinking water using activated alumina.
  Master's thesis, Indian Institute of Science.

\bibitem[{Krishna(1987)}]{Krishna87}
Krishna, R., 1987. Diffusion in multicomponent electrolyte systems. Chem. Eng.
  J. 35, 19--24.

\bibitem[{Krishna and Wesselingh(1997)}]{Krishna97}
Krishna, R., Wesselingh, J., 1997. The {M}axwell--{S}tefan approach to mass
  transfer. Chem. Eng. Sci. 52, 861--911.

\bibitem[{Lefevre et~al.(2002)Lefevre, Duc, Lepeut, Caplain, and
  F{\'e}doroff}]{Lefevre02}
Lefevre, G., Duc, M., Lepeut, P., Caplain, R., F{\'e}doroff, M., 2002.
  Hydration of $\gamma$-alumina in water and its effects on surface reactivity.
  Langmuir 18, 7530--7537.

\bibitem[{Liaw et~al.(1979)Liaw, Wang, Greenkorn, and Chao}]{Liaw79}
Liaw, C., Wang, J., Greenkorn, R., Chao, K., 1979. Kinetics of fixed-bed
  adsorption: {A} new solution. AIChE J. 25, 376--381.

\bibitem[{Mondal and George(2015)}]{Mondal15}
Mondal, P., George, S., 2015. A review on adsorbents used for defluoridation of
  drinking water. Rev. Environ. Sci. Biotechnol. 14, 195--210.

\bibitem[{Moreira et~al.(2006)Moreira, Soares, Casarin, and
  Rodrigues}]{Moreira06}
Moreira, R., Soares, J., Casarin, G., Rodrigues, A., 2006. Adsorption of
  \ce{CO2} on {H}ydrotalcite--like compounds in a fixed bed. Separ. Sci.
  Technol. 41, 341--357.

\bibitem[{Nagashima and Blum(1999)}]{Nagashima99}
Nagashima, K., Blum, F.~D., 1999. Proton adsorption onto alumina: {E}xtension
  of multisite complexation {(MUSIC)} theory. J. Colloid Interf. Sci. 217,
  28--36.

\bibitem[{Newman and Thomas-Alyea(2012)}]{Newman12}
Newman, J., Thomas-Alyea, K.~E., 2012. Electrochemical systems. John Wiley \&
  Sons, New Jersey.

\bibitem[{Nigussie et~al.(2007)Nigussie, Zewge, and Chandravanshi}]{Nigussie07}
Nigussie, W., Zewge, F., Chandravanshi, B., 2007. Removal of excess fluoride
  from water using waste residue from alum manufacturing process. J. Hazard.
  Mater. 147, 954 -- 963.

\bibitem[{Nordin et~al.(1999)Nordin, Sullivan, Phillips, and Casey}]{Nordin99}
Nordin, J.~P., Sullivan, D.~J., Phillips, B.~L., Casey, W.~H., 1999. Mechanisms
  for fluoride-promoted dissolution of bayerite [$\beta$-\ce{Al(OH)3} (s)] and
  boehmite [$\gamma$-\ce{AlOOH}]: \ce{^{19}F}-{NMR} spectroscopy and aqueous
  surface chemistry. Geochim. Cosmochim. Acta 63, 3513--3524.

\bibitem[{Okamoto and Imanaka(1988)}]{Okamoto88}
Okamoto, Y., Imanaka, T., 1988. Interaction chemistry between molybdena and
  alumina: {I}nfrared studies of surface hydroxyl groups and adsorbed carbon
  dioxide on aluminas modified with molybdate, sulfate, or fluorine anions. J.
  Phys. Chem. 92, 7102--7112.

\bibitem[{Ruthven(1984)}]{Ruthven84}
Ruthven, D.~M., 1984. Principles of Adsorption and Adsorption {P}rocesses. John
  Wiley \& Sons.

\bibitem[{Ryazanov and Dudkin(2003)}]{Ryazanov03}
Ryazanov, M., Dudkin, B., 2003. Acid--base properties of $\gamma$--\ce{Al2O3}
  suspension studied by p{K} spectroscopy. Colloid J+ 65, 761--766.

\bibitem[{Ryazanov and Dudkin(2004)}]{Ryazanov04}
Ryazanov, M., Dudkin, B., 2004. The p{K} spectroscopy study of the acid--base
  properties of hydrated aluminum oxide sols. Colloid J+ 66, 726--728.

\bibitem[{Seader and Henley(2006)}]{Seader06}
Seader, J.~D., Henley, E.~J., 2006. Separation process principles. Wiley India.

\bibitem[{Sircar and Hufton(2000)}]{Sircar00}
Sircar, S., Hufton, J., 2000. Why does the linear driving force model for
  adsorption kinetics work? Adsorption 6, 137--147.

\bibitem[{Snedecor and Cochran(1968)}]{Snedecor68}
Snedecor, G.~W., Cochran, W.~G., 1968. Statistical Methods. Oxford and IBH
  Publishing Co.

\bibitem[{Su and Suarez(1997)}]{Su97}
Su, C., Suarez, D., 1997. In situ infrared speciation of adsorbed carbonate on
  aluminum and iron oxides. Clays Clay Miner. 45, 814 -- 825.

\bibitem[{Tang et~al.(2009)Tang, Guan, Su, Gao, and Wang}]{Tang09}
Tang, Y., Guan, X., Su, T., Gao, N., Wang, J., 2009. Fluoride adsorption onto
  activated alumina: Modeling the effects of p{H} and some competing ions.
  Colloid. Surface. A 337, 33--38.

\bibitem[{Tefera et~al.(2014)Tefera, Hashisho, Philips, Anderson, and
  Nichols}]{Tefera14}
Tefera, D.~T., Hashisho, Z., Philips, J.~H., Anderson, J.~E., Nichols, M.,
  2014. Modeling competitive adsorption of mixtures of volatile organic
  compounds in a fixed--bed of beaded activated carbon. Environ. Sci. Technol.
  48, 5108--5117.

\bibitem[{Tjaden et~al.(2016)Tjaden, Cooper, Brett, Kramer, and
  Shearing}]{Tjaden16}
Tjaden, B., Cooper, S.~J., Brett, D.~J., Kramer, D., Shearing, P.~R., 2016. On
  the origin and application of the bruggeman correlation for analysing
  transport phenomena in electrochemical systems. Curr. Opin. Chem. Eng. 12,
  44--51.

\bibitem[{Wakao and Funazkri(1978)}]{Wakao78}
Wakao, N., Funazkri, T., 1978. Effect of fluid dispersion coefficients on
  particle-to-fluid mass transfer coefficients in packed beds: correlation of
  sherwood numbers. Chem. Eng. Sci. 33, 1375--1384.

\bibitem[{Wijnja and Schulthess(1999)}]{Wijnja99}
Wijnja, H., Schulthess, C., 1999. {ATR--FTIR} and {DRIFT} spectroscopy of
  carbonate species at the aged $\gamma$--\ce{Al2O3}/water interface.
  Spectrochim. Acta A 55, 861--872.

\bibitem[{Winkler and Thodos(1971)}]{Winkler71}
Winkler, B.~F., Thodos, G., 1971. Kinetics of orthophosphate removal from
  aqueous solutions by activated alumina. J. Water Pollut. Control Fed. 43,
  474--482.

\end{thebibliography}
\bibliographystyle{elsarticle-harv} 
\biboptions{authoryear}




\end{document}